\documentclass[a4paper,11pt]{article}
\pdfoutput=1 

\usepackage{bbold}
\usepackage{amsmath}
\usepackage{amssymb}
\usepackage{dsfont}
\usepackage{graphicx}
\usepackage{xspace}
\usepackage[multiple]{footmisc}
\usepackage{bbm} 
\usepackage{bm}
\usepackage{color}
\usepackage[utf8]{inputenc}
\usepackage{cancel}
\usepackage{slashed}
\usepackage{soul}
\usepackage{units}
\usepackage{enumitem}
\usepackage{makecell}
\usepackage{nicefrac}
\usepackage{mathtools}
\usepackage{subfig}
\usepackage{float}
\usepackage{cite}

\usepackage{tikz} 
\usetikzlibrary{chains,shapes,fit,calc}
\usetikzlibrary{positioning,arrows}

\usetikzlibrary{patterns,decorations.pathreplacing}
\usetikzlibrary{decorations.pathmorphing}
\usetikzlibrary{decorations.markings}
\usetikzlibrary{arrows,shapes}
\usetikzlibrary{shapes.misc}
\tikzset{
    gluon/.style={decorate, draw=black,
        decoration={coil,amplitude=4pt, segment length=4pt,aspect=0.7}} 
}
\tikzset{
    photon/.style={decorate, decoration={snake}},
}

\newcommand{\DRb}{\overline{\mbox{DR}}}
\newcommand{\al}{\alpha}
\newcommand{\beq}{\begin{eqnarray}}
\newcommand{\eeq}{\end{eqnarray}}
\newcommand{\ba}{\begin{eqnarray}}
\newcommand{\ea}{\end{eqnarray}}
\newcommand{\be}{\begin{equation}}
\newcommand{\ee}{\end{equation}}
\newcommand{\bpmatrix}{\begin{pmatrix}}
\newcommand{\epmatrix}{\end{pmatrix}}

\newcommand{\lsim}{\raisebox{-0.13cm}{~\shortstack{$<$ \\[-0.07cm]
      $\sim$}}~}

\newcommand\NMSSMCALC{{\tt NMSSMCALC}\xspace}
\newcommand\SARAH{{\tt SARAH}\xspace}

\newcommand\Fortran{{\tt Fortran}\xspace}
\newcommand\Mathematica{{\tt Mathematica}\xspace}

\newcommand\higgss{{\tt HiggsSignals}\xspace}

\newcommand\TARCER{{\tt TARCER}\xspace}
\newcommand\TSIL{{\tt TSIL}\xspace}
\newcommand\FeynArts{{\tt FeynArts}\xspace}
\newcommand\FeynCalc{{\tt FeynCalc}\xspace}

\newcommand{\order}[1]{\mathcal{O}(#1)}

\newcommand{\vev}{\textit{v}}

\newcommand{\vs}{\textit{v}_\textit{\tiny S}}
\newcommand{\MSbar}{\overline{\text{MS}}}
\newcommand{\DRbar}{\overline{\text{DR}}}
\newcommand{\OS}{\text{OS}}
\newcommand{\lnbar}{\overline{\log}}
\newcommand{\lnMR}{\overline{\log}  M_R^2 }

\newcommand{\del}{\partial}

\newcommand{\doublet}[2]{\begin{pmatrix} #1 \\ #2 \end{pmatrix}}

\newcommand{\ccdot}{\!\cdot\!}

\newcommand{\cw}{c_{\theta_W}}
\newcommand{\sw}{s_{\theta_W}}

\renewcommand{\Re}{\text{Re}}
\renewcommand{\Im}{\text{Im}}

\newcommand{\s}{\newline \vspace*{-3.5mm}}

\newcommand{\cb}{c_\beta}

\renewcommand{\sb}{s_\beta}

\newcommand{\tb}{t_{\beta}}

\newcommand{\mhpm}{M_{H^\pm}}
\newcommand{\mueff}{\mu_{\text{eff}}}

\newcommand{\tablegraph}[1]{\begin{minipage}{0.2\textwidth}\vspace{0.2mm}\includegraphics[width=\linewidth]{#1}\end{minipage}}

\newcommand{\diag}{\text{diag}}
\newcommand{\bea}{\begin{eqnarray}}
\newcommand{\eea}{\end{eqnarray}}
\newcommand{\crn}{\nonumber \\}
\newcommand{\braket}[1]{\left(#1\right)}
\newcommand{\sbraket}[1]{\left[#1\right]}
\newcommand{\tol}{{{\tiny(}\!{\tiny1}\!{\tiny)}\!}}
\newcommand{\ttl}{{{\tiny(}\!{\tiny2}\!{\tiny)}\!}}

\newcommand{\dnl}{\delta^{{\tiny(}\!{\tiny n}\!{\tiny)}\!}}

\newcommand{\deltatwo}{\delta^{ {\tiny(}\!{\tiny2}\!{\tiny)}}\!}
\newcommand{\deltaone}{\delta^{ {\tiny(}\!{\tiny1}\!{\tiny)}}\!}
\newcommand{\deltan}{\delta^{ {\tiny(}\!{\tiny n}\!{\tiny)}}\!}

\newcommand{\calR}{{\mathcal R}}
\newcommand{\calM}{{\mathcal M}}
\newcommand{\fr}{\frac}
\newcommand{\dol}{\delta^{{\tiny(}\!{\tiny1}\!{\tiny)}}\!}
\newcommand{\wave}{\dol\mathcal{Z}}
\newcommand{\wavetwo}{\deltatwo\mathcal{Z}}

\newcommand{\waven}{\dnl\mathcal{Z}}

\renewcommand{\eqref}[1]{Eq.~(\ref{#1})}

\RequirePackage[colorlinks=false
  ,linktocpage=true
  ,pdfproducer=medialab
  ,pdfa=true
]{hyperref}
\usepackage[capitalise, english]{cleveref}

\crefname{section}{Sec.}{Secs.}
\crefname{table}{Tab.}{Tabs.}

\graphicspath{{figs/}}
\makeatletter
\def\input@path{{chapters/}}
\makeatother

\begin{document}
\vspace{1cm}

\title{
\vspace*{-3cm}
\phantom{h} \hfill\mbox{\small IFIRSE-TH-2021-2} 
\\[-1.1cm]
\phantom{h} \hfill\mbox{\small KA-TP-11-2021}\\[-1.1cm]
	\phantom{h} \hfill\mbox{\small P3H-21-045}
	\\[1cm]
Two-Loop 
  $\order{\left(\alpha_t+\alpha_\lambda+\alpha_\kappa\right)^2}$
  Corrections to the Higgs Boson Masses in the CP-Violating NMSSM}

\newcommand{\AddrKAITP}{
Institute for Theoretical Physics (ITP), 
Karlsruhe Institute of Technology, \\
Wolfgang-Gaede-Stra{\ss}e 1, D-76131 Karlsruhe, Germany}

\author{
Thi Nhung Dao$^{1\,}$\footnote{E-mail:
  \texttt{dtnhung@ifirse.icise.vn}},
Martin
Gabelmann$^{2\,}$\footnote{E-mail: \texttt{martin.gabelmann@kit.edu}},
Margarete
M\"uhlleitner$^{2\,}$\footnote{E-mail: \texttt{margarete.muehlleitner@kit.edu}},
Heidi Rzehak$^{3\,}$\footnote{E-mail: \texttt{heidi.rzehak@itp.uni-tuebingen.de}}
\\[9mm]
{\small\it
$^1$Institute For Interdisciplinary Research in Science and Education, ICISE,}\\
{\small\it 590000 Quy Nhon, Vietnam.}\\[3mm]
{\small\it
$^2$Institute for Theoretical Physics, Karlsruhe Institute of Technology,} \\
{\small\it Wolfgang-Gaede-Str. 1, 76131 Karlsruhe, Germany.}\\[3mm]
{\small \it 
$^3$Institute for Theoretical Physics, University of T\"ubingen,}\\{\small \it Auf der Morgenstelle 14, 72076 T\"ubingen, Germany}\\[3mm]
}
\maketitle

\begin{abstract}
We present our computation of the $\order{(\alpha_t+\alpha_\lambda+\alpha_\kappa)^2}$ two-loop
corrections to the Higgs boson masses of the CP-violating
Next-to-Minimal Supersymmetric Standard Model (NMSSM) using the
Feynman-diagrammatic approach in the gaugeless 
limit at vanishing external momentum. We choose a mixed
$\overline{\mbox{DR}}$-on-shell (OS) renormalisation scheme for the
Higgs sector and apply both $\overline{\mbox{DR}}$ and OS 
renormalisation in the top/stop sector. For the treatment of the
infrared divergences we apply and compare three different
regularisation methods: the introduction of a regulator mass, the
application of a small momentum expansion, and the inclusion of the
full momentum dependence. Our 
new corrections have been implemented in the Fortran code {\tt
  NMSSMCALC} that computes the 
Higgs mass spectrum of the CP-conserving and CP-violating NMSSM as well as
the Higgs boson decays including the state-of-the-art higher-order
corrections.
Our numerical analysis shows that the newly computed
corrections increase with rising $\lambda$
and $\kappa$, remaining overall below about 3\%  compared to our
previously computed 
$\order{\alpha_t(\alpha_t+\alpha_s)}$ corrections, in the region
compatible with perturbativity below the GUT scale. The
renormalisation scheme and scale 
dependence is of typical two-loop order. The impact of the CP-violating phases
in the new corrections is small. We furthermore show that the Goldstone
Boson Catastrophe due to the infrared divergences can be treated in a
numerically efficient way by introducing a regulator mass that 
approximates the momentum-dependent results best for squared mass values in
the permille range of the squared renormalisation scale. Our results
mark another step forward in the program of increasing the precision
in the NMSSM Higgs boson observables.
\end{abstract}

\thispagestyle{empty}
\vfill
\pagebreak

\section{Introduction}
\label{sec:intro}
The Standard Model (SM) of particle physics belongs to the most
successful theories ever tested. Despite its success the SM
lacks explanations for a variety of open problems such as for example 
the nature of Dark Matter (DM) or the observed
baryon-antibaryon asymmetry in the universe. 
Models based on supersymmetry (SUSY)
\cite{Golfand:1971iw, Volkov:1973ix, Wess:1974tw, Fayet:1974pd,Fayet:1977yc, Fayet:1976cr, Nilles:1982dy,Nilles:1983ge, Frere:1983ag,Derendinger:1983bz,Haber:1984rc, Sohnius:1985qm,Gunion:1984yn, Gunion:1986nh}
are promising beyond-the-SM (BSM) candidates that offer solutions to
many problems the SM cannot address. In SUSY models the Higgs and gauge
sectors are related so that in the Minimal Supersymmetric Standard Model (MSSM) 
\cite{Gunion:1989we,Martin:1997ns,Dawson:1997tz,Djouadi:2005gj} the
tree-level mass $m_h$ of the lightest Higgs boson is bounded from above by the
$Z$-boson mass $m_Z$. Higher-order corrections, where those from the top/stop
sector play the dominant role, can shift the upper bound
to larger values. In order to reach the experimentally measured value
of 125.09 GeV \cite{Aad:2015zhl} for the SM-like Higgs boson mass, however, a
large soft-SUSY breaking mass scale $m_{\text{SUSY}}$ and/or a large mixing in
the stop sector is required so that naturalness arguments in favor of
supersymmetry become questionable. In the Next-to-MSSM (NMSSM) \cite{Barbieri:1982eh,Dine:1981rt,Ellis:1988er,Drees:1988fc,Ellwanger:1993xa,Ellwanger:1995ru,Ellwanger:1996gw,Elliott:1994ht,King:1995vk,Franke:1995tc,Maniatis:2009re,Ellwanger:2009dp}
the situation is more relaxed as it features additional $F$-term
contributions raising the tree-level Higgs mass to a higher value 
so that higher-order corrections can be smaller
compared to the MSSM and still lead to the required Higgs mass value. \s

In the meantime, the Higgs boson mass has turned into a precision
observable with an uncertainty of a few hundred MeV
\cite{Aad:2015zhl}. The precise knowledge of the Higgs boson mass is important 
as it is a crucial input parameter for all Higgs boson
observables~\cite{deFlorian:2016spz} and determines the stability of
the electroweak vacuum
\cite{Degrassi:2012ry,Buttazzo:2013uya,Bednyakov:2015sca}. Therefore,
in order to make meaningful interpretations of the experimental
results, the experimental accuracy has to be matched by 
the precision of the theory predictions. Only then sensible limits on
the still allowed parameter space of the model can be derived from the
experimental results and possibly distinguish between new physical
models in case of discovery, {\it cf.~e.g.}~Ref.~\cite{Muhlleitner:2017dkd}.
Consequently, a tremendous effort has been put in the computation of
the higher-order corrections to supersymmetric Higgs boson
masses. These calculations can be grouped into three classes that are
based on fixed-order (FO), effective field theory
(EFT) or hybrid techniques that make use of FO as well as EFT
results. A comprehensive and
complete overview on the status of the higher-order calculations in
the MSSM and NMSSM in the various approaches has been given in the recent
review \cite{Slavich:2020zjv}. In the following, we briefly review the most relevant 
studies related to the computations and implementations of the higher-order corrections to the
 Higgs boson masses in the NMSSM with a $\mathbb{Z}^3$ symmetry.
\s

In the CP-conserving NMSSM, the leading one-loop contributions to the
Higgs boson masses were presented in \cite{Ellwanger:1993hn,Elliott:1993ex,Elliott:1993uc,Elliott:1993bs,Pandita:1993hx,Pandita:1993tg,King:1995vk,Ellwanger:2005fh} while the
full one-loop corrections were provided in \cite{Degrassi:2009yq,Staub:2010ty} for the $\overline{\mbox{DR}}$ scheme
and in Refs. \cite{Ender:2011qh,Drechsel:2016jdg} for a mixed
$\DRb$-on-shell (OS) renormalisation scheme. Two-loop 
corrections at the order ${\cal O} (\alpha_t \alpha_s + \alpha_b
\alpha_s)$ in the $\overline{\mbox{DR}}$ scheme were obtained in the
effective potential approach in \cite{Degrassi:2009yq}. The authors of
\cite{Goodsell:2014pla} have provided the two-loop corrections beyond ${\cal O} (\alpha_t \alpha_s + \alpha_b
\alpha_s)$ in the gaugeless limit and 
in the $\overline{\mbox{DR}}$ scheme by differentiating numerically or analytically the generic two-loop effective potential presented in \cite{Martin:2001vx}. Including CP-violating phases,
the leading one-loop corrections were provided in 
~\cite{Ham:2001kf,Ham:2001wt,Ham:2003jf,Funakubo:2004ka,Ham:2007mt}. Subsequently, the full one-loop and logarithmically enhanced
 two-loop effects were computed in ~\cite{Cheung:2010ba} using the renormalisation group
 approach. Our group calculated the full one-loop corrections with momentum
 dependence in \cite{Graf:2012hh}
and the two-loop corrections of ${\cal 
  O}{(\al_t\al_s)}$ \cite{Muhlleitner:2014vsa}  and of ${\cal
O} (\al_t^2)$ \cite{Dao:2019qaz} in the approximation of
 vanishing external momentum in the gaugeless
limit.\footnote{Note that besides mass
  corrections we also provided higher-order corrections to the
  trilinear Higgs self-couplings, namely the 
  one-loop corrections to the trilinear Higgs self-couplings in the
  CP-conserving NMSSM \cite{Nhung:2013lpa} and the ${\cal O}(\alpha_t
  \alpha_s)$ corrections in the CP-violating NMSSM
  \cite{Muhlleitner:2015dua}.} The renormalisation is
based on a mixed $\DRb$-OS  scheme in the Higgs
  sector with the possibility to choose 
between the $\DRb$ or OS scheme in the renormalisation of the top/stop
sector. The independent calculations of the
full one-loop corrections were also presented  in \cite{Goodsell:2016udb} and in \cite{Domingo:2017rhb}  
applying the  $\overline{\mbox{DR}}$ and mixed $\overline{\mbox{DR}}$-OS schemes, respectively. 
 The issue of
residual gauge dependences in the higher-order Higgs mass corrections
was discussed in \cite{Dao:2019nxi,Domingo:2020wiy}.
\s 

There exist numerous codes implementing the higher-order corrections
to the NMSSM Higgs boson masses, some of them also partly calculate the Higgs
boson decays. The full one-loop and ${\cal O} (\alpha_t \alpha_s + \alpha_b
\alpha_s)$  correction in \cite{Degrassi:2009yq} were incorporated into {\tt NMSPEC} 
while  one-loop corrections from \cite{Domingo:2017rhb} were implemented  in {\tt NMHDECAY}\_{\tt CPV}
\cite{Domingo:2015qaa}, they are parts of {\tt NMSSMTools}
\cite{Ellwanger:2004xm,Ellwanger:2005dv,Ellwanger:2006rn}. The program
package can be interfaced with {\tt SOFTSUSY}
\cite{Allanach:2001kg,Allanach:2013kza}, which includes the
possibility of $\mathbb{Z}^3$ violation. The results of \cite{Staub:2010ty,Goodsell:2014pla,Goodsell:2016udb} were made available in {\tt SARAH}
\cite{Staub:2008uz,Staub:2010jh,Staub:2012pb,Staub:2013tta,Goodsell:2014bna,Goodsell:2014pla}
 with {\tt SPheno} \cite{Porod:2003um,Porod:2011nf}. 
 This is also possible by interfacing {\tt SARAH} with the package {\tt FlexibleSUSY}
\cite{Athron:2014yba,Athron:2017fvs}.  
The code {\tt FlexibleEFTHiggs} \cite{Athron:2016fuq} combines an effective field theory approach 
with a FO calculation to compute
the SM-like Higgs pole mass in various models,
including the NMSSM. We have implemented our FO 
calculations at one-loop and two-loop ${\cal O}(\alpha_t \alpha_s)$
and ${\cal O}(\alpha_t^2)$ in the program package {\tt
  NMSSMCALC}\footnote{The code can be downloaded from the url:
  https://www.itp.kit.edu/$\sim$maggie/NMSSMCALC/.}~\cite{Baglio:2013iia} which also computes the
Higgs boson decay widths and branching ratios both for the CP-conserving and
CP-violating case.\footnote{Recently, we published the code {\tt NMSSMCALCEW}
 \cite{Baglio:2019nlc,Dao:2020dfb} that includes besides the
 state-of-the-art QCD corrections 
 already included in {\tt NMSSMCALC} the SUSY-EW and SUSY-QCD
 corrections to the neutral and charged Higgs bosons in the
 CP-conserving and CP-violating case. One-loop corrected decay widths are also
  included in the code {\tt SloopS}
  \cite{Belanger:2016tqb,Belanger:2017rgu,Boudjema:2017ozm}. A generic 
implementation of the two-body partial decays widths at the full
one-loop level \cite{Goodsell:2017pdq} exists in the {\tt SARAH} and
{\tt SPheno} framework.}  Comparisons
of the NMSSM Higgs boson mass computations of the various codes were performed in \cite{Staub:2015aea} 
 for the $\DRb$ scheme and in \cite{Drechsel:2016htw} for the mixed
 $\DRb$-OS  scheme. \s

In the present paper, we provide the two-loop corrections controlled by the NMSSM superpotential parameters
$\lambda$ and $\kappa$ using the diagrammatic approach in the  mixed $\overline{\mbox{DR}}$-OS scheme of {\tt NMSSMCALC}, that was missing in our previous calculations of $\order{\alpha_t^2}$ ~\cite{Dao:2019qaz}. 
These corrections can be important for light singlet-like Higgs boson states as well as for the doublet-like states in case of large singlet-doublet mixings. In fact these corrections cannot be simply separated from the $\order{\alpha_t^2}$ corrections
due to Feynman diagrams proportional to $\order{\alpha_t(\alpha_t+\alpha_\lambda)}$ terms. Hence we combine them and denote them as $\order{(\alpha_t+\alpha_\lambda+\alpha_\kappa)^2}$ corrections.  As common practice, we use the combination of the gaugeless limit (the weak mixing angle $\theta_W$ is kept fixed and $e\to 0$ which leads to massless Goldstone bosons) and the vanishing external momentum approximations. On the one hand, these approximations are simple and good in practice. On the other hand they can give rise to the 
appearance of infrared (IR) divergences which are present in Feynman
diagrams with two and more   
Goldstone bosons in the internal lines. This problem is known as Goldstone Boson Catastrophe (GBC) \cite{Martin:2002iu,
Martin:2002wn,Martin:2014bca,Pilaftsis:2015bbs,Espinosa:2016uaw,Kumar:2016ltb,Espinosa:2017aew,
Braathen:2016cqe,Braathen:2017izn}. In this paper, we address this issue in detail and discuss our practical treatment in the code  {\tt NMSSMCALC}.  \s  
 
The paper is organised as follows. 
Section \ref{sec:model} introduces our notation as well as the NMSSM at
tree level. Section \ref{sec:renormalisation} discusses the different
renormalisation schemes and the derivation of all necessary one- and
two-loop counterterms. In section \ref{sec:self-energies} we describe
our treatment of the GBC. The set-up of the calculation and the
numerical analysis is given in Sec.~\ref{sec:pheno}. 
Section \ref{sec:results} is dedicated to the
numerical analysis of the newly calculated contributions. We conclude
in section \ref{sec:conclusions}. 

\section{The NMSSM Tree-Level Spectrum}
\label{sec:model}
We work in the $\mathbb{Z}_3$ symmetric NMSSM. For the two-loop corrections
of $\order{(\alpha_t+\alpha_\lambda+\alpha_\kappa)^2}$ we use the gaugeless limit and hence 
follow the same notation as in our previous calculations~\cite{Dao:2019qaz,Muhlleitner:2015dua}. 
We describe here only the Higgs, higgsino and top/stop sectors which are relevant for our renormalisation procedure. For the purpose of introducing model parameters, we present here all terms appearing in the NMSSM superpotential 
\begin{align}
    \mathcal{W}_{\text{NMSSM}} = 
	&
		\left[y_e \hat{H}_d\ccdot \hat{L} \hat{E}^c 
		+ y_d  \hat{H}_d \ccdot \hat{Q} \hat{D}^c 
		- y_u \hat{H}_u \hat{Q} \hat{U}^c
		\right]  - \lambda \hat{S} \hat{H}_d\ccdot \hat{H}_u 
		+ \frac{1}{3} \kappa \hat{S}^3  \;,
    \label{eq:wnmssm}
\end{align}
with the quark and lepton superfields $\hat{Q}$, $\hat{U}$, $\hat{D}$, $\hat{L}$, $\hat{E}$,
and the Higgs doublet
superfields $\hat{H}_d$, $\hat{H}_u$ and the singlet superfield
$\hat{S}$. Charge conjugated fields are denoted by the superscript
$c$. Color and generation indices have been suppressed. The
symplectic product $x\ccdot y= \epsilon_{ij}x^iy^j$ ($i,j=1,2$) is built with the
anti-symmetric tensor $\epsilon_{12}=\epsilon^{12}=1$. 
The parameters $\lambda,\kappa$ are complex in general. For simplicity, the Yukawa couplings $y_x$ ($x=e,d,u$) are chosen to be diagonal matrices. This setting has a negligible effect on the Higgs mass calculation since only the Yukawa couplings of the third generation are important.
Furthermore, we chose the convention that the $y_x$ ($x=e,d,u$) are
real by rephasing the left and right-handed Weyl-spinor fields as
$x_{L,R}\to x_{L,R} e^{i  \varphi_{\text{\tiny L,R}}}$.  
  The soft SUSY breaking Lagrangian reads
\begin{eqnarray}
{\cal L}_{\text{soft},\text{ NMSSM}} &=& -m_{H_d}^2 H_d^\dagger H_d - m_{H_u}^2
H_u^\dagger H_u -
m_{\tilde{Q}}^2 \tilde{Q}^\dagger \tilde{Q} - m_{\tilde{L}}^2 \tilde{L}^\dagger \tilde{L}
- m_{\tilde{u}_R}^2 \tilde{u}_R^* 
\tilde{u}_R - m_{\tilde{d}_R}^2 \tilde{d}_R^* \tilde{d}_R 
\nonumber \\\nonumber
&& - m_{\tilde{e}_R}^2 \tilde{e}_R^* \tilde{e}_R - (\epsilon_{ij} [y_e A_e H_d^i
\tilde{L}^j \tilde{e}_R^* + y_d
A_d H_d^i \tilde{Q}^j \tilde{d}_R^* - y_u A_u H_u^i \tilde{Q}^j
\tilde{u}_R^*] + \mathrm{h.c.}) \\
&& -\frac{1}{2}(M_1 \tilde{B}\tilde{B} + M_2
\tilde{W}_i\tilde{W}_i + M_3 \tilde{G}\tilde{G} + \mathrm{h.c.}) \nonumber
\\ 
\label{eq:lagrangiansoft}
&&- m_S^2 |S|^2 +
(\epsilon_{ij} \lambda 
A_\lambda S H_d^i H_u^j - \frac{1}{3} \kappa
A_\kappa S^3 + \mathrm{h.c.}) \;,
\end{eqnarray}
where again the summation over quark and lepton generation indices is
implicit. The $\tilde{Q}$, {$\tilde{u}_R$, $\tilde{d}_R$ and $\tilde{L}$, $\tilde{e}_R$} stand for the complex scalar
components of the corresponding quark and lepton superfields. 
The soft
SUSY breaking gaugino mass parameters $M_k$ ($k=1,2,3$) of the bino,
wino and gluino fields $\tilde{B}$, $\tilde{W}_l$ ($l=1,2,3$) and
$\tilde{G}$ as well as 
the soft SUSY breaking trilinear couplings
$A_x$ ($x=\lambda,\kappa,u,d,e$) are complex in the CP-violating
NMSSM whereas 
the soft SUSY breaking mass parameters of the scalar fields, 
$m_X^2$ ($X=S,H_d,H_u,\tilde{Q},\tilde{u}_R,\tilde{d}_R,\tilde{L},\tilde{e}_R$),
are real. 

\subsection{The Higgs Boson Sector \label{sec:higgssector}}
The tree-level Higgs boson potential inferred from 
${\mathcal{L}}_{\text{soft},\text{ NMSSM}}$ and the $F$-terms of
$\mathcal{W}_{\text{NMSSM}}$ reads
\begin{align}
    V_H  = &  
            (\left|\lambda S\right|^2 + m_{H_d}^2)H_d^\dagger H_d 
            + (\left|\lambda S\right|^2  + m_{H_u}^2)H_u^\dagger H_u 
            + m_S^2 |S|^2 \nonumber \\
           &
            + \left|\kappa S^2 -\lambda  H_d \ccdot H_u\right|^2
            + \left(\frac{1}{3} \kappa A_{\kappa} S^3 
            - \lambda A_\lambda S   H_d \ccdot  H_u  + \mathrm{h.c.}\right) \, ,
           \label{eq:higgspotential}
\end{align}
where we neglected the $D$-terms originating from the gauge sector as
they vanish in the gaugeless limit and hence are not needed in the following. Note that
gaugeless limit means that the $U(1)_Y$ and $SU(2)_L$ gauge couplings $g_1\to 0$ and 
$g_2\to 0$ while $\tan\theta_W =g_2/g_1$
is kept constant, where $\theta_W $ is the weak mixing angle. This is equivalent to the 
limit of vanishing electric charge and tree-level vector boson masses,
$e,M_W,M_Z\to 0$, while keeping $\tan\theta_W$ constant.
\s

Expanding the Higgs boson fields around their vacuum expectation
values (VEVs) $v_u$, $v_d$, and $v_s$, respectively, yields
\begin{equation}
    H_d = \doublet{\frac{v_d + h_d +i a_d}{\sqrt 2}}{h_d^-}, \,\, 
    H_u = e^{i\varphi_u}\doublet{h_u^+}{\frac{v_u + h_u +i a_u}{\sqrt 2}},\,\,
    S= \frac{e^{i\varphi_s}}{\sqrt 2} (v_s + h_s + ia_s)\, ,
   \label{eq:vevs}
\end{equation}
with the CP-violating phases $\varphi_{u,s}$. The three VEVs can be traded for $\tan\beta$,
the SM VEV $v$ and the effective $\mu$ parameter $\mueff$ as
\begin{align} \tb&\equiv\tan\beta= v_u/v_d \\
\vev^2&=v_u^2+v_d^2\approx \left(\unit[246]{GeV}\right)^2 \\
    \mueff&=\frac{e^{i \varphi_s}}{\sqrt 2}\vs \lambda\,.
    \label{eq:mueffcalc}
\end{align}
The MSSM limit can be obtained by taking the limit $\lambda,\kappa\to 0,~v_s\to\infty$ while
keeping $\mueff$ and   $\kappa/\lambda$ constant.
Using the Higgs potential given 
in Eq.~(\ref{eq:higgspotential}) we define tree-level tadpoles and
mass terms in the broken phase, 
\begin{align}
    V_H \supset &\quad \bm{t} {\bm{\phi}} +
                  {\bm{\phi}}^T\mathcal{M}_{\phi \phi}{\bm{\phi}}  
    + \bm{h}^{c,\dagger} \mathcal{M}_{h^+ h^-}\bm{h}^c \,\, , \\
               &\quad \text{with}\quad
               {\bm{\phi}}=(h_d,h_u,h_s,a_d,a_u,a_s)^T,\,\,
               \bm{h}^c=(h_d^{-*},h_u^+)\, ,\nonumber 
\end{align}
and the tadpole coefficients 
\begin{equation}
  (\bm{t})_j = t_{{\bm{\phi}_j}} = \frac{\del V_H}{\del {{\bm{\phi}_j}}}, \,\, j = 1, \dots, 6
  \,\, ,
\end{equation}
where only five of them are independent and
$t_{a_u}=t_{a_d}/t_\beta$. The tadpoles vanish at tree level but
affect the higher-order corrections. We keep them, however, for the
renormalisation procedure, and set them to zero afterwards. The
explicit expressions for the tadpoles and the squared mass matrices
$\mathcal{M}_{\phi \phi}$ and $\mathcal{M}_{h^+ h^-}$ can be found in
Ref. \cite{Dao:2019qaz}. The reference also contains 
a detailed discussion about the 
two-fold rotation of the neutral Higgs bosons first separating 
the Goldstone component with the rotation $\mathcal{R}^G(\beta_n)$,
{\it i.e.}~transforming from the basis
$(h_d, h_u, h_s, a_d, a_u,a_s)$ to 
$(h_d, h_u, h_s, a, a_s, G^0)$,
and second rotating into the mass basis
$(h_1, h_2, h_3, h_4, h_5, G^0)$
with the rotation matrix ${\cal R}$, 
\begin{align}
    \mathcal{M}_{hh}  = \,\, & \mathcal{R}^G(\beta_n)\mathcal{M}_{\phi\phi}(\mathcal{R}^G(\beta_n))^T  \label{eq:massmatrix} \\
  \mathcal{M}_{hh}^\prime  =\,\, & \mathcal{R} \mathcal{M}_{hh} \mathcal{R}^T
  \label{eq:massmatrix2}\\
     =\,\, &
             \text{diag}(m_{h_1}^2,m_{h_2}^2,m_{h_3}^2,m_{h_4}^2,m_{h_5}^2,m_{G^0}^2)
             \,, \nonumber
\end{align}
with
\beq
\label{eq:mg0}
m_{G^0}^2 = \frac{c_\beta t_{h_d} + s_\beta t_{h_u} }{v} \;.
\eeq
where we kept the dependence on the tadpole parameters explicitly for
$m_{G^0}^2$ for the later discussion of the cancellation of the IR
  divergences in \cref{sec:self-energies}.
It should be noted that $\mathcal{M}_{hh}^\prime$ is only diagonal for vanishing tadpole parameters.
For the charged Higgs fields a single rotation
$\mathcal{R}^{G^-}(\beta_c)$ is used,
\begin{align}
\mathcal{R}^{G^-}(\beta_c)\mathcal{M}_{h^+h^-}(\mathcal{R}^{G^-}(\beta_c))^T
=
\text{diag}(m_{G^\pm}^2, M_{H^\pm}^2)
    \, , \label{eq:chargedHiggsmass}
\end{align}
where the charged Goldstone boson mass
\begin{align}
    m^2_{G^\pm}  &= \frac{s_\beta t_{h_u} + c_\beta t_{h_d}}{v} \, \label{eq:MGpHm2}
\end{align}
as well as all off-diagonal elements vanish for vanishing tadpoles.
The rotation angles $\beta_n$ and $\beta_c$ coincide with $\beta$ at
tree level, $\beta_c=\beta_n=\beta$, which has been already applied in
Eqs.~(\ref{eq:mg0}) and (\ref{eq:MGpHm2}). We distinguish them since
$\beta_n$ and $\beta_c$  
as mixing angles do not need to obtain a counterterm while $\beta$ 
arising from the ratio of VEVs {has to be renormalised and receives a
non-vanishing counterterm}. 
After the renormalisation they are set equal to the tree-level value of $\beta$ again.
Note that we denote all masses apart from the charged Higgs boson mass
$M_{H^\pm}$ by small letters $m$ in order to indicate that they are
tree-level masses while we will denote loop-corrected masses by
capital $M$. As discussed later, $M_{H^\pm}$ will be
renormalised on-shell so that there the distinction between tree-level mass
and loop-corrected mass does not apply. 
\s

In accordance with the SUSY Les Houches Accord (SLHA)
\cite{Skands:2003cj,Allanach:2008qq} and for purpose of renormalisation, we decompose the complex parameters $A_\lambda$
and $A_\kappa$ into their imaginary and real parts and
$\lambda$ and $\kappa$ into their absolute values and phases
$\varphi_\lambda$ and $\varphi_\kappa$,
respectively.\footnote{Also $\lambda$ and $\kappa$ are
    read in by {\tt NMSSMCALC} 
in terms of their real and complex parts, in accordance with the
SLHA. For the numerical analysis, however, we choose a different, more convenient,
format in terms of absolute values and phases.} The phases
enter the tree-level Higgs mass matrix 
in two combinations together with $\varphi_u$ and $\varphi_s$,
\begin{eqnarray}
\varphi_y &=& \varphi_\kappa - \varphi_\lambda + 2 \varphi_s -
\varphi_u \label{eq:phase1} \\
\varphi_w &=& \varphi_\kappa + 3 \varphi_s \;, \label{eq:phase2}
\end{eqnarray}
where $\varphi_y$ is the only CP-violating phase at tree level in
the Higgs sector. 
In case of vanishing $\varphi_y$,  the CP-even components, $h_u,h_d,
h_s$, do not mix with the CP-odd ones, $a_d,a_u,a_s$.  We furthermore
trade $\Im A_{\lambda,\kappa}$ as well 
as $m_{H_{u,d},S}^2$ for the  tadpole parameters $t_{a_d, a_s}$ and
$t_{h_{d,u,s}}$, respectively, by using the tadpole conditions, {\it
  cf.}~Ref.~\cite{Dao:2019qaz} for details. In contrast to the MSSM,
the trilinear and quartic Higgs couplings in the NMSSM do not vanish
in the gaugeless limit, but involve $\lambda,\kappa,
A_\lambda,A_\kappa$. The
$\order{(\alpha_t+\alpha_\lambda+\alpha_\kappa)^2}$ corrections get
contributions from two-loop diagrams containing these Higgs
self-couplings. This causes the appearance of the GBC, which will be
discussed in detail in Sec.~\ref{sec:self-energies}.\s 

In \NMSSMCALC, we have two possibilities to choose the set of input parameters in the Higgs sector: either
\newcommand{\ReAkappa}{\Re A_{\kappa}}
\newcommand{\ReAlambda}{\Re A_{\lambda}}
\newcommand{\cosks}{c_{\varphi_\omega}}
\newcommand{\sinks}{s_{\varphi_\omega}}
\begin{align}
\left\{ t_{h_d},t_{h_u},t_{h_s},t_{a_d},t_{a_s},M_{H^\pm}^2,v,s_{\theta_W},
e,\tan\beta,|\lambda|,v_s,|\kappa|,\ReAkappa,\varphi_\lambda,\varphi_\kappa,\varphi_u,\varphi_s
\right\} \,, \label{eq:inputset1}
\end{align}
or
\begin{align}
\left\{ t_{h_d},t_{h_u},t_{h_s},t_{a_d},t_{a_s},v,s_{\theta_W},
e,\tan\beta,|\lambda|,v_s,|\kappa|,\ReAlambda,\ReAkappa,\varphi_\lambda,\varphi_\kappa,\varphi_u,\varphi_s
\right\} \,. \label{eq:inputset2}
\end{align}
In the first choice the charged Higgs mass is an input while $\ReAlambda$ is an input in the second one.
These parameters need to be renormalised at one- and two-loop level.

\subsection{The squark sector}
The relation between the top mass and the top quark Yukawa coupling is given by, 
\begin{equation}
    m_t = \frac{v_u y_t}{\sqrt 2} 
          e^{i(\varphi_u+\varphi_{\text{\tiny L}}-\varphi_{\text{\tiny R}})}\,,
          \label{eq:mttree}
\end{equation}
in which $m_t$ and $y_t$ are real by our convention. We use the freedom of choice of 
the phases $\varphi_{\text{\tiny L}}$, $\varphi_{\text{\tiny R}}$ of the left- and right-handed
top-quark fields and define $\varphi_{\text{\tiny
    L}}=-\varphi_{\text{\tiny R}}= -\varphi_u/2$. As result, the stop
mass matrix in the $(\tilde{t}_{L},\tilde{t}_{R})^T$ basis in the
gaugeless limit is given by
\begin{align}
 {\cal M}_{\tilde{t}}  &=
     \begin{pmatrix}
             m_{\tilde{Q}_3}^2 + m_t^2 & m_t
             \left(A_t^*e^{-i\varphi_u}-\frac{\mueff}{\tan\beta}\right) \\[2mm]
             m_t \left(A_t e^{i\varphi_u}
               -\frac{\mueff^*}{\tan\beta}\right) &
             m_{\tilde{t}_R}^2+m_t^2 
         \end{pmatrix}
          \;,\\
    \text{diag}( m_{\tilde{t}_1}^2, m_{\tilde{t}_2}^2)& =
    \mathcal{U}^{\tilde t}  {\cal M}_{\tilde{t}}  {\mathcal{U}^{\tilde{t}}}^\dagger\,,
\end{align}
where $\mathcal{U}^{\tilde{t}}$ rotates the left- and right-handed
stop fields $\tilde{t}_{L,R}$ into the mass eigenstates $\tilde{t}_{1,2}$. Similar to our previous calculations ~\cite{Dao:2019qaz,Muhlleitner:2015dua}, the bottom quark mass is set to zero
everywhere, hence the right-handed sbottom states decouple and only left-handed sbottom states are involved
  in the computation. The parameters in the squark sector that need to be renormalised at one-loop level are
\be m_t, \; m_{\tilde{Q}_3}, \; m_{\tilde t_R} \quad \mbox{and} \quad A_t \;.
\label{eq:stopparset} \ee 

\subsection{The Electroweakino Sector}
\label{sec:EWinoTree}
The mass generation for the wino, bino, higgsino and
singlino interaction states does not change significantly
w.r.t. Ref.~\cite{Dao:2019qaz} except that we do not assume
$\lambda=\kappa=0$. Therefore we only shortly repeat the used notation in this section.
Since the gauged Weyl-fermions do not couple to any other
  particles in the gaugeless limit we 
only consider the 3$\times$3 sub-matrix $\textbf{M}_{\chi^0}$ in the basis 
$(\tilde{H}^0_d,\tilde{H}^0_u,  \tilde{S})^T$ for the neutralinos,  
\begin{align}
\textbf{M}_{\chi^0} &=
    \begin{pmatrix}
        0        & \quad-\mueff     &  \quad -\frac{\lambda}{\sqrt 2}\vev\sb e^{i \varphi_u}\\[4mm]
        -\mueff  & \quad 0          &  -\frac{\lambda}{\sqrt 2}\vev\cb\\[4mm]
       \quad -\frac{\lambda}{\sqrt 2}\vev\sb e^{i \varphi_u}          &   -\frac{\lambda}{\sqrt 2}\vev\cb               & \sqrt{2} \kappa \vs e^{i\varphi_s} 
    \end{pmatrix}\,,\\
       \text{diag}(m_{\tilde{\chi}_3^0},m_{\tilde{\chi}_4^0},m_{\tilde{\chi}_5^0})
  &= N^* \textbf{M}_{\chi^0} N^\dagger 
    \label{eq:Mneutralino}
\end{align}
and the 1$\times$1 matrix for the charginos,
\begin{align}
    m_{\tilde{\chi}^\pm_2}  &=  \mueff V^*_{22} U^*_{22}
    \equiv |\mueff| \;, \label{eq:Mchargino}
\end{align}
where the neutralino masses are ordered as $|m_{\tilde{\chi}^0_3}|\leq
|m_{\tilde{\chi}^0_4}| \leq |m_{\tilde{\chi}^0_5}|$ and $U$, $V$
denote the $2\times 2$ unitary matrices for the rotation from the
gauge to the mass basis of the charginos.  
Note, that we absorbed the phase of $\mueff$ into the chargino mixing
matrix $V$ so that the higgsino couplings entering the two-loop
diagrams will depend on $e^{i\varphi_{\mueff}}$. In contrast to the
previous $\order{\alpha_t^2}$ corrections, the singlino and mixed
singlino-higgsino states now will also contribute in the two-loop
diagrams. The vertex and propagator counterterms
involving charginos and neutralinos enter one-loop counterterm inserted
diagrams. We therefore need to renormalise them at one-loop
level. However, all parameters in the electroweakino sector are also
present in the Higgs sector so that we do not need further
renormalisation conditions. 

\section{Renormalisation of the NMSSM Higgs Bosons at the Two-Loop Order}
\label{sec:renormalisation}
The loop corrected Higgs boson mass spectrum is obtained by
iteratively solving\footnote{We have
confirmed that the contributions from the Goldstone components 
are numerically negligible. Thus we drop them in the final calculation.}
\begin{equation}
    \text{det}\left(\mathbb{1}_{5\times 5}p^2-\mathcal{M}^\prime_{hh, 5\times 5} +\hat{\Sigma}_{hh}(p^2)\right)=0
    \label{eq:propagator}
\end{equation}
for the squared mass matrix $\mathcal{M}_{hh, 5\times 5}^\prime=m_{h_i}^2
\delta_{ij}$ ($i,j=1,...,5$), where the $m_{h_i}$ denote the tree-level
masses of the tree-level mass eigenstates $h_{i}$. The numerical
recipe for the iterative solution is described in Ref.~\cite{Dao:2019qaz}. 
The $\left(\hat{\Sigma}_{hh}\right)_{ij} \equiv  \hat{\Sigma}_{ij}$
stand for the renormalised self-energies 
 for the $h_i\to h_j$ transition and contain the one- and two-loop
 contributions which are denoted by the superscripts (1) and (2), respectively,
\begin{equation}
    \hat{\Sigma}_{ij} = \hat{\Sigma}_{ij}^{\tol}(p^2) +
    \hat{\Sigma}_{ij}^{\ttl}(p^2) \;.
    \label{eq:rensigmafull}
\end{equation}
The one-loop renormalised Higgs self-energies have already been
obtained with full momentum dependence in the CP-conserving and
CP-violating NMSSM in Refs.~\cite{Ender:2011qh,Graf:2012hh} to which
we refer for further details. The two-loop renormalised  Higgs
self-energies consist of the ${\cal O}(\alpha_t \alpha_s)$ corrections, which we 
computed in \cite{Muhlleitner:2014vsa}, and the
${\cal O}((\alpha_t+\alpha_\lambda+\alpha_\kappa)^2)$ contributions
computed in this paper, 
\beq
\hat{\Sigma}^{\tiny{(2)}} (p^2)_{ij} = \hat{\Sigma}^{\tiny{(2)},\alpha_t
  \alpha_s}_{ij} (0) + \hat{\Sigma}^{\tiny{(2)},(\alpha_t+\alpha_\lambda+\alpha_\kappa)^2}_{ij} (p^2) \;.
\eeq
Note that the ${\cal O}(\alpha_t
\alpha_s)$ corrections are evaluated in the approximation of vanishing external
momentum. In the ${\cal O}((\alpha_t+\alpha_\lambda+\alpha_\kappa)^2)$
corrections, however, we can choose between including the finite momentum
dependence or the Goldstone boson mass as regulator for the IR divergences (see Sec.~\ref{sec:self-energies}).
We therefore keep the momentum dependence in the following formulae.
We will drop the superscript $(\alpha_t+\alpha_\lambda+\alpha_\kappa)^2$
on the self-energies for simplicity of the expressions.  
The neutral Higgs renormalised self-energies\footnote{For the
  inclusion of Goldstone components, $i$, $j$ take the values 1 to 6
  where $h_6$ is identified with $G^0$.} are written as sum of 
the unrenormalised self-energies  $\Sigma_{ij}$ and the counterterms
at one-loop level as
\bea 
\hat \Sigma^\tol_{ij}(p^2) &=& \Sigma^\tol_{ij}(p^2) +\fr12  p^2
\left[\calR(  \wave ^\dagger +\wave)\calR^T\right]_{ij} \crn 
&&-\sbraket{ \calR\braket{ \fr12 \wave^\dagger  \calM_{hh} + \fr12
    \calM_{hh}\wave + \deltaone {\calM_{hh}} }\calR^T}_{ij } \;, \label{eq:rehiggsself1}
\eea
and at two-loop level 
\bea 
\hat \Sigma^\ttl_{ij}(p^2)=\Sigma^\ttl_{ij}(p^2)  +\fr12  p^2
\left[\calR\braket{\fr 12  (\wave)^\dagger\wave +  \wavetwo^\dagger
    +\wavetwo }\calR^T\right]_{ij}- \left( \deltatwo  M^2 \right) _{ij}  \;, \label{eq:rehiggsself2}
\eea
with
\begin{align} 
\left( \deltatwo  M^2 \right) _{ij}  =& \fr 12 \left[\calR\left(\fr12
(\wave)^\dagger \calM_{hh} \wave + \wave^\dagger \deltaone
\calM_{hh} +\deltaone \calM_{hh}\wave + \wavetwo^\dagger\calM_{hh}
\right. \right. \crn
& + \calM_{hh} \wavetwo\bigg)\calR^T\bigg]_{ij}
+\braket{ \calR \deltatwo {\calM_{hh}}
  \calR^T}_{ij} \;. \label{eq:ReHSelf}
\end{align}
In the above formulae, $\calM_{hh}$ and $\calR$ are the tree-level
Higgs mass matrix and the rotation matrix defined in
\eqref{eq:massmatrix} and \eqref{eq:massmatrix2}. The Higgs mass counterterm matrix at $n$-loop
level is denoted by $\deltan\calM_{hh}$ and  is obtained by replacing
the parameters {$P_i$} on which it depends by their renormalised quantities
plus corresponding counterterms $\delta P_i$ up to $n$-loop order, {\it i.e.} {$P_i
\to P_i + \delta^{(1)} P_i+ \cdots+\delta^{(n)} P_i$}, and expanding accordingly. Its explicit
expression can be found in the Appendix G of
Ref. \cite{Dao:2019qaz}.
The Higgs field renormalisation constant
matrix is given by
\be   
\delta^{(n)} {\cal Z}= \mathcal{R}^G \delta^{(n)} {\cal Z}^G \left(\mathcal{R}^G\right)^T
\ee
with
\be
\delta^{(n)} {\cal Z}^G =\diag (\Delta^{(n)} Z_{H_d},\Delta^{(n)} Z_{H_u},\Delta^{(n)}
Z_{S}, \Delta^{(n)} Z_{H_d},\Delta^{(n)}
Z_{H_u}, \Delta^{(n)} Z_{S})\,, \quad n=1,2 \;
\ee
where the renormalisation constants $\Delta^{(n)} Z_{\Phi}$, $\Phi=
H_u,H_d,S$ for the doublet and singlet fields will be given in
Sec.~\ref{sec:1and2ct}. \s

Similarly for the charged Higgs boson sector, the one-loop and two-loop
renormalised self-energies for the transitions 
$h^+_i\to h^+_j$ with $h_1^+ \equiv G^+$ and $h_2^+ \equiv H^+$ are
given by \eqref{eq:rehiggsself1} and
\eqref{eq:rehiggsself2}, respectively, with the following replacements 
\begin{align} 
\calM_{hh} &\to \calM_{h^+h^-}, \quad  &\calR &\to \calR^{G^-}\crn
   \deltan \calM_{hh} &\to \deltan\calM_{h^+h^-}, \quad  &\waven & \to
\waven_{h^+h^-}\,, \quad n=1,2 \;,
\end{align}
where the charged Higgs mass matrix $\calM_{h^+h^-}$ and the rotation
matrix $\calR^{G^-}$ are defined in \eqref{eq:chargedHiggsmass} and
the charged Higgs field renormalisation constant matrix is given by
\be  
\waven_{h^+h^-}^G =\diag (\Delta^{(n)} Z_{H_d},\Delta^{(n)} Z_{H_u}) \;. 
\ee

A list of all two-loop diagrams considered in this work to
calculate the unrenormalised self-energies for the charged and
neutral Higgs bosons is given in \cref{app:selfdiags}.

\subsection{One-Loop and Two-Loop Counterterms \label{sec:1and2ct}}
The Feynman integrals are calculated in dimensional regularisation in
$D=4-2\epsilon$ dimensions. Therefore, intermediate results will
contain ultraviolet (UV) divergences of the order $\order{\epsilon^{-2}}$ and
$\order{\epsilon^{-1}}$.
To render the renormalised self-energies UV-finite the relevant
parameters need to be renormalised 
to either one- or two-loop order depending on the explicit
dependence of the tree-level Higgs boson masses on the parameters. In
our case, this means that the top/stop and the electroweakino sector need
to be renormalised only at one-loop order while all other parameters are
required up to $\order{\epsilon^{-2}}$. Note that the chargino and
neutralino masses are derived quantities and depend on the (one-loop)
counterterms of the input parameters.  In the choice of the renormalisation schemes we follow
our previous two-loop calculations
\cite{Muhlleitner:2015dua,Dao:2019qaz}. Note that for parameters defined in the $\OS$ scheme, we also study the
dependence on $\order{\epsilon^1}$-terms in the corresponding one-loop
counterterms which are potentially multiplied with $\order{\epsilon^{-1}}$-terms from
loop-integrals and other one-loop counterterms, thereby generating additional finite
contributions. In this section we give explicit expressions for all
needed one-loop counterterms  
after applying our approximations. The remaining one-loop counterterms
have already been computed in Refs.~\cite{Ender:2011qh,Graf:2012hh}
which worked out the renormalisation of the
full NMSSM at the one-loop level in the $\DRbar$, OS and the mixed
$\DRb$-OS scheme.

\subsubsection{The Higgs Sector}
\label{sec:ren:higgs}
In the Higgs sector we apply a mixed $\DRb$-OS renormalisation
scheme. Working in the gaugeless limit at two-loop order the counterterm of the electric charge vanishes.
Furthermore, the counterterms of the phases
$\varphi_\alpha$ ($\alpha=s,u,\kappa,\lambda$) can be set to zero in
order to obtain a UV-finite result. Since the charged Higgs mass
$M_{H^\pm}$ can be traded for $\mbox{Re} A_\lambda$ and vice versa we have the
following two possible sets of input parameters together with the
applied renormalisation conditions,
\begin{eqnarray}
\underbrace{ t_{h_d},t_{h_u},t_{h_s},t_{a_d},t_{a_s},M_{H^\pm}^2,\vev,s_{\theta_W}}_{\mbox{on-shell
 scheme}},
\underbrace{\tan\beta,|\lambda|,v_s,|\kappa|,\mbox{Re} A_\kappa}_{\overline{\mbox{DR}} \mbox{ scheme}}\,,
\label{eq:mixedcond1}
\end{eqnarray}
in case $M_{H^\pm}^2$ is used as independent input, or
\begin{eqnarray}
\underbrace{ t_{h_d},t_{h_u},t_{h_s},t_{a_d},t_{a_s},\vev,s_{\theta_W}}_{\mbox{on-shell
 scheme}},
\underbrace{\tan\beta,|\lambda|,v_s,|\kappa|,\mbox{Re}
  A_\lambda,\mbox{Re} A_\kappa}_{\overline{\mbox{DR}} \mbox{ scheme}}\,,
\label{eq:mixedcond2}
\end{eqnarray}
for $\mbox{Re}A_\lambda$ as independent input. All above listed
parameters are renormalised at two-loop level except for the sine $s_{\theta_W}$ of
the Weinberg angle $\theta_W$ where only the non-vanishing one-loop
counterterm contributes. The matrix-valued Higgs field renormalisation
constants are needed up to two-loop level and are defined via $\DRb$
conditions as explained in the following.  
\\
\\
\underline{\textit{Higgs Boson Wave-Function Renormalisation Constants}}
\\
The field renormalisation of the Higgs boson gauge eigenstates\footnote{All off-diagonal renormalisation constants have been verified to vanish at one- and
two-loop order.} ($\Phi=H_{u,d},S$), 
\begin{equation}
    \Phi \to \left(1+
        \frac{1}{2}\Delta^{(1)}Z_\Phi 
        +\frac{1}{2}\Delta^{(2)}Z_\Phi
    \right)\Phi \;,
\end{equation}
with
\begin{equation} 
    \Delta^{(1)}Z_\Phi = \delta^{(1)}Z_\Phi \quad \mbox{and} \quad
    \Delta^{(2)}Z_\Phi = \delta^{(2)}Z_\Phi
    -\left(\frac{\delta^{(1)}Z_\Phi 
    }{2}\right)^2,\,\, 
    \label{eq:wfrs}
\end{equation}
is carried out in the $\DRbar$ scheme. We obtain the
  counterterms in two equivalent ways. They are computed by either
using Feynman diagrams or the  renormalisation group equations
(RGEs). For the former, they are given by the UV-divergent part of
the derivative of the unrenormalised self-energies with respect to
the momentum squared   
\begin{equation}
    \delta^{(n)} Z_{\Phi} =- \left. \frac{\del
        \Sigma^{(n)}_{\phi\phi}(p^2)}{\del p^2} 
\right|_{\text{\tiny div}}\,,\quad n=1,2 \quad \text{and} \quad \phi = h_d, h_u, h_s\;.
\end{equation}
For the latter, they can be written as \cite{Sperling:2013xqa,Sperling:2013eva}
\bea 
\delta^{(1)} Z_{\Phi} &=& \gamma_{\phi\phi}^{(1)} \fr{1}{\epsilon}\,,\\
\delta^{(2)} Z_{\Phi} &=& \fr12\gamma_{\phi\phi}^{(2)}
\fr{1}{\epsilon} + \fr12 \sbraket{(\gamma_{\phi\phi}^{(1)} )^2
  +\sum_{x}\beta^{(1)}(x) \fr{\del \gamma_{\phi\phi}^{(1)} }{\del
    x}}\fr{1}{\epsilon^2}\,, 
 \eea
where $\gamma_{\phi\phi}$ is the anomalous dimension of the
corresponding scalar field $\phi = h_d,h_u,h_s$, $x=\{y_t,\lambda,\kappa\}$ with 
$y_t=\sqrt{2}m_t/(\vev s_\beta)$ and $\beta^{(1)}(x)$ is the one-loop beta
function of the coupling $x$. The functions $\gamma_{\phi\phi}$ and 
$\beta(y_t)$ at one- and two-loop level can be obtained  from either
\cite{Sperling:2013xqa,Sperling:2013eva} or 
 the package $\SARAH$. Note that the RGE results are in the pure
 $\DRbar$ scheme which means that all parameters are renormalised in
 the $\DRbar$ scheme. In the following we will use the superscript
 $\DRbar$ on the wave-function renormalisation constants to indicate
 the pure $\DRbar$ scheme while we use the superscript $\OS$ for the
 scheme where $y_t$ and $v$ are renormalised in the OS scheme. Our 
diagrammatic results in the pure $\DRbar$ scheme are in full agreement
with the RGE results. At one-loop order, we find 
\begin{align}
    \delta^{(1)} Z_{H_d}^{\DRbar}& = - k |\lambda|^2 \fr{1}{\epsilon}\,,\\
    \delta^{(1)} Z_{H_u}^{\DRbar}& = - k \left( |\lambda|^2 + 3{y_t^{\DRbar}}^2 \right)\fr{1}{\epsilon} \,,\label{eq:ZHu1} \\
    \delta^{(1)} Z_{S}^{\DRbar}& = - 2k \left(|\lambda|^2 + |\kappa|^2
                                 \right)\fr{1}{\epsilon}\, , 
\end{align}    
while the two-loop results yield, 
\begin{align}                 
    \delta^{(2)} Z_{H_d}^{\DRbar}& = -\frac{k^2}{2} |\lambda|^2\left[2|\kappa|^2 + 3|\lambda^2|+  3{y_t^{\tiny \DRbar}}^2 \right]\braket{\fr{1}{\epsilon^2}-\fr{1}{\epsilon}}\,,  \\
    \delta^{(2)} Z_{H_u}^{\DRbar}& = -\frac{k^2}{2} |\lambda|^2\left[2|\kappa|^2 + 3|\lambda^2|+  9\frac{{y_t^{\tiny \DRbar}}^4}{|\lambda|^2}\right]\braket{\fr{1}{\epsilon^2}-\fr{1}{\epsilon}}\,, \label{eq:ZHu2}  \\
    \delta^{(2)} Z_{S}^{\DRbar}& = - 4k^2\left[|\kappa|^4+|\kappa|^2|\lambda|^2+ \frac{|\lambda|^4}{2} + \frac{3}{4} |\lambda|^2 {y_t^{\tiny \DRbar}}^2 \right]\braket{\fr{1}{\epsilon^2}-\fr{1}{\epsilon}} \,,
\end{align}
where $k=1/(4\pi)^2$.
A closer look at \cref{eq:ZHu1} shows that the scheme change from
$\DRbar$ to $\OS$ in the top/stop sector introduces additional
higher-order contributions via the one-loop field constant
$\delta^{(1)} Z_{H_u}$ 
\begin{align}
    \delta^{(2)} Z_{H_d}^{\OS} & = \delta^{(2)} Z^{\DRbar}_{H_d}    \\
    \delta^{(2)} Z_{H_u}^{\OS} & = \delta^{(2)} Z^{\DRbar}_{H_u}(y_t^{\OS}) + \left(\frac{\del}{\del y_t^{\OS}}\delta^{(1)} Z_{H_u}^{\DRbar}\right)\left(\delta^{(1)}y_t^{\OS}|_{\text{fin}}\right) \label{eq:dZHuOS}\\
    \delta^{(2)} Z_{H_s}^{\OS} & = \delta^{(2)} Z^{\DRbar}_{H_s} \;.
\end{align}
As noted above, the field constants denoted by
the superscript $\OS$ are actually still $\DRbar$-renormalised,
  {\it i.e.}~only the UV-divergent parts are taken into account, but only refer to the
additional UV-divergent sub-loop contributions from the top/stop
sector.
Note that we write the one-loop OS counterterm in the following form
\begin{equation}
    \delta^{(1)}y_t^{\OS} = \delta^{(1)}y_t^{\OS}|_{\text{fin}} + \frac{1}{\epsilon}\delta^{(1)}y_t^{\OS}|_{\epsilon^{-1}} + \epsilon\, \delta^{(1)}y_t^{\OS}|_{\epsilon}
    \label{eq:dytos}
\end{equation}
and similarly for all other one-loop $\OS$ counterterms.
Solving the top mass counterterm \cref{eq:deltamt} for
$\delta^{(1)}y_t$ and expanding \cref{eq:dZHuOS} to $\order{\epsilon^{-1}}$ we find 
\begin{equation}
    \delta^{(2)} Z_{H_u}^{\OS}- \delta^{(2)} Z_{H_u}^{\DRbar} = -\frac{1}{\epsilon}\frac{3 {m_t^{\OS}}^2}{4 \pi^2 \vev^2 \sb^2}\left(\frac{\delta^{(1)} m_t^{\OS}|_{\text{fin}}}{m_t^{\OS}}- \frac{\delta^{(1)}\vev^{\OS}|_{\text{fin}}}{\vev^{\OS}} \right) \;,
    \label{eq:zshift}
\end{equation}
where 
\DeclarePairedDelimiter\abs{\lvert}{\rvert}%
\newcommand{\ti}{\tilde  }
\newcommand{\mcha}{m_{\tilde \chi^\pm} }
\newcommand{\mchii}{m_{\ti \chi_i^0}}
\newcommand{\mchij}{m_{\ti \chi_j^0}}
\newcommand{\mhis}{m_{h_i}^2}
\newcommand{\mhjs}{m_{h_j}^2}
\newcommand{\mathcalR}{\mathcal{R}}
\bea
 \delta^{(1)} m_t^{\OS}|_{\text{fin}}&=&m_t^3\left( -c_{\beta}^2 B_1(m_t^2)0,\mhpm^2)+ \frac{s_{\beta}^2}{2} + B_0(m_t^2,m_{\tilde Q_3}^2, m_{\tilde\chi^+_2}^2) + B_1(m_t^2,m_{\tilde Q_3}^2, m_{\tilde\chi^+_2}^2) \right)\nonumber  \\
&& +m_t^3 \sum_{i=1}^5 \left((\mathcal R_{i2}^2 -\mathcal R_{i5}^2  )B_0(m_t^2,m_t^2, m_{h_i}^2)-
(\mathcal R_{i2}^2 +\mathcal R_{i5}^2  )B_1(m_t^2,m_t^2, m_{h_i}^2)\right) \nonumber\\
&&+ \sum_{i=1}^3\sum_{j=1}^2\bigg[ m_t^3B_1(m_t^2, m_{\tilde t_j}^2, m_{\tilde \chi^0_i}^2)|{N_{i2}}|^2+
m_t^2B_0(m_t^2, m_{\tilde t_j}^2, m_{\tilde \chi^0_i}^2) (m_t |{N_{i2}}|^2 \nonumber\\
&&+ e^{-i\varphi_u}m_{\tilde \chi^0_i} (N_{i2}^*)^2
\mathcal{U}^{\tilde t}_{i2} \mathcal{U}^{*\tilde t}_{i1}+
e^{i\varphi_u}m_{\tilde \chi^0_i} N_{i2}^2
\mathcal{U}^{\tilde t}_{i1} \mathcal{U}^{*\tilde t}_{i2} )\bigg]\, \\
\delta^{(1)}\vev^{\OS}|_{\text{fin}}&=& \fr{3}{32 \pi^2 s_{\theta_W}^2 v} \bigg(c_{2\theta_W}
|\mathcal{U}^{\tilde t}_{11}|^2 F_0(m_{\tilde t_1}^2,m_{\tilde Q_3}^2)
+  c_{2\theta_W} |\mathcal{U}_{\tilde t_{21}}|^2 F_0(m_{\tilde
  t_2}^2,m_{\tilde Q_3}^2)  \crn 
&&- c_{\theta_W}^2 |\mathcal{U}^{\tilde t}_{11}|^2 |\mathcal{U}^{\tilde 
  t}_{12}|^2 F_0(m_{\tilde t_1}^2,m_{\tilde t_2}^2) \bigg)  + \frac{1}{16\pi^2 vs_{\theta_W}^2}\Delta\vev\,,  
\eea
with
\bea
\Delta\vev&=& -c_{2\theta_W} A_0(\mhpm^2) + c_{2\theta_W}\sum_{i=1}^3\bigg[ (\abs{N_{i1}}^2+
\abs{N_{i2}}^2) F_1(\mcha^2 , \mchii^2 )\crn
&& +8 \mcha \mchii \Re(e^{i\varphi_s} N^*_{i1}N_{i2}) B_0(0,\mcha^2, \mchii^2)\bigg]\crn
&&-\frac14 c_{2\theta_W} \sum_{i,j=1}^3\bigg[ \abs{(N^*_{i1}N_{j1}-N^*_{i2}N_{j2})}^2 F_1(\mchij^2 ,\mchii^2 )\crn
&& +4\mchii \mchij B_0(0,\mcha^2, \mchii^2)(N^*_{i1}N_{j1}-N^*_{i2}N_{j2})^2\bigg]\crn
&&+\frac14\sum_{i=1}^5\bigg[ 2 s_{\theta_W}^2 (\mathcalR_{i1}^2+\mathcalR_{i2}^2+
\mathcalR_{i4}^2)A_0(\mhis)
+ c_{2\theta_W} F_2(\mhis,\mhpm^2)\crn
&&\times \left( (s_\beta \mathcalR_{i1} - c_\beta \mathcalR_{i2})^2 +  \mathcalR_{i4}^2 \right)+ F_2(\mhis,0)\left( -s_{\theta_W}^2(c_\beta \mathcalR_{i1} + s_\beta \mathcalR_{i2})^2 \right)\bigg]\crn
&&-\frac18 c_{2\theta_W}\sum_{i,j=1}^5 F_2(\mhis,\mhjs) \left(
(s_\beta \mathcalR_{j1}- c_\beta \mathcalR_{j2})\mathcalR_{i4}+
(s_\beta \mathcalR_{i1}- c_\beta \mathcalR_{i2})\mathcalR_{j4}
\right)
\eea
and
\bea 
F_0(x,y) &=& x+y -\fr{2xy}{x-y}\log\fr{x}{y}\,, \\
F_1(x,y) &=& x+y -2\fr{x^2\overline{\log}(x)-y^2\overline{\log}(y)}{x-y}\,, \\
F_2(x,y) &=& 3x+3y -2\fr{x^2\overline{\log}(x)-y^2\overline{\log}(y)}{x-y}\,.
\eea 
and where $A_0$ and $B_0$ denote the scalar
  one-loop one-point and two-point functions and $B_1$ the tensor
  one-loop two-point function \cite{thooft} and $\overline{\log}(x) = \log\left( x/\mu_R^2\right)$ with the renormalization scale~$\mu_R$. \s

In the Feynman diagrammatic approach we have computed $\delta^{(2)}
Z_{\Phi}$ in two different ways. 
In one computation we kept the full momentum dependence in the
UV-divergent parts of 
all diagrams. We then evaluated each contribution with non-zero
momentum and found the sum of all contributions being
independent of $p^2$. In another computation we took 
the limit $p^2\to 0$ right after taking the derivative. Here, the
coefficients of intermediate results of the single poles feature
logarithmic and quadratic IR divergences. As will be discussed later in
Sec.~\ref{sec:self-energies}, a mass regulator can be introduced to
deal with the IR divergences. We found full agreement with the finite
$p^2$-result and no dependence on the mass regulator in the sum of all
Feynman diagrams when using the IR-save loop functions defined in
\cref{app:irsafeloop}, which gives us yet another possibility to
verify if a mass-regularisation scheme is actually useful. \\ \\
\underline{\textit{The VEV and the Weak Mixing Angle Counterterm}}\\
The VEV countertem in the OS scheme is given by
\begin{align}
    \frac{\delta^{(n)}\vev^{\OS}}{\vev ^{\OS}} &= \frac{\cw^2}{2\sw^2}\left(
\frac{\delta^{(n)}M_Z^2}{M_Z^2} - \frac{\delta^{(n)}M_W^2}{M_W^2} 
\right) +
\frac{\delta^{(n)}M_W^2}{2 M_W^2} + \delta_{n2} \delta^{(2)} \bar{v}
, \,\,\, n=1,2\, ,\\ \nonumber
        \delta^{(2)} \bar{v} &= -\frac{1}{8\sw^4}\left[ 
        \left(\frac{\delta^{(1)}M_W^2}{M_W^2}\right)^2
        -2\cw^2(1+2\sw^2)\frac{\delta^{(1)}M_W^2}{M_W^2} \frac{\delta^{(1)}M_Z^2}{M_Z^2} \right. \\ \nonumber
    &\left. \qquad\qquad\quad      +\cw^2(1+3\sw^2)\left(\frac{\delta^{(1)}M_Z^2}{M_Z^2}\right)^2 \right] 
\end{align}
where $\delta_{nm}$ is the Kronecker delta. 
The weak mixing angle counterterm at one-loop order reads
\be 
\delta^{(1)} s_{\theta_W} = \fr{ c_{\theta_W}^2}{ 2s_{\theta_W}} \left(
\frac{\delta^{(1)}M_Z^2}{M_Z^2} - \frac{\delta^{(1)}M_W^2}{M_W^2} 
\right) \;,
\ee
where  $M_{W/Z}^2$ are the
squared vector-boson masses. The vector bosons are renormalised OS
with the corresponding counterterms given by
\begin{eqnarray}
\frac{\delta^{(n)}M_W^2}{M_W^2} = \frac{\Sigma_W^{(n),T} (0)}{M_W^2}
\quad \mbox{and} \quad
\frac{\delta^{(n)}M_Z^2}{M_Z^2} = \frac{\Sigma_Z^{(n),T} (0)}{M_Z^2} \;,
\end{eqnarray}
with $\Sigma_{V}^{(n),T}$ ($V=W,Z$) denoting the transverse part of the
unrenormalised $n$-loop vector boson self-energy evaluated at zero
external momentum. Note that whereas $\delta^{(n)}M_V^2$ and $M_V^2$
are separately zero in the gaugeless limit, their ratio entering the
counterterms of the VEV and $\sin \theta_W$ is
non-zero. \s

In the pure $\DRb$ scheme, the one-loop counterterm $\delta^{(1)}
s_{\theta_W}^{\DRbar} $ vanishes while the explicit evaluation of the UV-divergent part of
the VEV counterterm is found to be 
\begin{eqnarray}
    \frac{\delta^{(n)} \vev^{\DRbar}}{\vev^{\DRbar}}
    = \frac{s_\beta^2}{2} \Delta^{(n)} Z_{H_u}^{\DRbar} + \frac{c_\beta^2}{2} \Delta^{(n)} Z_{H_d}^{\DRbar} + \delta_{2n} \frac{s^2_{2\beta}}{32}\left(\delta^{(1)} Z_{H_d}^{\DRbar} - \delta^{(1)}Z_{H_u}^{\DRbar} \right)^2 \;.
\label{eq:vevrelation}
\end{eqnarray}
This is in accordance with the relation given in
Refs.~\cite{Sperling:2013xqa,Sperling:2013eva}
\begin{equation}
    \frac{\delta^{(n)}v_i^{\DRbar}}{v_i^{\DRbar}} = \frac{1}{2}\Delta^{(n)}Z_{H_i}^{\DRbar}  \, \quad 
i=u,d,s \;,
    \label{eq:vevzfac}
\end{equation}
which connects the counterterm of the VEV $v_i$ to the field
renormalisation constant of the respective field $H_i$. Exploiting
$\vev=\sqrt{v_u^2+v_d^2}$ yields Eq.~(\ref{eq:vevrelation}). \s

Note that since the ${\Sigma_{Z,W}^{(2),T}}$ are evaluated at vanishing external
momentum, we encounter intermediate IR divergences due to the appearance of
massless Goldstone boson propagators. 
However, these divergences cancel in the sum of all two-loop
self-energy diagrams. This will be discussed in detail in Sec.~\ref{sec:self-energies}.
\\
\\
\underline{\textit{Tadpole Parameters}}
\\
Requiring the tree-level minimum of the potential to be the
\textit{true} minimum, higher-order tadpole contributions must be
fully compensated by their counterterms, {\it i.e.}\footnote{Note that there is a typo in $\delta^{(2)}t_\phi$
  in \cite{Dao:2019qaz}.}
\begin{subequations}
\begin{align}
    \delta^{(1)}t_\phi &= t^{(1)}_\phi  \\
    \delta^{(2)}t_\phi &= t^{(2)}_\phi - \frac{\delta^{(1)}{\cal Z}_{\phi\phi}}{2} \delta^{(1)}t_{\phi}
\, ,
\end{align}
\label{eq:tadren}%
\end{subequations}
where $t_{\phi}^{(n)}$ is the $n$-loop tadpole contribution of the external field $\phi$, {$\phi=h_{d},h_u,h_s,
a_d,a_s$}.
In Eqs.~(\ref{eq:mixedcond1}) and (\ref{eq:mixedcond2}), we call this
in slight abuse of the language an $\OS$ condition even though the tadpoles are strictly
speaking not associated with any on-shell field. 
The full set of all two-loop tadpole diagrams considered in this work
is given in \cref{app:tads}. 
\\
\\
\underline{\textit{Charged Higgs Boson Mass}}
\\
As mentioned earlier, we have the option to choose either $\Re
A_\lambda$ in the $\DRbar$ scheme as input parameter or $\mhpm^2$ in the
OS scheme. \s

In the approximation of vanishing external momentum, the mixing
between the charged Higgs boson and the charged Goldstone boson is
negligible. If the external momentum squared is set equal to the
charged Higgs boson mass squared one may have to consider this mixing
effect, however. We follow our definition of the charged Higgs mass counterterms at
one- and two-loop order given in Ref.~\cite{Dao:2019qaz}. For convenience, we
present here the most important formulae. 
At one-loop order, the charged Higgs boson mass counterterm in the OS
scheme is given by 
\begin{equation}
    \delta^{(1)}M_{H^\pm}^{2} = \Sigma^{(1)}_{H^- H^-}(p^2=0) -
    M_{H^\pm}^2 \delta^{(1)}\mathcal{Z}_{H^-H^-} \;,
\end{equation}
while at two-loop order we have
\begin{align}
    \delta^{(2)} M_{H^\pm}^{2} =& \Sigma^{(2)}_{H^- H^-}(p^2=0) - M_{H^\pm}^2\left[ \fr 14\left(\delta^{(1)}\mathcal{Z}_{H^-H^-}\right)^2 + \delta^{(2)}\mathcal{Z}_{H^-H^-}  \right] \nonumber \\
                                       & - \delta^{(1)}\mathcal{Z}_{H^-H^-}\delta^{(1)}{M_{H^\pm}^{2}} - \delta^{(1)}\mathcal{Z}_{H^-G^-}\delta^{(1)}m_{H^-G^-}
    \, ,
    \label{eq:charged2LCT}
\end{align}
with
\bea 
\delta^{(n)} {\cal Z}_{H^-H^-}&=& \cos^2\!\beta \Delta^{(n)} Z_{H_u}+\sin^2\!\beta
\Delta^{(n)} Z_{H_d}\\ 
\wave_ {H^-G^-} &=& \cos\beta \sin\beta(-\deltaone
Z_{H_d} + \deltaone Z_{H_u})\\ 
\deltaone m_{H^-G^-} &=&\frac{-c_\beta^2 M_{H^\pm}^2 v \deltaone \,t_\beta
  +c_\beta \deltaone \,t_{h_u}-\deltaone \,t_{h_d}s_\beta}{v}+\frac{i \deltaone
  \,t_{a_d}}{s_\beta v}\;.
\eea

In case $\Re A_\lambda$ is given as independent input parameter, 
the charged Higgs mass counterterms have to be obtained as functions
of all other counterterms by inserting their respective loop
expansions in the formula for the charged Higgs boson mass. For the
explicit formulae of the counterterms and details on the calculation
of the loop-corrected charged Higgs boson mass, we refer to
\cite{Dao:2019qaz}. \s

Some of the two-loop charged Higgs boson self-energy diagrams suffer
from IR divergences, but the sum of all contributions is indeed
IR-finite and does not dependent on the regulator mass. 
\\
\\
\underline{\textit{Ratio of the VEVs $\tan\beta$}}
\\
The parameter $\tan\beta$ is given by the ratio of the VEVs $v_u$ and $v_d$.
Using \cref{eq:vevzfac} its counterterm can be related to the field
renormalisation constants which are calculated in the $\DRbar$ scheme 
so that the one-loop $\DRbar$ counterterm is given by
\begin{equation}
    \delta^{(1)}\tan\beta =
    \frac{\tan\beta}{2}\left(\delta^{(1)}Z_{H_u}- \delta^{(1)}Z_{H_d}
    \right) =    \frac{3}{2}k\tan\beta y_t^2
      \fr{1}{\epsilon} \,, \label{eq:delta1TB}
\end{equation}
while the two-loop expansion yields
\begin{equation}
    \delta^{(2)}\tan\beta = \frac{\tan\beta}{2}\left( \Delta^{(2)}Z_{H_u}- \Delta^{(2)}Z_{H_d} \right) +\frac{\tan\beta}{4}\left[ \left( \delta^{(1)}Z_{H_d} \right)^2 - \delta^{(1)}Z_{H_d} \delta^{(1)}Z_{H_u} \right]
    \, , \label{eq:delta2tanbeta}
\end{equation}
where $\Delta^{(2)}Z_{H_{u,d}} $ were defined in \eqref{eq:wfrs}. 
As can be inferred from Eq.~(\ref{eq:delta2tanbeta}) the two-loop
counterterm of $\tan\beta$ also depends on $\delta^{(2)}
Z_{H_u}$ which receives additional UV-divergent shifts when we change
the renormalisation scheme of the top/stop sector. Accordingly,
$\delta^{(2)}\tan\beta$ will be affected by such a scheme change.
\\
\\
\underline{\textit{Superpotential Parameters, Soft-SUSY-Breaking Parameters and
    Singlet VEV}}
\\
Due to SUSY-non-renormalisation theorems
\cite{Salam:1974jj,Grisaru:1979wc,West:1990rm}, the superpotential
parameters are renormalised through the field renormalisation
constants. Therefore, we have two possibilities to construct the
one-loop counterterms: (i) using RGEs for the 
superpotential parameters
\cite{Staub:2008uz,Staub:2010jh,Staub:2012pb,Staub:2013tta,Goodsell:2014bna,Goodsell:2014pla},
(ii) using the calculated field renormalisation constants together
with the non-renormalisation theorems. At one-loop order
we verified that the two methods 
yield consistent results, resulting in the $\DRbar$ counterterms 
    \begin{align}
        \delta^{(1)} |\lambda| &= \frac{1}{2}\beta_{|\lambda|}^{(1)}\fr{1}{\epsilon} = -\frac{|\lambda|}{2}\sum_i^{H_u,H_d,S} \delta^{(1)}Z_{i}  =  \frac{k |\lambda|}{2} \left(2|\kappa|^2 + 4|\lambda|^2 + 3 {y_t}^2  \right)\fr{1}{\epsilon}
        \label{eq:lambdact1L} \\
        \delta^{(1)} |\kappa| &= \frac{1}{2}\beta_{|\kappa|}^{(1)}
                                \fr{1}{\epsilon} = -
\frac{3|\kappa|}{2}\delta^{(1)}Z_{S}  = 3 k |\kappa| \left(|\kappa|^2
                                + |\lambda|^2 \right)\fr{1}{\epsilon}
\label{eq:kappdact1L}
\, .
\end{align}
The $\DRbar$ counterterm of the singlet VEV can be obtained by using
\cref{eq:vevzfac},
\begin{equation}
        \delta^{(1)} \vs = \frac{\vs}{2}\delta^{(1)}Z_{S} 
        \, .
       \label{eq:vsct1L}
\end{equation}
The $\DRbar$ counterterm of the soft SUSY breaking coupling $\Re
A_\kappa$ can be obtained from either the one-loop RGE or a
diagrammatic calculation and reads 
    \begin{align}
        \delta^{(1)} \Re A_\kappa &= 6k\left(|\kappa|^2\Re A_\kappa +
                                    |\lambda|^2 \Re A_\lambda
                                    \right)\fr{1}{\epsilon} 
        \, .
        \label{eq:akapct}
    \end{align}
Likewise, the $\DRbar$ counterterm for $\delta^{(1)}\Re A_{\lambda}$
can be extracted from the RGEs, if $\Re A_{\lambda}$ is chosen as
input instead of the charged Higgs boson mass,
    \begin{equation}
        \delta^{(1)} \Re A_\lambda^{\DRbar}  = 
		k\left( 2|\kappa|^2\Re A_\kappa + 4|\lambda|^2\Re
                  A_\lambda + 3 \Re A_t y_t^2 \right) 
\fr{1}{\epsilon}
        \, .
    \end{equation}
Performing a one-loop counterterm expansion of the following
expression\footnote{The phases $\varphi_w$ and $\varphi_y$ have been
  defined in Eqs.~(\ref{eq:phase1}) and (\ref{eq:phase2}),
  respectively.} 
\begin{align}
\Re A_\lambda=&-\frac{|\kappa| v_s \cos (\varphi_w)}{\sqrt{2}}
+ \frac{s_{2\beta}\cos
        (\varphi_y-\varphi_w) }{\sqrt{2}|\lambda| c_{\beta-\beta_c}^2 v_s 
} \bigg( m_{H^\pm}^2 + \frac{1}{2} |\lambda| ^2 
c_{\beta-\beta_c}^2 \vev^2
\nonumber\displaybreak[0]\\[2mm]&
-\frac{s_\beta \left(c_\beta c_{\beta_c}^2 t_{h_u}+s_\beta s_{\beta_c}^2
                                  t_{h_d}\right)+
                                  c_{\beta-\beta_c}^2 t_{a_d} \tan (\varphi_y-\varphi_w)}{c_\beta s_\beta^2 \vev}\bigg)
\;,
\displaybreak[0]\, \label{eq:charalam}
\end{align}
in the gaugeless limit
yields a relation between the $\DRbar$ counterterm of $m_{H^\pm}^2$ 
and $\Re A_\lambda$ which was used to cross-check the UV-pole of the
charged Higgs boson self-energy.  \s

Furthermore, \cref{eq:akapct,eq:charalam} reveal the implicit
dependence of the $\DRbar$ counterterm 
$\delta^{(1)} \Re A_{\kappa}$ on  OS defined
parameters. This parametrisation of $\delta^{(1)} \Re A_{\kappa}$ is useful to study the generation
of additional two-loop contributions when performing a scheme change
by expanding  $\delta^{(1)}
A_{\kappa}$ about $\delta^{(1)} X^{\OS}$,  $X=\{\vev, \mhpm, t_i\}$ leading to terms of the form $\left(\partial_{X^{\OS}} \delta^{(1)}
A_{\kappa}\right) \left(\delta^{(1)} X^{\OS}|_{\text{fin}} 
\right)
$ which will be discussed in
\cref{sec:oeps}. \s

At two-loop order the missing two-loop
counterterms are constructed by demanding UV-finiteness in some
components of the renormalised neutral self-energies, 
\begin{subequations} 
\begin{align}
    \delta^{(2)} |\lambda|    &:&   \left. \left[\cb \hat{\Sigma}_{h_1 h_1} + \sb \hat{\Sigma}_{h_1 h_2} \right]\right|_{\text{UV-div}} &= 0 \\ 
    \delta^{(2)} \vs          &:&   \left. \left[\cb \hat{\Sigma}_{h_1 h_3} - \sb \hat{\Sigma}_{h_2 h_3} \right]\right|_{\text{UV-div}} &= 0 \\ 
    \delta^{(2)} |\kappa|     &:&   \left. \left[  \hat{\Sigma}_{h_2 h_3} \right]\right|_{\text{UV-div}} &= 0 \\ 
    \delta^{(2)} \Re A_\kappa &:&   \left. \left[  \hat{\Sigma}_{h_3 h_3} \right]\right|_{\text{UV-div}} &= 0 
    \, ,
\end{align}
    \label{eq:selfsolve}
\end{subequations}
which were verified to also render all other components UV-finite.
We found that the solutions to this system of equations are in agreement
with the following expressions of the counterterms for $|\lambda|$, $|\kappa|$ and
$\vs$ in the pure $\DRbar$ scheme, 
\begin{align}
    \delta^{(2)} |\lambda| &= -\frac{|\lambda|}{2}\left[ \sum_{i,j}^{H_u,H_d,S} \left(\Delta^{(2)}Z_{i} - \frac{1}{4}\delta^{(1)}Z_{i}\delta^{(1)}Z_{j}\left(1+\delta_{ij} \right) \right)
                                                    \right]
    \label{eq:lambdact2L}\\
    \delta^{(2)} |\kappa| &=  -3\frac{|\kappa|}{2} \left(\Delta^{(2)}Z_{S}  -\left(\delta^{(1)} Z_{S}\right)^2 \right)
    \label{eq:kappact2L}\\
    \delta^{(2)}\vs & = \frac{\vs}{2}\Delta^{(2)}Z_{S}
    \label{eq:vSct2L}
    \, ,
\end{align}
where \cref{eq:vSct2L} is in agreement with \cref{eq:vevzfac} while \cref{eq:lambdact2L,eq:kappact2L} can 
also be derived from the NMSSM superpotential, \cref{eq:wnmssm}, using the SUSY-non-renormalisation theorem.\s

Even though the superpotential parameters are
renormalised in the $\DRbar$ scheme, \cref{eq:lambdact2L} changes w.r.t. the
single-pole when changing from $\DRbar$ to $\OS$ in $\vev$ and/or
the top/stop sector. This is due to the change in $\delta^{(2)}Z_{H_u}$, {\it cf.}~Eq.~(\ref{eq:zshift}).

\subsubsection{The Higgsino Sector}
The chargino and neutralino masses are derived quantities that depend
on the input parameters $\vev,\,\vs,\, \beta,\, \lambda$ and $\kappa$,
\textit{cf.} \cref{sec:EWinoTree}. These parameters appear already in
the Higgs sector, we therefore do not need additional renormalisation
conditions. The chargino mass counterterm is found by expanding the
tree-level mass in the gaugeless limit and reads 
\begin{align}
    \delta^{(1)} m_{\tilde{\chi}^\pm_2} &= \delta|\mu_\text{eff}|=\frac{1}{\sqrt{2}} \left( \vs \delta^{(1)}|\lambda| + |\lambda| \delta^{(1)} \vs \right) \;,
\end{align}
while the neutralino mass counterterm matrix\footnote{This matrix is not diagonal in general.} is given by, 
\be N^* \delta^{(1)} \textbf{M}_{\chi^0} N^\dagger \,, \ee
where the non-vanishing components of the symmetric matrix
$\delta^{(1)} \textbf{M}_{\chi^0}$ are 
\begin{align}
    \left(\delta^{(1)} \textbf{M}_{\chi^0}\right)_{12} &= -e^{i (\varphi_\lambda+\varphi_s)} \delta|\mu_\text{eff}|\\ 
    \left(\delta^{(1)} \textbf{M}_{\chi^0}\right)_{13} &=  -\fr{e^{i\varphi_u}
\vev \lambda}{\sqrt2} \left( c_\beta^3 \delta^{(1)}\tan\beta + s_\beta\fr{\delta^{(1)}|\lambda|}{|\lambda|} + s_\beta \fr{\delta^{(1)}\vev}{\vev}   \right) \\
    \left(\delta^{(1)} \textbf{M}_{\chi^0}\right)_{23} &=
    -\frac{\cb \vev \lambda}{\sqrt{2}} \left(
   \fr{ \delta^{(1)}|\lambda| }{|\lambda|}+  \fr{\delta^{(1)}\vev}{\vev} -\cb \sb  \delta^{(1)}\tan\beta \right)\\
    \left(\delta^{(1)} \textbf{M}_{\chi^0}\right)_{33} &= \sqrt{2}
    e^{i \varphi_s} |\kappa|\vs \left( \fr{\delta^{(1)}|\kappa|}{|\kappa|} +\fr{
\delta^{(1)}\vs}{\vs}\right) \;.
\end{align}
They only enter in one-loop diagrams with a mass counterterm
insertion. As a further cross-check, we verified that these
counterterms render the renormalised chargino/neutralino self-energies
in the gaugeless limit UV-finite.

\subsubsection{The Squark Sector} 
In the squark sector we apply both the OS and the $\DRb$
renormalisation scheme for the following set of parameters at one-loop
level, 
\be m_t, \; m_{\tilde{Q}_3}, \; m_{\tilde t_R} \quad \mbox{and} \quad A_t \;.
 \ee 
 Expanding the OS and
$\overline{\mbox{DR}}$ counterterms of the parameters $X=m_t,
m_{\tilde{Q}_3},m_{\tilde{t}_R}, A_t$ in terms of the dimensional
regulator $\varepsilon$, we have 
\bea 
\delta X^{\OS} &=& \fr{1}{\varepsilon} \delta X_{\varepsilon^{-1}} +  \delta
X_{\text{fin}}  +\varepsilon \delta X_{\varepsilon} \label{eq:OScounterterm}\\
 \delta X^{\DRbar} &=& \fr{1}{\varepsilon} \delta X_{\varepsilon^{-1}}\,, 
\label{eq:DRbarcounterterm}
\eea
where we have kept the terms proportional to $\varepsilon$ in the
expansion of the $\OS$  
one-loop counterterms, since we want to investigate the dependence of
our results on these terms. 
We follow our definitions of the one-loop counterterms of the squark
sector in Ref.~\cite{Dao:2019qaz}. 
We therefore do not repeat the squark counterterm definitions
here, but only present analytic
expressions\footnote{These results are 
  consistent with the results from the RGEs for $y_t$ obtained by $\SARAH$
  \cite{Staub:2008uz,Staub:2010jh,Staub:2012pb,Staub:2013tta,Goodsell:2014bna,Goodsell:2014pla}.}
for these counterterms in the $\DRbar$ scheme. The one-loop
counterterm for top quark mass reads 
\begin{equation}
    \delta^{(1)} m_t^  = 
        \frac{1}{\sqrt{2}} \left[
            \vev \sb \delta^{(1)}y_t + y_t \left(\sb \delta^{(1)}\vev + \cb^3 \vev \delta^{(1)}\tan\beta \right)
        \right]\, ,
    \label{eq:deltamt}
\end{equation}
where $\delta^{(1)}\vev$ is obtained from
\eqref{eq:vevrelation}, $\delta^{(1)}\tan\beta$ is defined in \eqref{eq:delta1TB} and 
the counterterm for top quark Yukawa coupling reads
\begin{align}
\delta^{(1)}y_t &= \frac{1}{2}\beta^{(1)}_{y_t} =
y_t\frac{k}{2}\left( |\lambda|^2 + 6
{y_t}^2\right)  \fr{1}{\varepsilon}\, . 
\end{align}
The one-loop counterterm for the top-quark trilinear coupling is given by
\begin{align}
\delta^{(1)}A_t & = k\left( |\lambda|^2 A_\lambda + 6 A_t y_t^2\right) \fr{1}{\varepsilon}\;,
\end{align} 
while the counterterms for $m_{\tilde{Q}_3}$ and $m_{\tilde t_R}$ read
\begin{align}
\delta^{(1)} (m_{\tilde{Q}_3}^2)&=  \frac{1}{2} \beta_{m_{\tilde{Q}_3}^2} \fr{1}{\varepsilon}= \fr{k}{\varepsilon} y_t^2\left(|A_t|^2 + m_{\tilde{Q}_3}^2 +m_{\tilde t_R}^2 
+ c_\beta^2 M_{H^\pm}^2 -\fr12|\lambda|^2v_s^2\right) \\ 
\delta^{(1)} (m_{\tilde t_R}^2)^{\DRbar}&= 2 \delta^{(1)}( m_{\tilde{Q}_3}^2)\,.
\end{align} 
We have found full agreement between the UV-divergent parts of all
top/stop counterterms of the OS and the $\DRb$ scheme.

\subsection{Independence of $\order{\epsilon^1}$ Counterterms}
\label{sec:oeps}
At  $\order{\alpha_t\alpha_s}$
\cite{Muhlleitner:2014vsa,Muhlleitner:2015dua} and
$\order{\alpha_t^2}$ in the MSSM limit \cite{Dao:2019qaz} it was shown
that finite contributions 
generated by the $\order{\epsilon^1}$-terms of the top/stop counterterms can
be compensated by a finite shift in the two-loop wave-function
renormalisation constant $\delta^{(2)}Z_{H_u}$.\s

However, in the present
$\order{(\alpha_t+\alpha_\kappa+\alpha_\lambda)^2}$ calculation
additional finite contributions are generated in the renormalised 
self-energies by $\order{\epsilon^1}$-terms of on-shell counterterms of $\vev$, the charged Higgs
boson mass and the tadpoles. In order to keep the parameters $\tan
\beta$, $\lambda$, and $\Re A_\kappa$ as pure
$\DRbar$ parameters, which also corresponds to the aforementioned
and applied RGEs, finite parts have to be taken into account in the
respective counterterms. These finite parts can be derived from
\cref{eq:ZHu1,eq:ZHu2,eq:delta2tanbeta,eq:lambdact2L,eq:akapct},
\begin{subequations}
\begin{align}
    \left. \delta^{(2)} Z_{H_u}^{\OS} \right|_{\text{fin}} &= 
       \sum_{\alpha}^{\vev,m_t} \left(\frac{\partial}{\partial \alpha^\OS}\delta^{(1)}Z_{H_u}\right)
        \left(\epsilon\, \delta^{(1)}\alpha^{\text{OS}}|_{\epsilon^{1}}\right)\\
    \left. \delta^{(2)} \tan^{\OS}\beta \right|_{\text{fin}} &= 
        \frac{\tan\beta}{2}\left. \delta^{(2)} Z_{H_u}^{\OS} \right|_{\text{fin}}\\
    \left. \delta^{(2)} |\lambda^{\OS}| \right|_{\text{fin}} &= 
        -\frac{|\lambda|}{2} \delta^{(2)} Z_{H_u}^{\OS}|_{\text{fin}} \\
    \left. \delta^{(2)} \Re A_\kappa^{\OS} \right|_{\text{fin}} &= 
        \sum_{\alpha} \left(\frac{\partial}{\partial \alpha}\delta^{(1)}\Re A_\kappa\right)
         \left(\epsilon\, \delta^{(1)}\alpha^{\text{OS}}|_{\epsilon^{1}}\right)
         \,, \alpha = \{\vev,M_{H^\pm}^2,t_{h_d,h_u,a_d}\}
         \, .
\end{align}
         \label{eq:DRfinshifts}%
\end{subequations}
The superscript "$\OS$" on the left-hand side refers to the $\OS$-nature of the
implicit parameters $\vev, \, m_t,\, t_i$ and $M_{H^\pm}^2$.
In addition, also the SM VEV counterterm receives a finite contribution, 
$\left(\partial_{v} \delta^{(1)} \vev|_{\epsilon^{-1}}\right)\left( \delta^{(1)}\vev|_{\epsilon^{1}} \right)$,
from the $\order{\epsilon}$ expansion at two-loop order.
These finite parts in the counterterms cancel the $\mathcal
O(\epsilon^1)$ contributions from the one-loop counterterm insertion
diagrams in the self-energies. \s

Therefore, the implementation of the calculation in $\NMSSMCALC$ for
simplicity does not contain the $\order{\epsilon}$ contributions in
the $\OS$ one-loop counterterms and does not include any finite terms
in the two-loop $\DRbar$ counterterms. This leads to the same results
for the renormalised self-energies but requires less computational
resources.

\section{Two-Loop Corrections in the Gaugeless Limit}
\label{sec:self-energies}
We are working in the gaugeless limit where one neutral ($G^0$) 
and two charged Goldstone bosons ($G^\pm$) are massless. In this limit, in contrast
to the MSSM, the couplings between two Goldstone bosons with one/two
neutral Higgs bosons, two charged Higgs bosons, one/two $W$ and
one/two $Z$ bosons do not vanish in the NMSSM. In the computation of
the tadpoles and self-energies at vanishing external momentum IR
divergences appear due to the massless Goldstone bosons and the
non-vanishing  couplings.  In this section we discuss our approaches
to treat these  IR divergences in the computation of the tadpoles,
charged Higgs boson, $W$ and $Z$ boson self-energies and the neutral
Higgs boson self-energies. \s 

Before this study, there was only the code \SARAH which implemented
the two-loop corrections controlled 
by the NMSSM superpotential parameters $\lambda$ and $\kappa$
\cite{Braathen:2017izn}. The code makes use of the two-loop effective
potential for a general renormalisable theory computed in
\cite{Martin:2001vx}. A solution of the IR divergences, which are also
known as the Goldstone Boson Catastrophe\footnote{Originally, the term
  \textit{Goldstone Boson Catastrophe} goes back to 
spurious IR divergences and imaginary parts encountered in the SM effective
potential in the Landau gauge and its first derivative
\cite{Martin:2013gka}.
}, has been presented in
Ref. \cite{Braathen:2016cqe}.  In the effective potential approach,
the tadpoles and Higgs self-energies are derived from 
the first and second field derivatives of the effective potential
$V_H^{\text{eff}}(x^{\text{run}})$ where $x^{\text{run}}$ denotes
$\MSbar$ or $\DRb$ parameters in the model. The minimum of the
effective potential  
is obtained by solving the tadpole equations order by order. At the
minimum of the tree-level potential the running Goldstone boson
squared-mass\footnote{For 
  simplification, we do not distinguish between the masses of the
  neutral, positively and negatively charged Goldstone bosons here. In
  general, their running masses can be different.}
$(m_G^{\text{run}})^2$  is zero, but is non-zero in general (it can be
very small or negative \cite{Martin:2014bca}). In
$V_H^{\text{eff}}(x^{\text{run}})$, there are terms which are 
proportional to $(m_G^{\text{run}})^2
\overline{\log}(m_G^{\text{run}})^2$, where 
\beq
\overline{\log} (X^2) = \log \left(
  \frac{X^2}{\mu_R^2}\right) \;,
\eeq
with $\mu_R$ denoting the renormalisation scale. Therefore the tadpole equations
contain terms 
proportional to $\overline{\log}(m_G^{\text{run}})^2$ which are divergent in the limit
$(m_G^{\text{run}})^2\to 0$ and have an unphysical imaginary part if $(m_G^{\text{run}})^2< 0$. The 
 solution\footnote{It is close to the solutions worked out in Refs.
\cite{Martin:2014bca,Elias-Miro:2014pca,Pilaftsis:2015bbs,Espinosa:2016uaw}
for the SM and extended for the MSSM in Ref. \cite{Kumar:2016ltb}. The Goldstone
contributions are resummed by integrating out all heavy
degrees of freedom when calculating the $n-1$ loop-corrected Goldstone boson
mass and using this \textit{effective} Goldstone boson mass in the
minimization of the $n$-loop effective potential. It was shown in Ref.
\cite{Espinosa:2017aew}, that this procedure resums the spurious
IR divergences to all orders in perturbation theory.} proposed in \cite{Braathen:2016cqe} at two-loop order is in fact to replace  
\be  
(m_G^{\text{run}})^2 = (M_G^{\text{OS}})^2 - \Sigma_{GG}^{(1)} \,
\ee
where the OS mass is set to be zero and $\Sigma_{GG}^{(1)}$ denotes the one-loop unrenormalised self-energy for the Goldstone boson component. Using the 
above relation directly in the expressions of the tadpoles and
self-energies given in \cite{Martin:2003it} a cancellation is found of the divergent terms
$\overline{\log}(m_G^{\text{OS}})^2$ in the tadpoles at two-loop
order.  However, the divergences remain in some sets of neutral Higgs
boson self-energy diagrams. These diagrams are then identified and
calculated using a small momentum expansion. \s

In contrast to the previous study, we use the Feynman diagrammatic
approach to directly calculate scalar one- and two-point functions at
the two-loop order and do not use the available general expressions
for the self-energies. Our intermediate results contain UV and IR
divergences. The UV divergences are canceled by the counterterms 
of the parameters and fields introduced via the renormalisation procedure
as discussed in the previous section.  We are flexible in our choice of the
renormalisation schemes for different parameters. In particular, we
renormalise the tadpoles in the OS scheme. This means that the masses
of all Goldstone bosons are always on-shell and zero at all orders in
the perturbation theory. Using this OS scheme we find the full
cancellation of the IR divergences in the tadpoles, charged Higgs, $W$
and $Z$ boson self-energies, but only a partial cancellation in the neutral Higgs
boson self-energies. For the sets of diagrams that 
contain Goldstone bosons but do not have IR divergences we set the mass
of Goldstone boson equal to zero and do not mention them furthermore
in this section. We now discuss the sets of one- and/or two-loop
diagrams where IR divergences appear.  In \cref{sec:irtopos},
\cref{tab:irtoposTAD,tab:irtoposSELF}, we give the 
complete list of all IR-divergent two-loop tadpole and self-energy
topologies together with the loop integrals causing the IR
divergences. In the last column of these tables we specify whether the
IR divergence is spurious, {\it i.e.}~cancels against contributions of other diagrams, or
whether a non-zero external momentum needs to be included for the
regularisation.\s

We regulate all IR divergences in three different ways: 
$(i)$ By using a mass regulator everywhere. This allows us to study
which IR divergences are actually spurious. We expect a residual
dependence on the mass regulator only in the subset of diagrams that
require momentum regularisation for a physical result but
which can in principle also be treated by mass regularisation, see
Sec.~\ref{sec:gbcmassreg}. 
$(ii)$ By using a mass regulator only in the subset
which features spurious IR divergences. In the
remaining subset with the genuine IR divergences, we use non-zero 
external momentum and apply analytically known results for the 
small momentum expansion of the loop integrals. This is equivalent to
the generalised effective potential approximation introduced in
Refs.~\cite{Braathen:2016cqe,Braathen:2017izn} and is described in
detail in Sec.~\ref{sec:partialp}.
$(iii)$ By including the  full external momentum dependence in the
computation of all Feynman
diagrams of $\order{\left(\alpha_t+\alpha_\lambda+\alpha_\kappa
\right)^2}$ making use of \TSIL \cite{Martin:2005qm}. This requires only
the regularisation of the spurious IR divergences, see Sec.~\ref{sec:fullp}.
\subsection{Infrared Mass Regulator}
\label{sec:gbcmassreg}
Using a mass regulator $M_R^2$ in IR-divergent loop integrals, {\it
  cf.}~\cite{Braathen:2017izn} for the NMSSM, induces not only
$\overline{\log}^n M_R^2$-terms which diverge in the limit $M_R\to 0$,
but also $\order{M_R^2}$-terms which actually vanish in the IR-limit.
Therefore, for an arbitrary loop integral
$f(m_1^2,...,m_i^2,m_j^2,...,m_n^2)$ which diverges for any combination 
of $m_i,m_j\to 0$, we expand $f(m_1^2,...,M_R^2,M_R^2,...,m_n^2)$ around
$M_R^2=0$ up to first non-vanishing order in $\overline{\log}^n M_R^2$
and $M_R^{-2 n \leq -2}$. To 
some extend, this is equivalent to the 
expansion in the small Goldstone boson mass used in the resummation
procedure of the effective potential \cite{Braathen:2016cqe,Kumar:2016ltb}.
However, since we also want to regulate those divergences that do not
cancel out in the sum of all diagrams, we 
require a larger set of expanded loop functions which is given in 
\cref{app:irsafeloop}. Performing the expansion instead of simply
setting the Goldstone mass equal to a finite mass value should reduce the
dependence on the regulator mass further. \s

In order to investigate the cancellation of the IR divergences we
divide the topologies that contain IR divergences into five sets, {\it
cf.}~\cref{tab:irtoposTAD,tab:irtoposSELF} in the appendix. Set A includes
topologies without external momentum flowing into the loop,
\textit{i.e.} the tadpole topologies 1-3 in \cref{tab:irtoposTAD} 
and the self-energy topologies 8, 10 and 13 in
\cref{tab:irtoposSELF}. It is evident that these must form IR-finite
subsets. Set B contains the self-energy topologies 4, 7, 11  with two
Goldstone bosons with the same momentum in
\cref{tab:irtoposSELF}. This set B is also IR-finite. The self-energy
topologies 4, 7 and 11 with two Goldstone bosons that couple with one
external line belong to set C while the self-energy topologies with
three Goldstone bosons belong to set E. Set D contains the self-energy
topologies 5, 6, 9 and 12. Tadpoles contain only topologies of
set A, while the charged Higgs self-energy contains topologies of
the sets A and B. The $W$ and $Z$ boson self-energies contain
topologies of the sets A, B, D, and E where each of these sets is
separately IR-finite. The cancellation happens due to the
contributions from the finite parts of the neutral and charged
Goldstone boson mass counterterms $\left(\delta^{(1)}
  m_{G^0}^2\right)$ and $\left(\delta^{(1)}
  m_{G^\pm}^2\right)$ which are related to the finite part of
the one-loop counterterms as, 
\begin{align} 
\left(\delta^{(1)} m_{G^0}^2\right)_{\text{fin}}
= \left(\delta^{(1)} m_{G^\pm}^2\right)_{\text{fin}} 
&=\frac{s_\beta (\delta^{(1)} t_{h_u})_{\text{fin}} + c_\beta (\delta^{(1)}t_{h_d})_{\text{fin}} }{v} \,.
 \end{align}
\s

The neutral Higgs boson self-energies contain all five sets of diagrams. The
first derivative of the UV parts of the self-energies with respect to
the momentum squared are used to compute  
the wave function renormalisation constants (WFRs)
$\delta^{(n)}Z_{\Phi}$. We find that the  
WFRs are IR finite. This is consistent with \cite{Chetyrkin:1997fm,Vladimirov:1980xy}
where it was shown that the WFRs do not depend on the
regulator mass as long as appropriate counterterms are introduced, 
{\it cf.}~\cite{Brod:2020lhd} for a recent application. For the
neutral Higgs boson self-energies
we do not see such a cancellation in the sets D, and E but
only in the sets A, B and C. Individual contributions to the sets D
and E may contain divergences proportional to $1/M_R^2$, however in the sum they are
canceled out. The remaining IR divergence depends on the mass
regulator as $\overline{\log} M_R^2$ and $\overline{\log}^2 M_R^2$. A
detailed study of the phenomenological impact of the residual 
$M_R^2$ dependence on the Higgs boson mass corrections is performed in
\cref{sec:gbccomparison}. 

\subsection{Partial Momentum Dependence}
\label{sec:partialp}
A solution of the GBC not involving a mass regulator and without the need
of the time consuming numerical evaluation of all loop integrals
including external momentum is given by the generalised effective
potential approximation \cite{Braathen:2016cqe}. Here only the
subset of diagrams of \cref{fig:selfirfin}  with a residual
dependence on $M_R^2$ is computed at $p^2\neq 0$. Additionally, the
number of independent mass scales is significantly reduced for this
particular subset making it possible to evaluate it analytically
either with exact $p^2$ dependence or approximately around $p^2 =
0$. \s

As mentioned in the previous section, there exist residual IR
divergences in the sets D and E of the neutral Higgs boson self-energies.
These IR divergences can be avoided by using non-zero external
momentum. The obvious and fastest way is to use a small momentum
expansion only in these sets which is similar to the implementation of
the effective potential approximation in \SARAH \cite{Braathen:2017izn}.
For the sake of comparison we also apply this method in our code. For
sets D and E we calculate Feynman diagrams with full
momentum dependence and use the expansion of the loop integral around
$p^2=0$. In this evaluation the masses of the Goldstone bosons
are set to zero.
In the expansion we removed terms of ${\cal O}(p^2)$ but kept terms
proportional to $1/p^2,$ $\overline{\log}(p^2)$ and terms independent of $p^2$.  The
necessary special cases for the loop integrals have already been worked out in Refs.
\cite{Braathen:2016cqe,Braathen:2017izn,Martin:2003qz,Martin:2005qm,Broadhurst:1987ei,Davydychev:1992mt,Scharf:1993ds}. \s

Even when using finite external momentum, the individual diagrams in
set C, D and E still feature an IR divergence $\order{\lnMR}$ originating from the
integrals $\textbf{V}(x,0,z,u)$ and $\textbf{C}(x,0,0)$ which cancels
in the sum of all contributions.
The cancellation can be obtained by making use of the identity 
\begin{equation}
    \textbf{V}(x,0,z,u) = -\tilde{V}(x,z,u) -
    \left. \textbf{B}(z,u) \right|_{p^2=0}\textbf{C}(x,0,0).
\end{equation}
The integral $\tilde{V}(0,z,u)$ is IR-finite for $p^2\neq
0$ (since it scales with $\log p^2$) and has been calculated
in Ref. \cite{Braathen:2017izn}\footnote{Note that our notation
slightly differs from appendix A.1.1 in Ref. \cite{Braathen:2017izn}.
Their $\textbf{B}(x,y^\prime)$ and $\textbf{P}(z,u)$ corresponds to
our $\textbf{C}(x,y,y)$ and $-\left. \textbf{B}(z,u)
\right|_{p^2=0}$.}\footnote{There is a sign mistake in the
corresponding identity in Ref. \cite{Braathen:2017izn}
Eq. (A.5). However, the implementation in \SARAH has the correct
sign.}. Therefore, the choice of regulating $\textbf{C}(x,0,0)$ is not
important as there is no dependence on this function in the final
result. The small momentum approximation, however, breaks down latest
near the various thresholds involved in the diagrams, making the full
momentum dependence necessary for a reliable result.\s

At the one-loop order, it is easy to prove that a strict 
expansion of the massless scalar two-point integral around $p^2=0$ 
yields the same result as starting with $p^2=0$ and expanding around a
small Goldstone boson mass. This might raise the question whether the
two expansions are also connected at the two-loop level.
However, we only expand the IR-divergent two-loop integrals around $p^2=0$ but use the exact
analytic result for the one-loop integral (without expanding it). This leads to additional
constant and $\log p^2$ terms which are also present in the
full-momentum calculation.

\subsection{Full Momentum Dependence \label{sec:full}}
\label{sec:fullp}
In this approach the full momentum dependence is taken into account in all two-loop
diagrams of the $\order{\left(\alpha_t+\alpha_\lambda+\alpha_\kappa
\right)^2}$ corrections in the gaugeless limit. This
has not been done in the literature for the NMSSM before our study. The
momentum dependence for the 
$\order{\alpha_t \alpha_s}$ corrections has been studied in
Refs.~\cite{Degrassi:2014pfa,Borowka:2014wla,Domingo:2020wiy} within
the MSSM using differential equations and sector decomposition  
for the numerical evaluation of the loop integrals. We expect that the momentum
dependence in the NMSSM is comparable to the one of the MSSM
$\order{\alpha_t \alpha_s}$ corrections which were found to be at most
about one GeV for the loop-corrected  SM-like Higgs boson mass
compared to the zero momentum approximation. \s

Applying the integral basis used by \TSIL the generalisation of our framework to
non-zero external  momentum is straightforward.
There is only one class of diagrams, \cref{fig:allselfgeneric} (h), 
which requires more care as it contains additional $1/p^{2}$-terms that
do not allow to numerically take the limit $p^2\to 0$. Therefore, we
set $p^2=0$ before invoking \TARCER in this particular diagram when
using the zero/partial-momentum approximation and assume arbitrary
$p^2$ when reducing the integral for the full-momentum calculation.
Using $p^2\neq 0$ in all two-loop diagrams requires the inclusion of
additional wave-function renormalisation constants in
\eqref{eq:rehiggsself2}. We have checked that this indeed restores
UV-finiteness in the full momentum approach. Further modifications to
the two-loop counterterms are not necessary. However, the use of an
$\OS$ scheme for the charged Higgs boson mass would in principle also
generate additional finite shifts originating from the momentum
dependence of the charged Higgs boson self-energy. For simplicity, we
do not include momentum dependence in the calculation of the $\OS$
charged Higgs mass counterterm and the VEV counterterm. \s

The Higgs boson masses are obtained by iteratively solving
for the pole of the (two-loop) propagator, \cref{eq:propagator}.
However, so far the $p^2$-dependence
was only taken into account at the one-loop level in \NMSSMCALC such
that the result of the two-loop self-energies had to be calculated
only once. The inclusion of external momentum in the new two-loop self-energy
corrections would require them to take part in the iterative procedure
and thus slow down the overall runtime by several orders of
magnitude.
Therefore, we chose a fixed value of
$p^2=(m_{h_i}^2+m_{h_j}^2)/2$, where $m_{h_{i,j}}$ are the
tree-level Higgs boson masses, for the calculation of the new self-energy
corrections $\hat{\Sigma}^{(2)}_{ij}(p^2)$. This approximation would
in principle require a more detailed study to estimate its numerical
impact. However, in Ref.
\cite{Braathen:2017izn} the same situation was studied for the SM with
the result that the Higgs boson mass prediction varies only by a few
MeV if the external momentum is varied by several orders of magnitude.
\s

Note that results for the NMSSM Higgs mass corrections including the
momentum-dependent contributions while at the
same time applying the gaugeless limit should be taken with
care, {\it cf.}~also the discussion in \cite{Slavich:2020zjv}. While in the MSSM the momentum-dependent and the electroweak gauge
contributions are of similar size, in the NMSSM there are additional $F$-term
contributions of $\order{\lambda,\kappa}$ which could yield an additional
enhancement of the momentum-dependent corrections if
$\lambda,\kappa\propto \order{1}$, especially through the mixing of
heavy and light Higgs components in the self-energies. Therefore, the new
momentum-dependent results should be taken with care as their
contributions could be either comparable or sub-dominant to the
missing electroweak gauge coupling contributions depending on the
considered parameter point.  We
leave the inclusion of the gauge-dependent contributions for future
work.

\section{Set-up of the Calculation and of the Numerical Analysis}
\label{sec:pheno}
\subsection{Tools, Checks and {\tt NMSSMCALC} Release\label{sec:tools}}
We performed two independent calculations to derive the here presented
new two-loop corrections to the  NMSSM Higgs boson masses and
cross-checked the results against each other. Both of them used \SARAH
4.14.3 \cite{Staub:2008uz,Staub:2010jh,Staub:2012pb,Staub:2013tta,Goodsell:2014bna,Goodsell:2014pla} to generate the model file including the vertex
coun\-ter\-terms. This file was used in \FeynArts 3.1 
\cite{Kublbeck:1990xc,Hahn:2000kx} to generate all required one- and
two-loop Feynman diagrams for the calculation of the mass
corrections. We used \FeynCalc 9.2.0
\cite{Mertig:1990an,Shtabovenko:2016sxi} for the evaluation of the
fermion traces and the tensor reduction of the one- and two-loop integrals
and the amplitudes with the counterterm-inserted diagrams. For the
reduction to the two-loop master integrals including the full momentum
dependence we additionally used \TARCER 2.0 \cite{Mertig:1998vk}, a
patched version that comes with \FeynCalc. We use the loop integrals  
  defined in \TSIL \cite{Martin:2005qm}. They are the basis integrals
  of \TARCER extended by a few convenient functions. \s

The two implementations differ in the way the tensor reduction is
performed, namely: $(i)$ setting $p^2=0$ before the reduction and writing the 
result in terms of one- and two-loop tadpole integrals and  $(ii)$ using
general $p^2$-dependence during the whole procedure.
While method $(i)$ is only able to regulate the GBC using a mass
regulator as described in \cref{sec:gbcmassreg}, method $(ii)$ is more
flexible and also able to include partial as well as full external momentum
dependence as described in \cref{sec:partialp,sec:fullp}.
In addition to the consistency checks regarding UV-finiteness
discussed in the previous chapters, the two implementations have been
cross-checked against each other in the limit $p^2=0$. \s

In \cref{sec:results} we also investigate the $p^2$ dependence of our
results. 
The computation at non-zero $p^2$, however, significantly increases
the complexity and runtime of the code due to the dependence on the external library
\TSIL. Even though $\TSIL$ was specifically designed for the
evaluation of two-loop self-energy integrals, the runtime of one
parameter point with a naive \Fortran implementation can be of
$\order{\text{hours-days}}$ because of the large amount of different
mass scales and diagrams entering at the considered order.
Therefore, the second implementation is not part of the public
\NMSSMCALC release but only consists of private \Mathematica notebooks
that make heavy use of caching and parallelisation in order to
speed up the computation. Non-zero $p^2$ results can be provided on
request. \s

The updated version of {\tt NMSSMCALC} including the
  new two-loop corrections to the NMSSM Higgs boson masses in the
CP-conserving and CP-violating NMSSM can be downloaded from the url:
\begin{center}
https://www.itp.kit.edu/$\sim$maggie/NMSSMCALC/
\end{center}
On this webpage we give a description of the new files that have been
included. In the input file {\tt inp.dat}, besides the option to
choose the computation of the new 
two-loop corrections in the block {\tt MODSEL} we added a new block {\tt
REGFACTOR} that allows to choose the size of the regulator mass which
by default is set to $M_R^2 = 10^{-3} \mu_0^2$ ({\it cf.}~the
discussion in Sec.~\ref{sec:gbccomparison}). Note, that our new
two-loop computation becomes numerically unstable for nearly degenerate Higgs
mass values. Therefore, in {\tt NMSSMCALC} we 
automatically switch to the computation of the ${\cal O}(\alpha_t
(\alpha_s + \alpha_t))$ corrections for $|m_{H_i}-m_{H_{i\pm1}}| \le 10^{-3}$~GeV where the
$m_{H_{i,i\pm1}}$ denote the tree-level Higgs masses in the gaugeless
limit. This is also done in case the user has chosen
in the input file to compute the ${\cal
  O}((\alpha_t + \alpha_\lambda + \alpha_\kappa)^2 + \alpha_t
\alpha_s))$ corrections. In the output file the actually computed
loop-order will be stated.

\subsection{The Parameter Scan}
For the numerical discussion of our results we performed a scan in the
NMSSM parameter space in order to obtain parameter scenarios that are
in accordance with the most recent experimental constraints. We
checked the parameter points of our random scan against compatibility
with experimental constraints from the Higgs data by using {\tt
   HiggsBounds} 5.9.0~\cite{Bechtle:2008jh,Bechtle:2011sb,Bechtle:2013wla} 
and {\tt HiggsSignals} 2.6.1~\cite{Bechtle:2013xfa}. The required effective
NMSSM Higgs boson couplings normalised to the corresponding
SM values have been obtained with the Fortran code {\tt NMSSMCALC} 
\cite{Baglio:2013iia}. 
One of the neutral CP-even Higgs bosons, called $h$ from now on,
is required to behave as the SM-like Higgs boson and have a mass in the
range 
\begin{eqnarray}
122 \mbox{ GeV } \le m_h \le 128 \mbox{ GeV} \;,
\end{eqnarray}
when including all previous and the newly calculated two-loop
corrections of this paper where we use per default the mixed
$\DRbar$-OS scheme specified above and OS renormalisation in the
  top/stop and charged Higgs boson sector and an infrared
  mass regulator $M_R$ with $M_R^2 = 10^{-3}\mu_R^2$.  
The SM input values have been chosen
as~\cite{PhysRevD.98.030001,Dennerlhcnote}    
\begin{equation}
\begin{tabular}{lcllcl}
\quad $\alpha(M_Z)$ &=& 1/127.955, &\quad $\alpha^{\MSbar}_s(M_Z)$ &=&
0.1181\,, \\
\quad $M_Z$ &=& 91.1876~GeV\,, &\quad $M_W$ &=& 80.379~GeV \,, \\
\quad $m_t$ &=& 172.74~GeV\,, &\quad $m^{\MSbar}_b(m_b^{\MSbar})$ &=& 4.18~GeV\,, \\
\quad $m_c$ &=& 1.274~GeV\,, &\quad $m_s$ &=& 95.0~MeV\,,\\
\quad $m_u$ &=& 2.2~MeV\,, &\quad $m_d$ &=& 4.7~MeV\,, \\
\quad $m_\tau$ &=& 1.77682~GeV\,, &\quad $m_\mu$ &=& 105.6584~MeV\,,  \\
\quad $m_e$ &=& 510.9989~keV\,, &\quad $G_F$ &=& $1.16637 \cdot 10^{-5}$~GeV$^{-2}$\,.
\end{tabular}
\end{equation} 
We follow the SLHA format~\cite{Skands:2003cj} in which the soft SUSY
breaking masses and trilinear couplings are understood as $\DRb$ parameters at 
the scale
\begin{eqnarray} 
\mu_0 = M_{\text{SUSY}}= \sqrt{m_{\tilde{Q}_3} m_{\tilde{t}_R}} \;.  \label{eq:renscale}
\end{eqnarray}
This is also the renormalisation scale that we use in the computation
of the higher-order corrections. 
In Tab.~\ref{tab:scanranges} we summarize the ranges applied in the 
parameter scan. In order to roughly ensure perturbativity below the GUT scale we
  require that both $\lambda$ and $\kappa$ remain below 0.7.
According to the SLHA format, also $\lambda$, $\kappa$, $\mueff$ and
$\tan\beta$ are understood to be $\DRbar$ parameters at the scale
$M_{\text{SUSY}}$. 
Note that in the scan we kept all CP-violating phases equal to
zero. For the investigation of the impact of CP violation we will then
turn on individual phases. 
\begin{table}[h]
\centering
\begin{minipage}[t]{0.5\textwidth}
\centering
    \begin{tabular}[t]{ll}
        parameter              & scan range [TeV]                \\ \hline
$M_{H^\pm}$            & [0.5, 1]               \\
$M_1,M_2$              & [0.4, 1]                    \\
$M_3$                  & 2               \\
$\mueff$               & [0.1, 1]              \\
$m_{\tilde{Q}_3}, m_{\tilde{t}_R}$      & [0.4, 3]               \\
$m_{\tilde{X}\neq \tilde{Q}_3,\tilde{t}_R}$      & 3        
\end{tabular}
\end{minipage}\hfill
\begin{minipage}[t]{0.5\textwidth}
\centering
    \begin{tabular}[t]{ll}
parameter              & scan range                 \\ \hline
$\tan\beta$            & [1, 10]                    \\
$\lambda$              & [0.01, 0.7]                 \\
$\kappa$               & $\lambda\cdot \xi$          \\
$\xi$                  & [0.1,1.5]                   \\
$A_t$                  & [-3, 3] TeV                 \\
$A_{i\neq t}$          & [-2, 2] TeV 
\end{tabular}
\end{minipage}
\caption{Scan ranges for the random scan over the NMSSM parameter
  space. Values of $\kappa=\lambda\cdot\xi>0.7$ are omitted. All soft
  breaking masses $m_{\tilde{X}}$ with $\tilde{X}=\ti b_R,\ti L,\ti \tau$
  and trilinear couplings $A_i$ with $i= b,\tau, \kappa$, are set
  equal to \unit[3]{TeV}.
 }
\label{tab:scanranges}
\end{table}
We retain scan points with a $\chi^2$ computed by {\tt
  HiggsSignals}-2.6.1 that is consistent with  
an SM $\chi^2$ within $2\sigma$.\footnote{In {\tt HiggsSignals}-2.6.1,
  the SM $\chi^2$ obtained with the latest data set is 84.44. We
  allowed the NMSSM $\chi^2$ to be in the range $[78.26,90.62]$.}  
We omit parameter points with any of the following mass configurations,
\begin{align}
 (i)\quad   &
 m_{\chi_i^{(\pm)}},m_{h_i}>\unit[1]{TeV},m_{\tilde{t}_2}>\unit[2]{TeV}, \nonumber\\
 (ii)\quad  &  m_{h_i}-m_{h_j}<\unit[0.1]{GeV}, m_{\chi_i^{(\pm)}}-m_{\chi_j^{(\pm)}}<\unit[0.1]{GeV} \nonumber\\
 (iii)\quad &  m_{\chi^\pm_1}<\unit[94]{GeV},  m_{\tilde{t}_1}<\unit[1]{TeV}\nonumber \;.
\end{align}
The first constraint ensures that no large logarithms appear that
would jeopardize the validity of a fixed-order calculation. The second condition $(ii)$
excludes degenerate mass configurations for which the two-loop
part of the \NMSSMCALC code is not yet optimised\footnote{The
limit $\lambda,\kappa\to 0$ is also numerically difficult. In this limit we
advise to use the $\order{\alpha_t(\alpha_s+\alpha_t)}$ corrections
instead.}. The third condition takes into account model-independent lower 
limits for the lightest chargino and stop masses. 

\section{Results \label{sec:results}}
In the subsequent numerical analysis we show scatter plots summarising
the overall behaviour of our new results and perform specific
investigations for two sample parameter points. The first point, {\tt P1OS}, has
been chosen among our allowed parameter points and is
defined as follows:\s

\noindent
{\bf Parameter Point {\tt P1OS}:} All complex phases are set to zero and
the remaining input parameters are given by 
\begin{eqnarray}
      |\lambda| &=& 0.46 \,, \; |\kappa|=0.43 \,, \; \mbox{Re}
              (A_\kappa) = -4 \mbox{ GeV}\,, \; |\mu_{\text{eff}}| =
              200 \mbox{ GeV} \,, \; \tan\beta = 3.7 \,, \nonumber \\
M_{H^\pm} &=& 640 \mbox{ GeV} \,,\; 
m_{\tilde{Q}_3} =1 \mbox{ TeV} \,, \; m_{\tilde{t}_R}= 1.8 \mbox{ TeV}
  \,, \; m_{\tilde{X}\neq \tilde{Q}_3,\tilde{t}_R}= 3 \mbox{ TeV}
  \,,\; \nonumber \\
A_t&=&2 \mbox{ TeV} \,, \;  A_{i\neq t,\kappa}=0 \mbox{ GeV} \,, \;
|M_1| = 2 |M_2|= 800 \mbox{ GeV} \,,\; M_3= 2 \mbox{ TeV} \;. \label{eq:P1OS}
\end{eqnarray}
In accordance with the SLHA format $\mu_{\text{eff}}$ is taken as input
parameter, from which $v_s$ and $\varphi_s$ can be computed using
Eq.~(\ref{eq:mueffcalc}). We call this point {\tt
    P1OS} in order to mark that the SM-like
Higgs boson mass value around 125~GeV is obtained for the OS
renormalisation in the top/stop sector. Table~\ref{tab:massvaluesP1OS}
summarizes the mass values that we obtain for {\tt P1OS} at tree level, at
one-loop order and at two-loop level including the previously
available ${\cal O}(\alpha_t \alpha_s)$ and ${\cal O}(\alpha_t
(\alpha_s+\alpha_t))$ corrections and finally the corrections including
our new results, the ${\cal O}((\alpha_t + \alpha_\lambda +
\alpha_\kappa)^2 + \alpha_t \alpha_s)$ corrections. From now on, we
denote these by $\alpha^2_{\text{new}}$, {\it i.e.}
\begin{eqnarray}
\alpha^2_{\text{new}} \equiv (\alpha_t + \alpha_\lambda +
\alpha_\kappa)^2 + \alpha_t \alpha_s \;.
\end{eqnarray}
The numbers in brackets are the values that we obtain for $\DRb$
renormalisation in the top/stop sector. \s

\begin{table}[t]
\begin{center}
 \begin{tabular}{|l||c|c|c|c|c|}
\hline
                                                    & ${h_1}$   & ${h_2}$ & ${h_3}$ & ${a_1}$ & ${a_2}$ \\ \hline \hline
tree-level                                          & 87.64     & 365.32  & 646.65  & 103.09  & 639.83  \\
main component                                      & $h_u$     & $h_s$   & $h_d$   & $a_s$   & $a_d$   \\ \hline
one-loop                                            & \makecell{133.97\\ (115.21)}    & \makecell{359.42\\ (359.35)}  & \makecell{646.67\\ (646.4)}   & \makecell{116.51\\ (116.8)}   & \makecell{639.78\\ (639.8)}  \\ \hline
two-loop ${\cal O}(\alpha_t \alpha_s)$              & \makecell{119.09\\ (119.98)}    & \makecell{359.36\\ (359.37)}  & \makecell{646.5 \\ (646.43)}  & \makecell{116.76\\ (116.69)}  & \makecell{639.81\\ (639.79)}  \\ \hline
two-loop ${\cal O}(\alpha_t(\alpha_s+ \alpha_t))$   & \makecell{125.58\\ (120.15)}    & \makecell{359.36\\ (359.37)}  & \makecell{646.6 \\ (646.43)}  & \makecell{116.76\\ (116.69)}  & \makecell{639.81\\ (639.79)}  \\ \hline
two-loop ${\cal O}(\alpha_{\text{new}}^2)$          & \makecell{125.03\\ (120.18)}    & \makecell{359.68\\ (359.59)}  & \makecell{646.62\\ (646.47)}  & \makecell{116.58\\ (116.63)}  & \makecell{639.77\\ (639.78)}  \\ \hline
 \end{tabular}
\caption{{\tt P1OS}: Mass values in GeV and main components of the neutral Higgs
  bosons at tree-level, one-loop, two-loop $\order{\alpha_t\alpha_s}$,
  two-loop $\order{\alpha_t(\alpha_s +  \alpha_t)}$ and at two-loop
  ${\cal O}(\alpha^2_{\text{new}})$
  obtained by using OS renormalisation in the top/stop sector. Numbers in brackets are results obtained in the $\DRb$ scheme.  The
  main component stays the same at all orders in both schemes.}
\label{tab:massvaluesP1OS}
\end{center}
\end{table}

The table also contains the information on the main singlet/doublet and
scalar/pseu\-do\-sca\-lar component of the respective mass eigenstate. 
The stop masses in the OS and the $\DRbar$ scheme are given by
\begin{eqnarray}
\begin{array}{llclllcll}
\mbox{OS:} & m_{\tilde{t}_1}^{\OS} &=& 1022.64 & \mbox{ GeV} \,, & \quad
                              m_{\tilde{t}_2}^{\OS} &=&
                              1815.54 & \mbox{ GeV} \,, \nonumber \\
\overline{\mbox{DR}}: & 
    m_{\tilde{t}_1}^{\DRbar} &=& 991.64 & \mbox{ GeV}\,, & \quad
    m_{\tilde{t}_2}^{\DRbar}&=& 1815.40 & \mbox{ GeV} \,.
\end{array}
\end{eqnarray}
In this scenario the $h_u$-like Higgs boson with mass
around \unit[125.03]{GeV} at ${\cal O}(\alpha_{\text{new}}^2)$ (OS
renormalisation in the top/stop sector)
is given by the lightest Higgs boson $h_1$. Being $h_u$-like it behaves SM-like (as it
couples maximally to top quarks and hence the LHC Higgs
signal strengths are reproduced). The remaining spectrum features an
$a_s$-like and $h_s$-like Higgs boson in the low to intermediate mass
range with mass values around 117 and 360~GeV, respectively, and a
doublet-like scalar and pseudoscalar around 640~GeV 
at ${\cal O}(\alpha_{\text{new}}^2)$.
Since the admixture determines the Higgs coupling strengths
and consequently the size of the loop corrections, in the following
plots we label the Higgs bosons according to their dominant admixture
and not by their mass ordering unless stated otherwise.\footnote{Note
  that the dominant admixture of a 
  specific Higgs mass eigenstate can change when loop corrections are
  included.} In this way we make sure to 
consistently compare and interpret the impact of the loop
corrections. \s

Defining the absolute value of the relative correction to the mass
value when going successively from loop order $a$ in the tables to
loop order $b$ in the next row by $|m^b-m^a|/m^a$, we see that the
$h_u$-like mass changes considerably upon inclusion of the one-loop
corrections, by 53\% in the OS and by 31\% in the $\DRbar$
scheme. Since in the $\DRbar$ scheme we already partly resum
higher-order corrections, the relative ${\cal O}(\alpha_t
\alpha_s)$ correction compared to the one-loop result is only 4\%
while in the OS scheme we have a reduction by 11\%, moving the
obtained mass values in the two schemes close to each other. The inclusion of the ${\cal
  O}(\alpha_t^2)$ correction in addition increases the discrepancy again, as
already observed in our publication \cite{Dao:2019qaz}. We have a relative correction
of 5\% in the OS and of close to 0\% in the $\DRbar$ scheme. Our newly
calculated corrections move the two values a little bit closer
again. The $\DRbar$ result is increased by a very small 
amount and the OS value is slightly reduced. Still the absolute
difference between the two results amounts to about 5 GeV. As expected
the overall size of the 
two-loop corrections is much smaller than the one-loop corrections and
amounts to a few percent. This behaviour is also reflected in
Fig.~\ref{fig:ATcomp} that we will discuss in the next section.\s
    
While the parameter point {\tt P1OS} is characterized
by a small singlet admixture to the $h_u$-like Higgs mass we also
present results for a parameter point {\tt P2OS} which features 
large singlet admixture to the $h_u$-like mass in order to investigate
the impact of our newly computed corrections. It is defined by\s

\noindent
{\bf Parameter Point {\tt P2OS}:} All complex phases are set to zero
and the remaining input parameters are given by
\begin{eqnarray}
      |\lambda| &=& 0.59 \,, \; |\kappa|=0.23 \,, \; \mbox{Re}
              (A_\kappa) = -546 \mbox{ GeV}\,, \; |\mu_{\text{eff}}| =
              397 \mbox{ GeV} \,, \; \tan\beta = 2.05 \,, \nonumber \\
M_{H^\pm} &=& 922 \mbox{ GeV} \,,\; 
m_{\tilde{Q}_3} =1.2 \mbox{ TeV} \,, \; m_{\tilde{t}_R}= 1.37 \mbox{ TeV}
  \,, \; m_{\tilde{X}\neq \tilde{Q}_3,\tilde{t}_R}= 3 \mbox{ TeV}
  \,,\; \label{eq:P1OSnew} \\
A_t&=&-911 \mbox{ GeV} \,, \;  A_{i\neq t,\kappa}=0 \mbox{ GeV} \,, \;
|M_1| =656 \mbox{ GeV} \,, \; |M_2|= 679 \mbox{ GeV} \,,\; M_3= 2
\mbox{ TeV} \;. \nonumber
\end{eqnarray}
For the stop masses we obtain
\begin{eqnarray}
\begin{array}{llclllcll}
\mbox{OS:} & m_{\tilde{t}_1}^{\OS} &=& 1212.54 & \mbox{ GeV} \,, & \quad
                              m_{\tilde{t}_2}^{\OS} &=&
                              1402.77 & \mbox{ GeV} \,, \nonumber \\
\overline{\mbox{DR}}: & 
    m_{\tilde{t}_1}^{\DRbar} &=& 1190.44 & \mbox{ GeV}\,, & \quad
    m_{\tilde{t}_2}^{\DRbar}&=&  1392.33 & \mbox{ GeV} \,.
\end{array}
\end{eqnarray}

\begin{table}[t]
\begin{center}
 \begin{tabular}{|l||c|c|c|c|c|}
\hline
                                                    & ${h_1}$   & ${h_2}$ & ${h_3}$ & ${a_1}$ & ${a_2}$ \\ \hline \hline
tree-level                                          & 96.86     & 112.10  & 926.25  & 511.34  & 925.86  \\
main component                                      & $h_u$     & $h_s$   & $h_d$   & $a_s$   & $a_d$   \\ \hline
one-loop                                            & 129.01    & 135.09  & 926.69   & 512.55   & 925.08  \\
main component                                      & $h_s$     & $h_u$   & $h_d$   & $a_s$   & $a_d$   \\ \hline
two-loop  ${\cal O}(\alpha_t \alpha_s)$             & 121.36    & 129.7  & 926.37   & 512.62   & 925.11  \\
main component                                      & $h_u$     & $h_s$   & $h_d$   & $a_s$   & $a_d$   \\ \hline
two-loop ${\cal O}(\alpha_t(\alpha_s+ \alpha_t))$  & 126.09    & 130.04  & 926.49   & 512.62   & 925.11  \\
main component                                      & $h_u$     & $h_s$   & $h_d$   & $a_s$   & $a_d$   \\ \hline
two-loop ${\cal O}(\alpha_{\text{new}}^2)$             & 125.28    & 129.92  & 926.63   & 511.92   & 925.08  \\
main component                                      & $h_u$     & $h_s$   & $h_d$   & $a_s$   & $a_d$   \\ \hline
 \end{tabular}
 \caption{{\tt P2OS}: Mass values in GeV and main components of the neutral Higgs
  bosons at tree-level, one-loop, two-loop $\order{(\alpha_t\alpha_s)}$,
  two-loop $\order{(\alpha_t(\alpha_s +  \alpha_t))}$ and at two-loop
  ${\cal O}((\alpha_t+\alpha_\kappa+\alpha_\lambda)^2 + \alpha_t \alpha_s)$
  obtained by using OS renormalisation in the top/stop sector.}
\label{tab:massvalues5}
\end{center}
\end{table}
\begin{table}[t]
\begin{center}
 \begin{tabular}{|l||c|c|c|c|c|}
\hline
                                                    & ${h_1}$   & ${h_2}$ & ${h_3}$ & ${a_1}$ & ${a_2}$  \\ \hline \hline
tree-level                                          & 96.86     &
                                                                  112.10  & 926.25  & 511.34  & 925.86 \\ main component                                      & $h_u$     & $h_s$   & $h_d$   & $a_s$   & $a_d$     \\ \hline
one-loop                                            & 116.3    & 130.1  & 926.33   & 512.66   & 925.18   \\ \hline
two-loop  ${\cal O}(\alpha_t \alpha_s)$             & 121.65    & 130.39  & 926.46   & 512.61   & 925.15 \\ \hline
two-loop ${\cal O}(\alpha_t(\alpha_s+ \alpha_t))$   & 121.54    & 130.38  & 926.45   & 512.61   & 925.15 \\ \hline
two-loop ${\cal O}(\alpha_{\text{new}}^2)$          & 121.69    & 130.2  & 926.53   & 512.12   & 925.15  \\ \hline
 \end{tabular}
\caption{{\tt P2OS}: Same as Tab.~\ref{tab:massvalues5} but using
  $\DRbar$ renormalisation in the top/stop sector. The main component stays the same at all considered orders.} 
\label{tab:massvalues6}
\end{center}
\end{table}
Comparing the mass values at the various loop orders in the OS scheme
(Tab.~\ref{tab:massvalues5}) and in the $\overline{\mbox{DR}}$ scheme
(Tab.~\ref{tab:massvalues6}) we observe the same behaviour as for the
point {\tt P1OS}. However, in the OS scheme, the
nature of the $h_1$ and $h_2$ mass eigenstates, respectively, change
when moving from tree level to one-loop level and again when including
the two-loop corrections. Thus $h_1$ ($h_2$) is $h_s$-like
($h_u$-like) at one-loop order but $h_u$-like ($h_s$-like) at two-loop
order. This is due to large mixing effects between the $h_u$- and
$h_s$-like states with a large singlet component in the $h_u$-like
state. In the $\overline{\mbox{DR}}$ scheme, on the other hand, $h_1$ ($h_2$) is always
$h_u$-like ($h_s$-like). The partial resummation of higher-order
corrections through $\overline{\mbox{DR}}$ renormalisation in the
top/stop sector implies smaller one- and two-loop corrections as compared to
the OS scheme and thereby less sensitivity to possible singlet
admixture effects in the mass corrections.\s

The impact of our newly computed corrections with respect to the
already available two-loop corrections at 
$\order{\alpha_t(\alpha_t+\alpha_s)}$
is less than 1\% for the two considered parameter points.
However, we will show in the following subsections that the corrections can be
enhanced for large values of $\lambda$ and $\kappa$.
\subsection{Impact of the New Two-Loop Corrections}
\label{sec:loopcorrections}
In the following we discuss in more detail the impact of our newly
computed two-loop corrections, both for the point {\tt P1OS} with
small singlet admixture and for the point {\tt P2OS} with large
singlet admixture. \s

\begin{figure}[b!]
    \centering
        \includegraphics[width=0.49\textwidth]{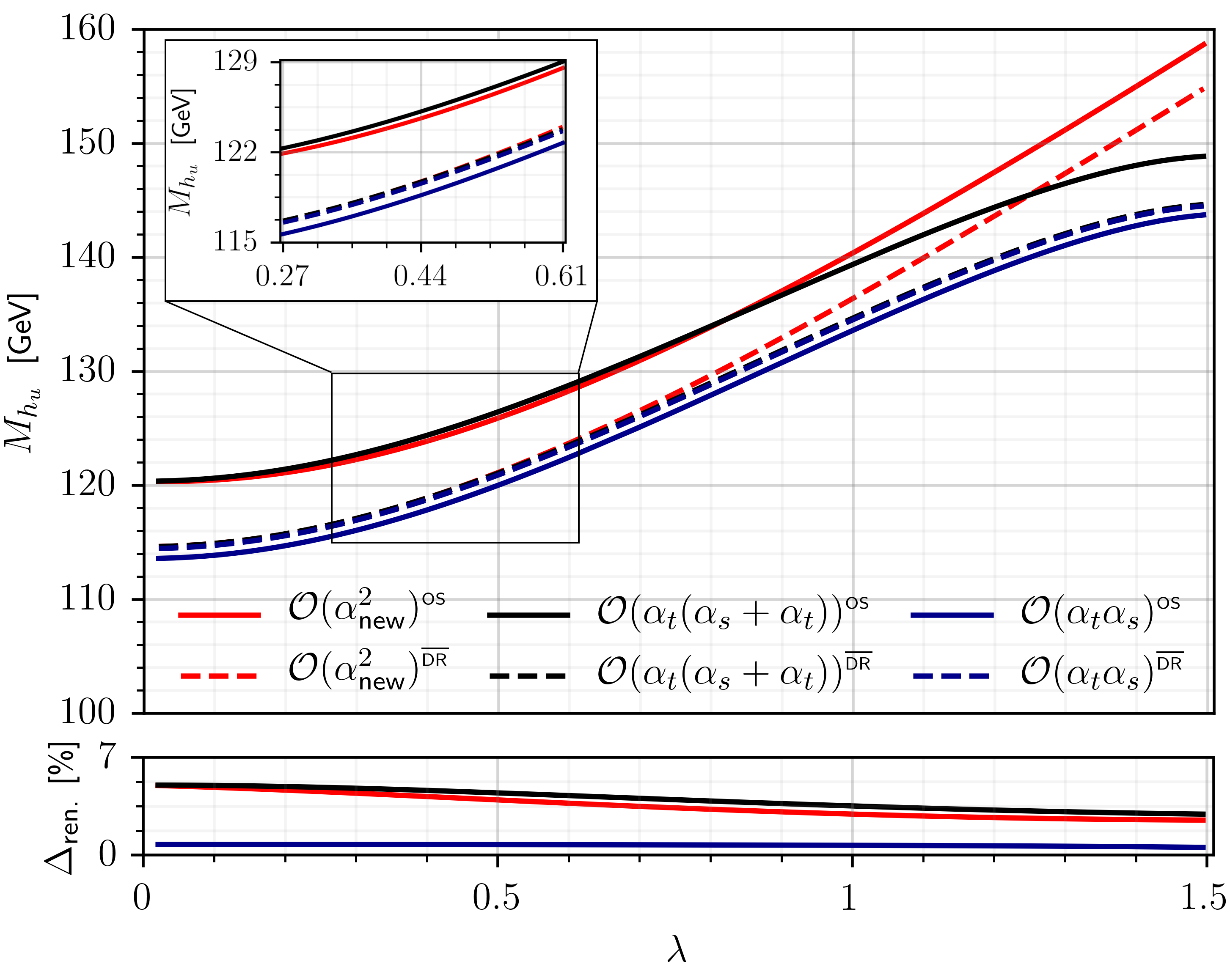}
        \centering
        \includegraphics[width=0.49\textwidth]{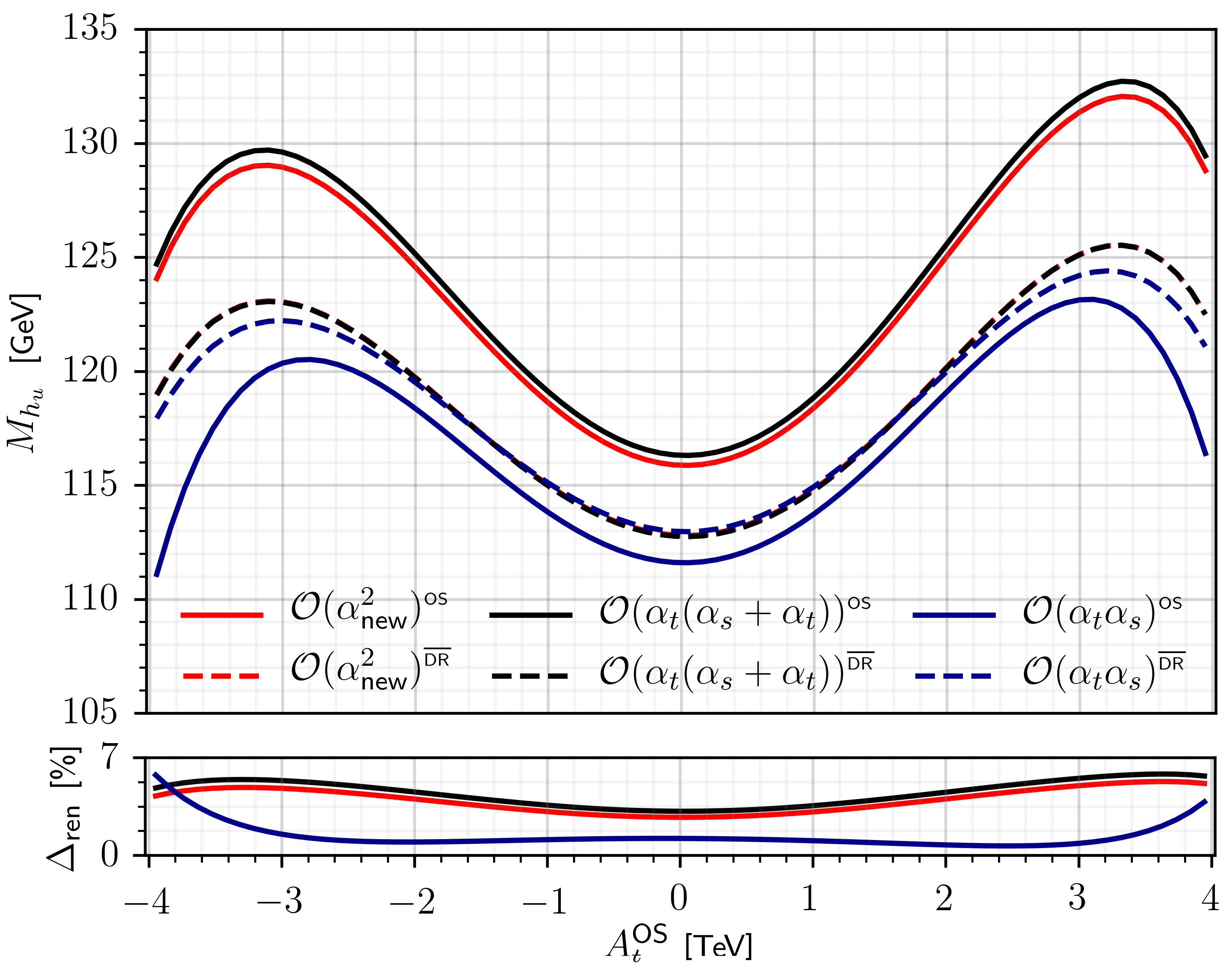}
    \caption{Upper panels: Mass values of the $h_u$-like Higgs boson for the
      parameter point {\tt P1OS} at two-loop ${\cal O}(\alpha_t \alpha_s)$
      (blue),  ${\cal O}(\alpha_t(\alpha_s+ \alpha_t))$ (black)
      and ${\cal O}(\alpha^2_{\text{new}})$ (red) in the OS (full) and
      $\DRbar$ (dashed) renormalisation scheme as a function of
      $\lambda$ (left) and $A_t^{\text{OS}}$ (right).
      The black and blue dashed lines (left) lie on top of each other.
      Lower panels: Relative change $\Delta_{\text{ren}}$ ({\it cf.}~text) due to
      renormalisation scheme change in the top/stop sector at ${\cal
        O}(\alpha^2_{\text{new}})$ (red), ${\cal O}(\alpha_t(\alpha_s+\alpha_t))$ (black) and
      ${\cal O}(\alpha_t \alpha_s)$ (blue).
      The zoomed region shows the range of $|\lambda|$ that is
      compatible with $\higgss$ when using the new
      correction. 
  } 
    \label{fig:ATcomp}
\end{figure}
\paragraph{Small singlet admixture}
In the upper panels of Fig.~\ref{fig:ATcomp} we show for the parameter point {\tt P1OS}
the two-loop corrected mass values at
${\cal O}(\alpha_t \alpha_s)$ (blue), ${\cal O}(\alpha_t(
\alpha_s+\alpha_t))$ (black) and including our newly computed
corrections at ${\cal O}(\alpha^2_{\text{new}})$ (red) both for OS (full) and
$\overline{\mbox{DR}}$ (dashed) renormalisation in the top/stop
sector. The left figure shows the dependence on the NMSSM-specific parameter
$\lambda$ and the right figure the one on the soft-SUSY breaking
trilinear coupling $A_t^{\text{OS}}$. The small insert in
Fig.~\ref{fig:ATcomp} (left) enlarges the parameter region compatible
with the experimental Higgs data at 
${\cal O}(\alpha^2_{\text{new}})$.
For this and all other benchmark points we simultaneously vary
$\kappa=\lambda\cdot \kappa_0/\lambda_0$ where $\kappa_0,\lambda_0$
are the starting values of the respective input parameters. This
enables us to vary $\lambda$ and $\kappa$ over large ranges without
encountering negative mass squares.\s

For a given renormalisation scheme, we define the relative size of the new two-loop
corrections at ${\cal O}(\alpha_{\text{new}}^2)$ to the mass $M_h$ of the Higgs boson $h$ with respect to
the already available two-loop corrections at ${\cal O}(\alpha_i^2)$, as
\begin{eqnarray}
\Delta^{\alpha_{\text{new}}^2}_{\alpha_i^2} = \frac{\left|
    M_h^{\alpha^2_{\text{new}}}-  M_h^{\alpha^2_i}\right| }{
  M_h^{\alpha^2_i}} \;, \label{eq:corrdef}
\end{eqnarray}
with $\alpha^2_i=\alpha_t(\alpha_s + \alpha_t)$ and $\alpha_i^2 =
\alpha_t \alpha_s$, respectively.
Here and in the following, loop-corrected Higgs mass values are always denoted
by capital $M$. We find that in the $\DRbar$ scheme
the relative impact of our new corrections with respect to the
previous two-loop orders is about the
same and varies between 0 and 7\% for $\lambda=0$ to 1.5. 
In the OS scheme the relative impact with respect to the ${\cal
O}(\alpha_t(\alpha_s+ \alpha_t))$ corrections is in the same range as in the $\DRbar$ scheme 
while it varies between 6\% and more than 10\% for 
$\Delta^{\alpha_{\text{new}}^2}_{\alpha_t \alpha_s}$. Overall, the new
corrections increase with $\lambda$. When varying
$A_t^{\text{OS}}$ in the range -4 to 4 TeV, the relative corrections
in the $\overline{\mbox{DR}}$ scheme are less than 1\%. In the OS
scheme this is also the case for
$\Delta^{\alpha_{\text{new}}^2}_{\alpha_t (\alpha_s+\alpha_t)}$ while
$\Delta^{\alpha_{\text{new}}^2}_{\alpha_t\alpha_s}$
varies between 4\% and more than 10\% for $A_t^{\text{OS}}=0$ and
$A_t^{\text{OS}}=\pm 4$~TeV, respectively. Overall, the impact of the
new two-loop corrections is of the order of a few percent and
increases for very large values of $\lambda$ and $\kappa$ as
expected. From Fig.~\ref{fig:ATcomp} (right) we furthermore infer that
the corrections are asymmetric with respect to the sign of $A^{\text{OS}}_t$. \s

The lower panels in Fig.~\ref{fig:ATcomp} show the
relative change in the mass corrections at fixed loop order when
switching the renormalisation scheme in the top/stop sector,
\begin{equation}
\Delta_{\text{ren}} = \frac{\left| M_h^{m_t(\DRbar)}-
    M_h^{m_t(\text{OS})}\right| }{ M_h^{m_t(\DRbar)} } \;. \label{eq:rendelta}
\end{equation}
The comparison of the results in the two different
renormalisation schemes gives one ingredient for the estimate of the uncertainty
on the Higgs mass values due to missing higher-order corrections. 
In the whole plotted $\lambda$ range at 
${\cal O}(\alpha_t \alpha_s)$ the impact is less than 1\%, while it
increases to values between about 3 and more than 5\% upon inclusion
of the ${\cal O}(\alpha_t (\alpha_s + \alpha_t))$ and slightly less in the new ${\cal
  O}(\alpha^2_{\text{new}})$ corrections, respectively. Also for the
plotted $A_t^{\text{OS}}$ values the renormalisation scheme dependence
is larger for these loop orders (with values between 2.5 and more than
5\%) than for ${\cal O}(\alpha_t \alpha_s)$ except for large negative
$A_t^{\text{OS}}$ values. \s

In general one needs to be careful in drawing
conclusions on the remaining theoretical uncertainty at a given loop
order as long as not all existing contributions at the investigated
loop order are included. Since the scheme dependence induced
by the top/stop sector is also not significantly reduced upon
inclusion of the $\order{\alpha_t(\alpha_\lambda+\alpha_\kappa)}$ corrections, the
3-loop corrections of $\order{\alpha_t\alpha_s^2}$ or even beyond
might be required. However, in the rest of this section we show that
there are cases with an $h_u$-like Higgs boson at \unit[125]{GeV}, where the
scheme dependence is significantly reduced.

\begin{figure}[t!]
    \centering
        \includegraphics[width=0.49\textwidth]{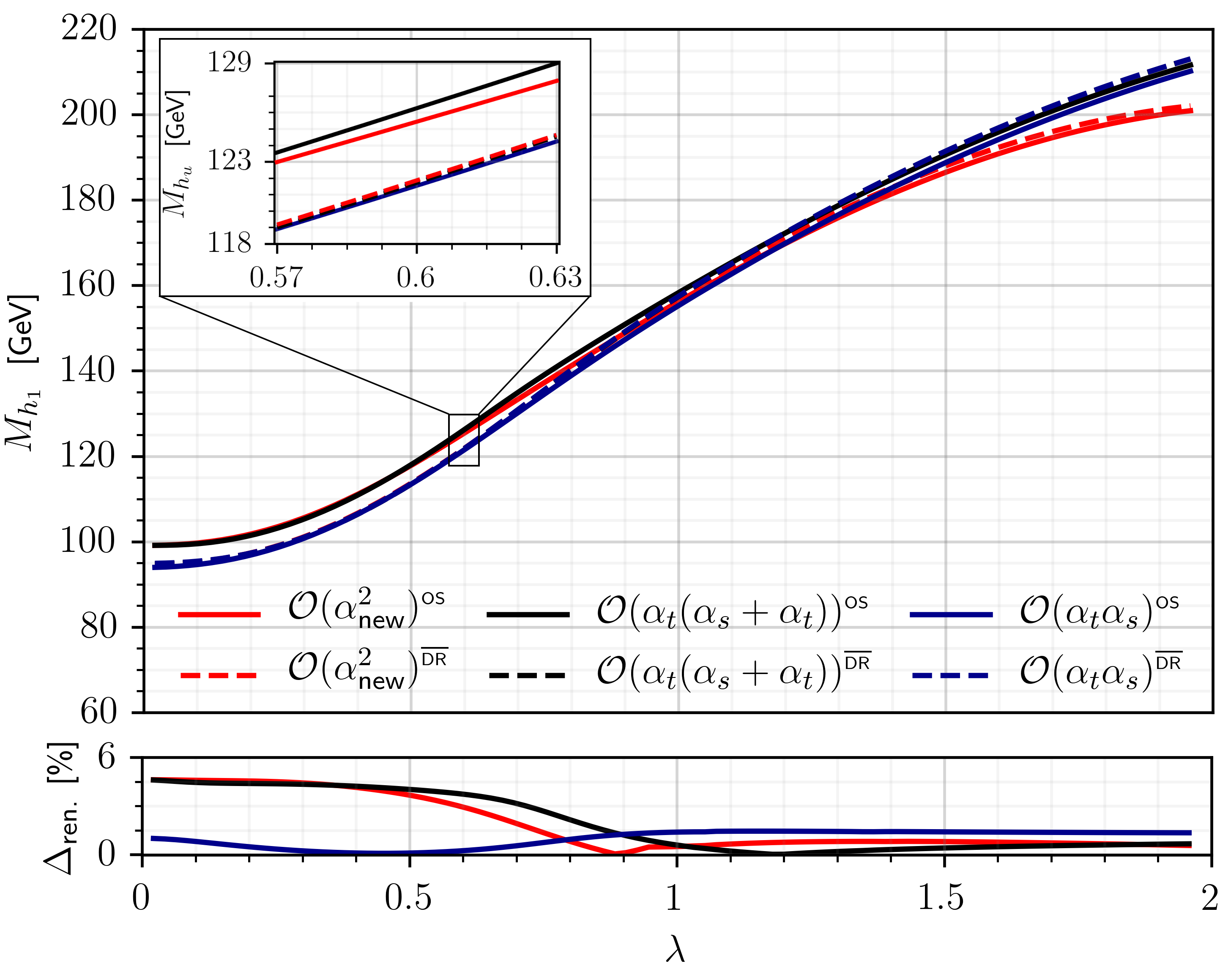}
    \caption{Same as Fig.~\ref{fig:ATcomp} (left) but for the
      parameter point {\tt P2OS}.
      The black and blue dashed lines lie on top of each other.
    } 
    \label{fig:P1OSnew}
\end{figure}

\paragraph{Large singlet admixture}
We now turn to the impact of our corrections for the benchmark point
{\tt P2OS} which is characterized by a large singlet admixture to the
$h_u$-like Higgs state. In Fig.~\ref{fig:P1OSnew} we show
the absolute mass values as a function of $\lambda$ (upper) 
and the dependence on the renormalisation scheme (lower) for {\tt P2OS}.
The notation is the same as in Fig.~\ref{fig:ATcomp} (left).
Like in the case with small singlet admixture, all two-loop
corrections are close to each other in the $\overline{\mbox{DR}}$
scheme for $\lambda \le 1$, for $\lambda \ge 1$ the new corrections
start to deviate from the previous ones reaching a relative correction
$\Delta^{\alpha_{\text{new}}^2}_{\alpha_i^2} \approx 5$\%
($\alpha_i^2=\{ \alpha_t (\alpha_s + \alpha_t), \alpha_t \alpha_s \}$). In
the OS scheme the impact is slightly more pronounced.
We find non-zero relative corrections
$\Delta^{\alpha_{\text{new}}^2}_{\alpha_t(\alpha_s+\alpha_t)}$ in the
OS scheme starting for $\lambda \ge 0.5$ increasing to up to 5\% for $\lambda=2$.
The impact of the new corrections  with respect to
$\order{\alpha_t\alpha_s}$, however, varies from 6\% at $\lambda=0$ to zero for $\lambda$ around 1.2 and up
to 5\% again at $\lambda=2$.
As for the renormalisation scheme dependence, 
{\it cf.}~Fig.~\ref{fig:P1OSnew} (lower), for 
$\lambda\lsim 0.75$ it is largest for the 
${\cal O}(\alpha_{\text{new}}^2)$ and the 
${\cal O}(\alpha_t(\alpha_s+\alpha_t))$ corrections while for larger
$\lambda$ values the ${\cal O}(\alpha_t\alpha_s)$
corrections show the largest scheme
dependence. Overall, the scheme dependence is of the order of a few
percent, specifically in the region allowed by current collider
constraints it is reduced by 1-2\% w.r.t.~the $\order{\alpha_t(\alpha_s+\alpha_t)}$ result.
This shows that an estimate of the residual uncertainty due to missing
higher-order corrections based on the renormalisation scheme variation
in the top/stop sector cannot be made without taking into account the
computation of the complete two-loop corrections since the top/stop sector
contributes with mixed contributions such as
$\order{\alpha_t(\alpha_\lambda+\alpha_\kappa)}$. \s

\begin{figure}[t!]
    \centering
        \includegraphics[width=0.49\textwidth]{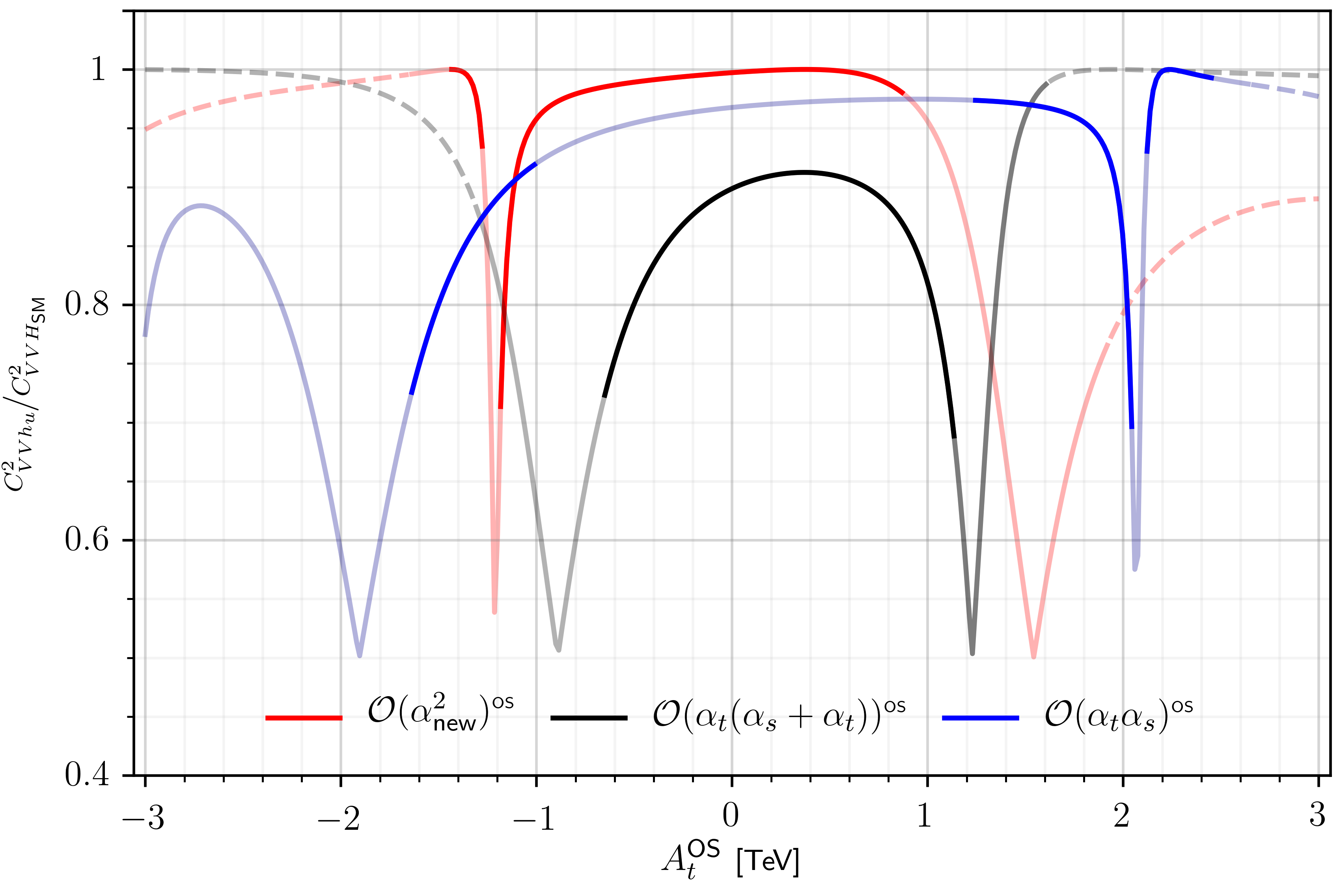}
        \includegraphics[width=0.49\textwidth]{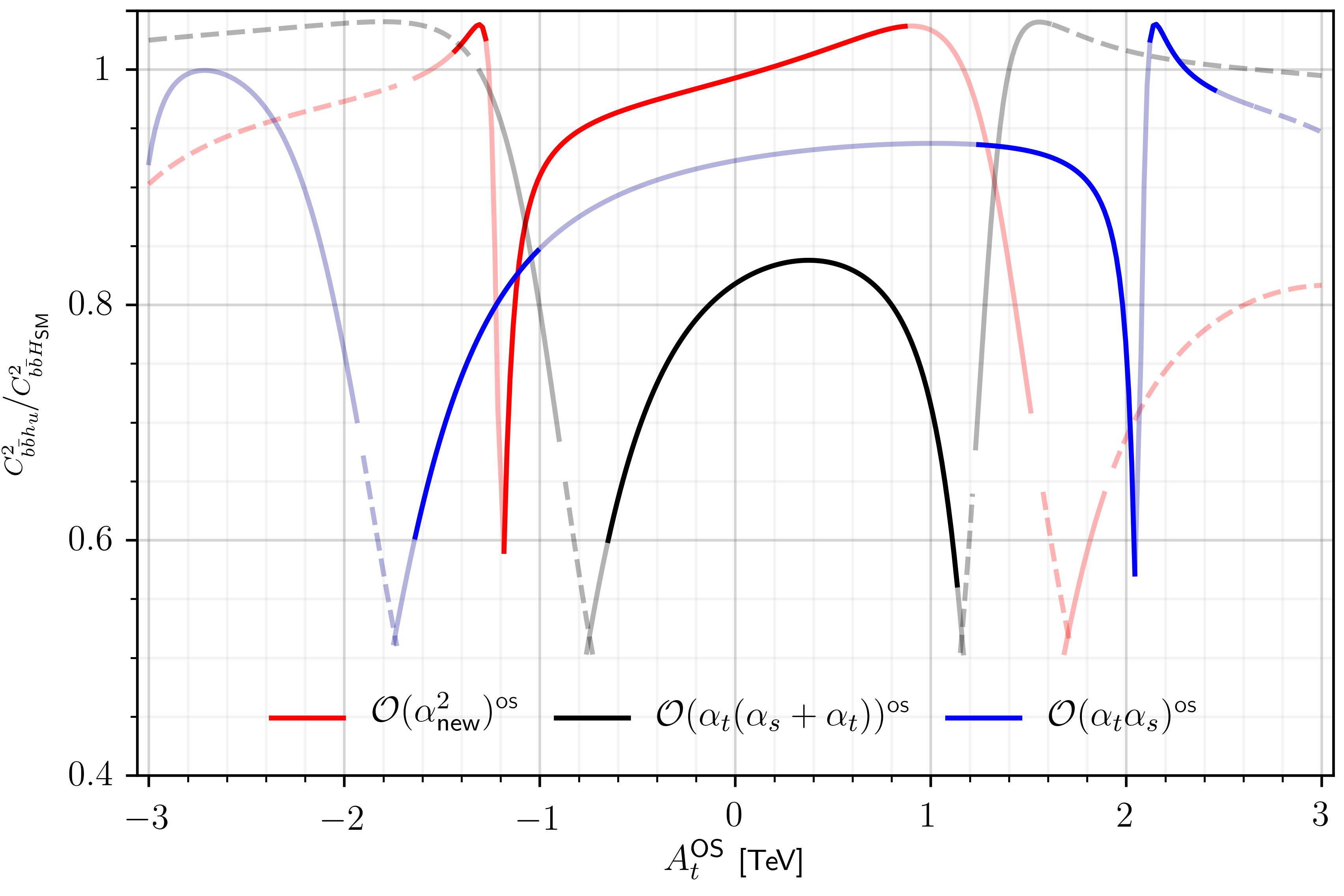}
    \caption{The couplings squared of the $h_u$-like Higgs boson 
        to the massive vector bosons $V$ ($V=W,Z$) (left) and to the
        bottom quarks (right) normalized  to the corresponding SM
        values as a function of $A_t^{\text{OS}}$ for
        {\tt P2OS} in the OS scheme at ${\cal O}(\alpha_{\text{new}}^2)$ (red), ${\cal
          O}(\alpha_t (\alpha_s+ \alpha_t))$ (black), and ${\cal 
        O}(\alpha_t \alpha_s)$ (blue).
        Transparent lines are either excluded by \higgss or do not
        fulfill the Higgs mass constraint. Full lines correspond
        to $h_u$ being the lightest Higgs $h_1$, dashed lines to 
        $h_u= h_2$. 
    } 
   \label{fig:couplings} 
\end{figure}

In Fig.~\ref{fig:couplings} (left) we show, as a function of
$A_t^{\text{OS}}$, the impact of the two-loop 
corrections on the phenomenologically important squared coupling
$C_{VVh_u}^2$
of the $h_u$-like Higgs boson to the massive gauge bosons $V$ ($V=Z,W$)
normalized to the corresponding SM value $C^2_{VVH_{\text{SM}}}$. The
right plot shows the squared bottom-quark coupling to the $h_u$-like 
Higgs boson $C_{b\bar{b}h_u}^2$ normalized to the corresponding SM value
$C_{b\bar{b}H_{\text{SM}}}^2$. We have chosen
OS renormalisation in the top/stop sector for this plot. The color
code corresponds to the loop order, ${\cal O}(\alpha_{\text{new}}^2)$
(red), ${\cal O}(\alpha_t (\alpha_s+ \alpha_t))$ (black), ${\cal
  O}(\alpha_t \alpha_s)$ (blue). The style
change of the lines signals the following. Transparent lines denote
the $A_t^{\text{OS}}$ range where the experimental Higgs signal
constraints are not fulfilled any more. Full lines indicate the
parameter region where the $h_u$-like Higgs boson is the lightest
Higgs state in the spectrum, dashed lines correspond to the
second-lightest Higgs boson being $h_u$-like. We have already seen in
Tab.~\ref{tab:massvalues5} that the nature of the Higgs mass
eigenstate can change depending on the loop order. We also see dips
 in the plots. Here the singlet-doublet admixture
of the two lightest Higgs states becomes large inducing nearly same
values for their respective coupling values to the SM particles. The
comparison of the coupling values for the various two-loop orders
(comparison of the red, blue and black lines) clearly shows that the
Higgs couplings and hence the Higgs boson phenomenology is strongly
affected by the order of included loop corrections. The comparison of the
full and transparent regions shows how the allowed parameter range is
impacted by the loop order. This underlines the need of precision
calculations in order to be able to delineate the underlying parameter
range through the measurement of the Higgs properties. 

\paragraph{Whole parameter sample} 
\begin{figure}[t!]
    \centering
    \includegraphics[width=0.59\textwidth]{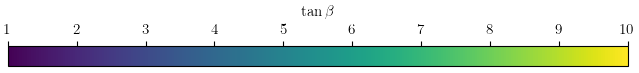}\\
    \includegraphics[width=0.49\textwidth]{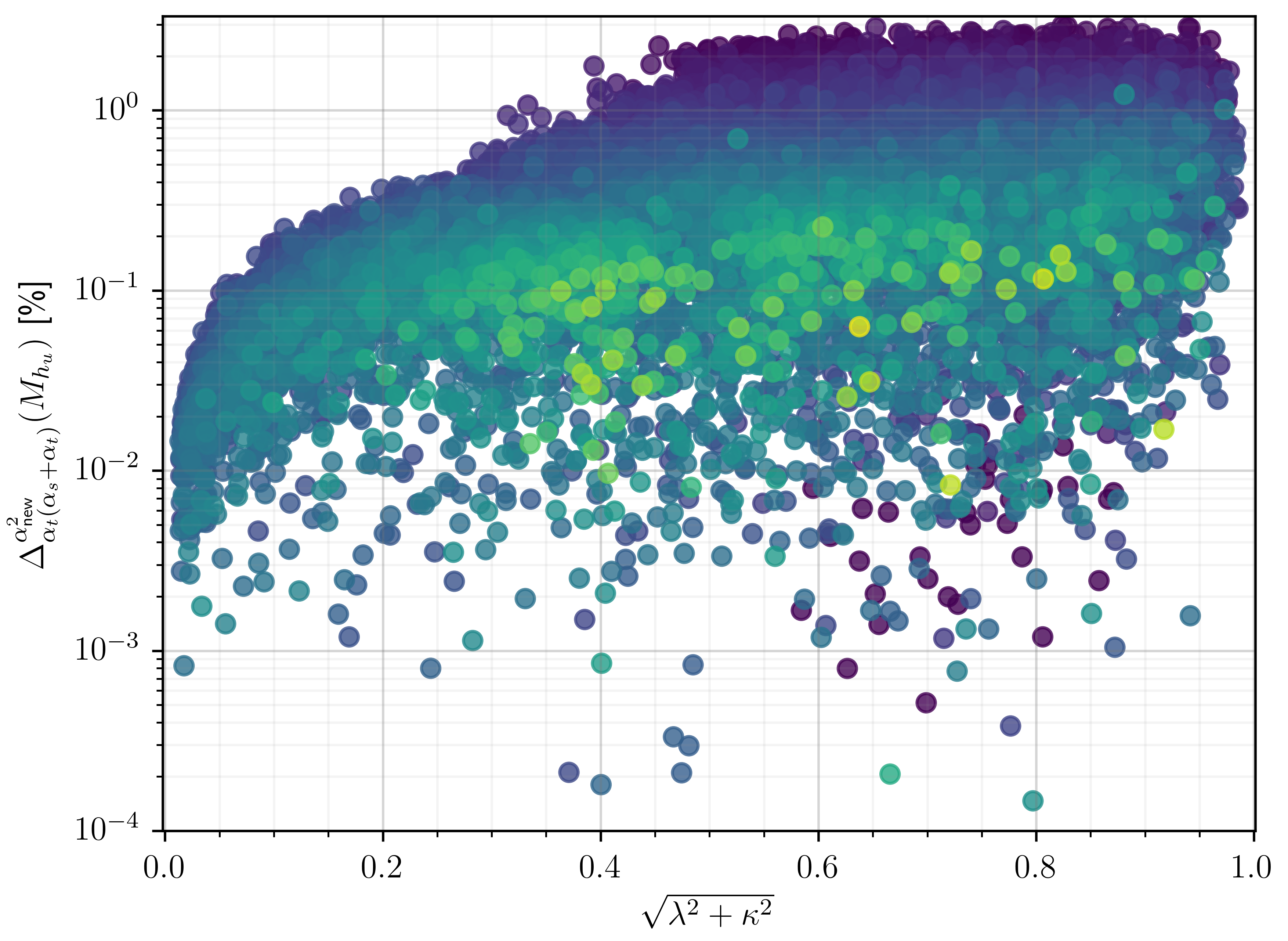}
    \includegraphics[width=0.49\textwidth]{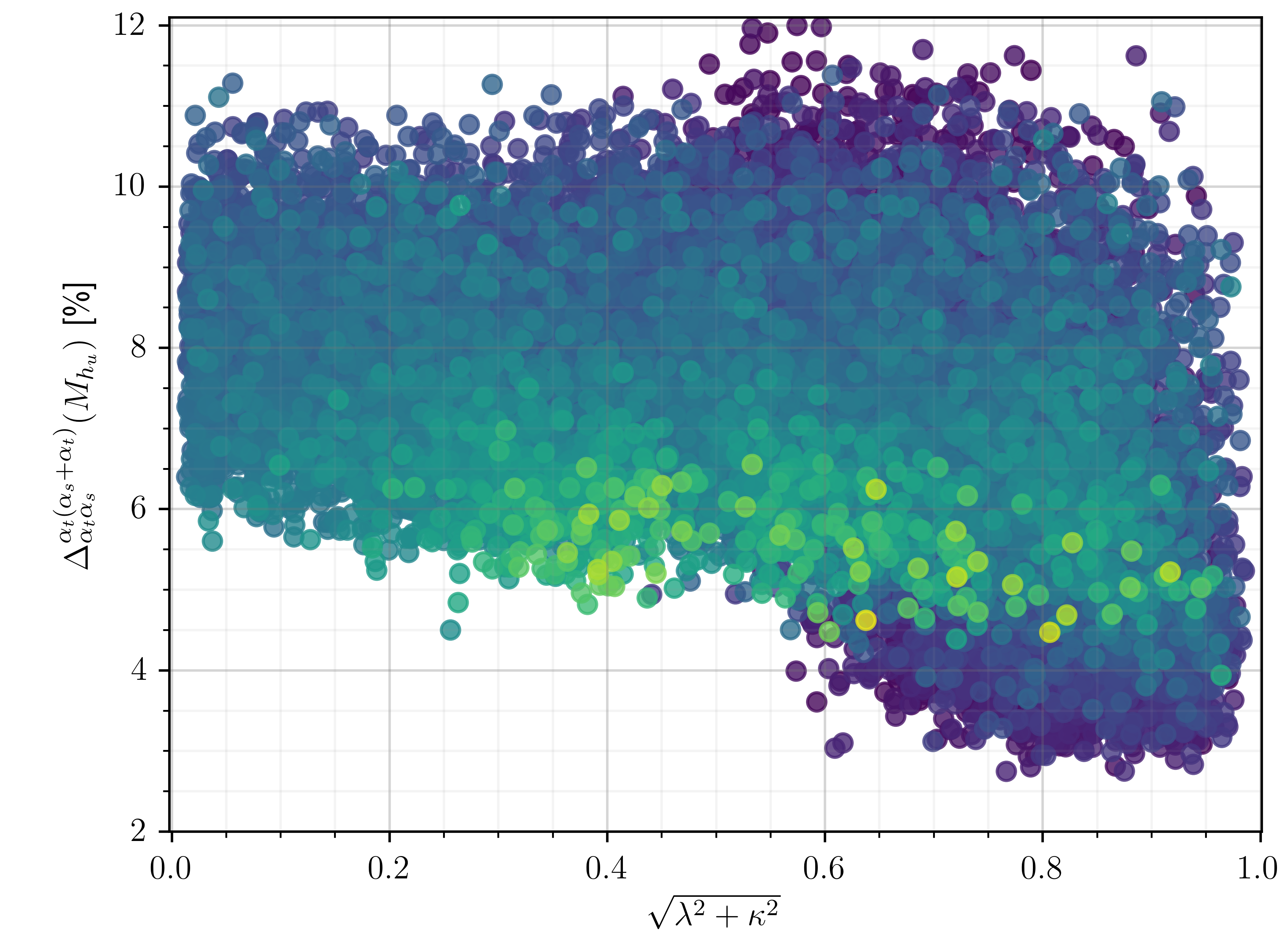}
    \caption{For all allowed parameter points: 
     Relative size ({\it cf.}~text)
     $\Delta_{\alpha_t(\alpha_t+\alpha_s)}^{\alpha^2_{\text{new}}}$
     (left) and $\Delta^{\alpha_t(\alpha_s+\alpha_t)}_{\alpha_t \alpha_s}$ (right)
     of the two-loop corrections to $M_{h_u}$ as a function of $\sqrt{\lambda^2
     + \kappa^2}$. The color code indicates the value of
   $\tan\beta$. }
    \label{fig:overall1}
\end{figure}
In Figs.~\ref{fig:overall1} and \ref{fig:overall2} we investigate the
overall impact of our corrections by looking at the whole parameter
sample. In Fig.~\ref{fig:overall1} we show for all allowed parameter
points obtained in our scan the relative sizes
$\Delta_{\alpha_t(\alpha_s+\alpha_t)}^{\alpha^2_{\text{new}}}$ of our
new two-loop corrections ${\cal O}(\alpha^2_{\text{new}})$ to
$M_{h_u}$ with 
respect to the previous ones at ${\cal O}(\alpha_t (\alpha_s +
\alpha_t))$ (left) and compare them to the impact of our previously computed two-loop
corrections in \cite{Dao:2019qaz}, {\it i.e.}~the relative sizes 
$\Delta^{\alpha_t(\alpha_s+\alpha_t)}_{\alpha_t \alpha_s}$ of 
the two-loop corrections ${\cal O}(\alpha_t(\alpha_s+\alpha_t))$ with
respect to the ${\cal O}(\alpha_t \alpha_s)$ corrections (right). Note that in
our scan we applied OS renormalisation in the top/stop sector. The points are
displayed as a function of the NMSSM-specific parameter combination 
$\sqrt{\lambda^2 + \kappa^2}$ with the color code indicating the
$\tan\beta$ value. As expected, the maximally obtained values for the
relative sizes of our new corrections increase with 
$\sqrt{\lambda^2 + \kappa^2}$, remaining overall below about 3\% since
we did not consider too large values of $\lambda$ and $\kappa$ in our
scan to ensure perturbativity below the GUT scale. 
The impact of
${\cal O}(\alpha_t (\alpha_s+\alpha_t))$ with respect to 
${\cal O}(\alpha_t \alpha_s)$ is larger with maximum relative
corrections of up to about 12\%. The upper bound of the corrections
does not depend on $\sqrt{\lambda^2 + \kappa^2}$. The smallest corrections
are obtained for large  $\sqrt{\lambda^2 + \kappa^2}$ and small
$\tan\beta$. Larger singlet admixtures induce smaller couplings 
of the $h_u$-like Higgs boson to the top
quarks and hence a smaller impact of the additional 
${\cal O}(\alpha_t^2)$ but larger impact of the
$\order{(\lambda+\kappa)^2}$ corrections. 
For the singlet-like scalar Higgs bosons the effect of the newly
computed corrections is smaller: for  $h_s$ ($a_s$) (not shown here) we find for most
cases, when going from ${\cal O}(\alpha_t (\alpha_s+\alpha_t))$ to
${\cal O}(\alpha^2_{\text{new}})$, a relative increase in the mass
values below 0.4\% (0.1\%) with some outliers up to 6\%
(3\%). \s

\begin{figure}[t!]
    \centering
    \hspace{-7.2cm}\includegraphics[width=0.5\textwidth]{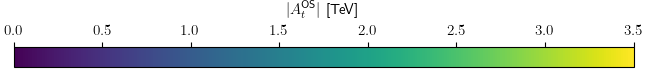}\\
    \includegraphics[width=0.49\textwidth]{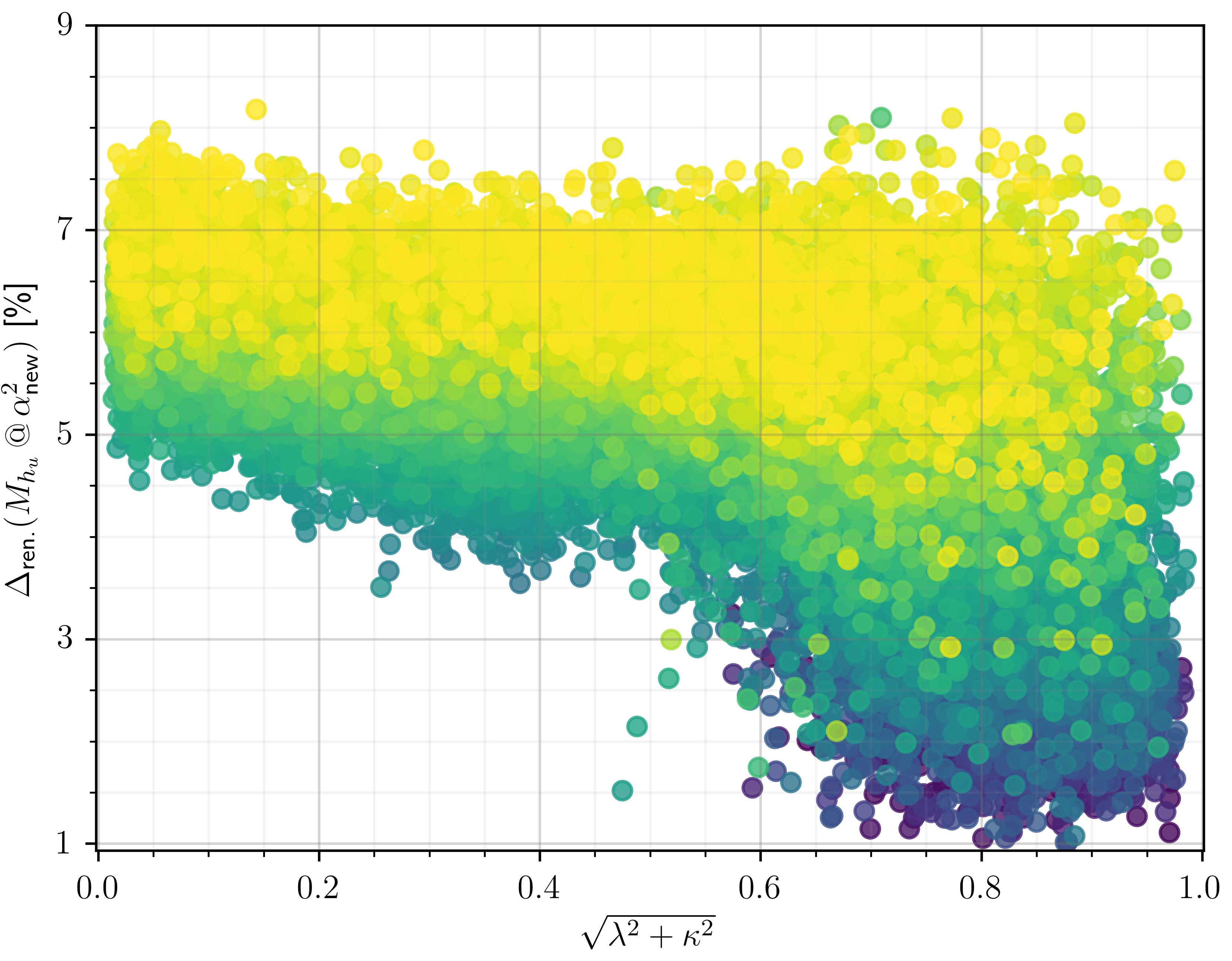}
    \includegraphics[width=0.49\textwidth]{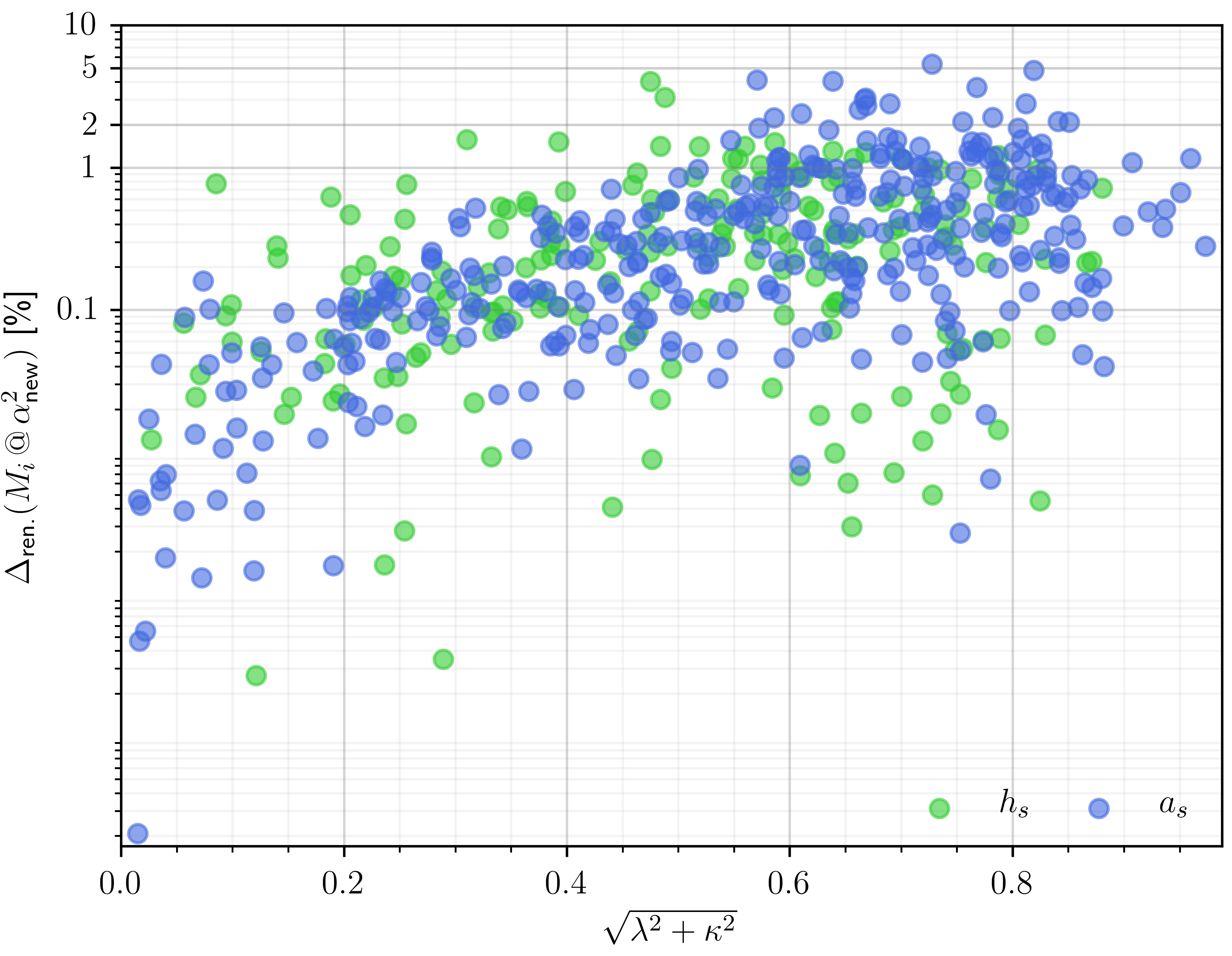}
    \caption{Scheme dependence $\Delta_{\text{ren}}$ at ${\cal
        O}(\alpha_{\text{new}}^2)$ when using $\OS$ or $\DRb$ conditions in the top/stop 
      sector: Left: For $M_{h_u}$ for all allowed parameter
      points. Right: For $M_{a_s}$ (blue) and $M_{h_s}$ (green) for all
      parameter points where they are lighter than
      $M_{h_u}$. The color code in the left plot
        denotes the value of $|A_t^{\text{OS}}|$.}
    \label{fig:overall2}
\end{figure}
In Fig.~\ref{fig:overall2} (left) we show the relative change
$\Delta_{\text{ren}}$, see Eq.~(\ref{eq:rendelta}), of our ${\cal
  O}(\alpha_{\text{new}}^2)$ corrections to the $h_u$ like mass
$M_{h_u}$ for all allowed parameter points as a function of the
NMSSM-specific combination $\sqrt{\lambda^2+\kappa^2}$ where the color
code indicates the value of $|A_{t}^{\text{OS}}|$. The change of the
renormalisation scheme requires a conversion of the involved top/stop
sector parameters, so that $\Delta_{\text{ren}}$ clearly depends on
the value of $|A_t^{\text{OS}}|$. This is also observed in the plot, where the
largest effects from the change of the 
renormalisation scheme in the top/stop sector are found for large $|A_t^{\text{OS}}|$
values. The smallest renormalisation scheme dependence is obtained for
small  $|A_t^{\text{OS}}|$ and large $\sqrt{\lambda^2+\kappa^2}$. In
this regime the impact of the new corrections from the Higgs- and electroweakino
sectors becomes more pronounced while the top/stop sector contributes
less. Therefore, the renormalisation scheme dependence introduced by
the top/stop sector is reduced further.
Overall the renormalisation scheme
dependence at ${\cal O}(\alpha_{\text{new}}^2)$ is larger than 5\% if
$|A_t^{\text{OS}}|>\unit[2]{TeV}$. This is not surprising since we only
consider points with $m_{\tilde{t}_2}<\unit[2]{TeV}$ in the scan.
Corrections of the order
$\order{A_t/m_{\tilde{t}_i}}$ (and higher powers) become large
and introduce a large scheme dependence if $|A_t|>m_{\tilde{t}_i}$.
However, for $|A_t^{\text{OS}}|\lesssim\unit[1-2]{TeV}$ the scheme
dependence is under good control (i.e. smaller than the overall size
of the new two-loop corrections) in the NMSSM-specific parameter
region $\sqrt{\lambda^2+\kappa^2}>0.6$.\\

The right plot of Fig.~\ref{fig:overall2} shows the renormalisation
scheme dependence $\Delta_{\text{ren}}$ at ${\cal
  O}(\alpha^2_{\text{new}})$ for the singlet-like scalar $h_s$ (green) and
pseudoscalar $a_s$ (blue) for all of the considered scenarios where they are lighter than
$h_u$. A tendency of increasing maximum values for the renormalisation
scheme dependence with rising NMSSM-specific couplings can be inferred
from the plot. Since the states are mostly $h_s$-like, one would
assume that they are less affected by higher-order corrections involving $\alpha_t$ in the new
corrections. However, we find renormalisation scheme dependences of up to about
5\% for very large $\lambda$ and $\kappa$. This shows that corrections
of $\order{\alpha_t(\alpha_\lambda+\alpha_\kappa)}$ 
can indeed be important for singlet-like states if $\lambda$ and
$\kappa$ become large.
Overall the renormalisation scheme dependence is of typical sizes expected at
two-loop order. \s

\subsection{Renormalisation Scale Dependence}
\begin{figure}[b!]
    \centering
        \includegraphics[width=0.49\textwidth]{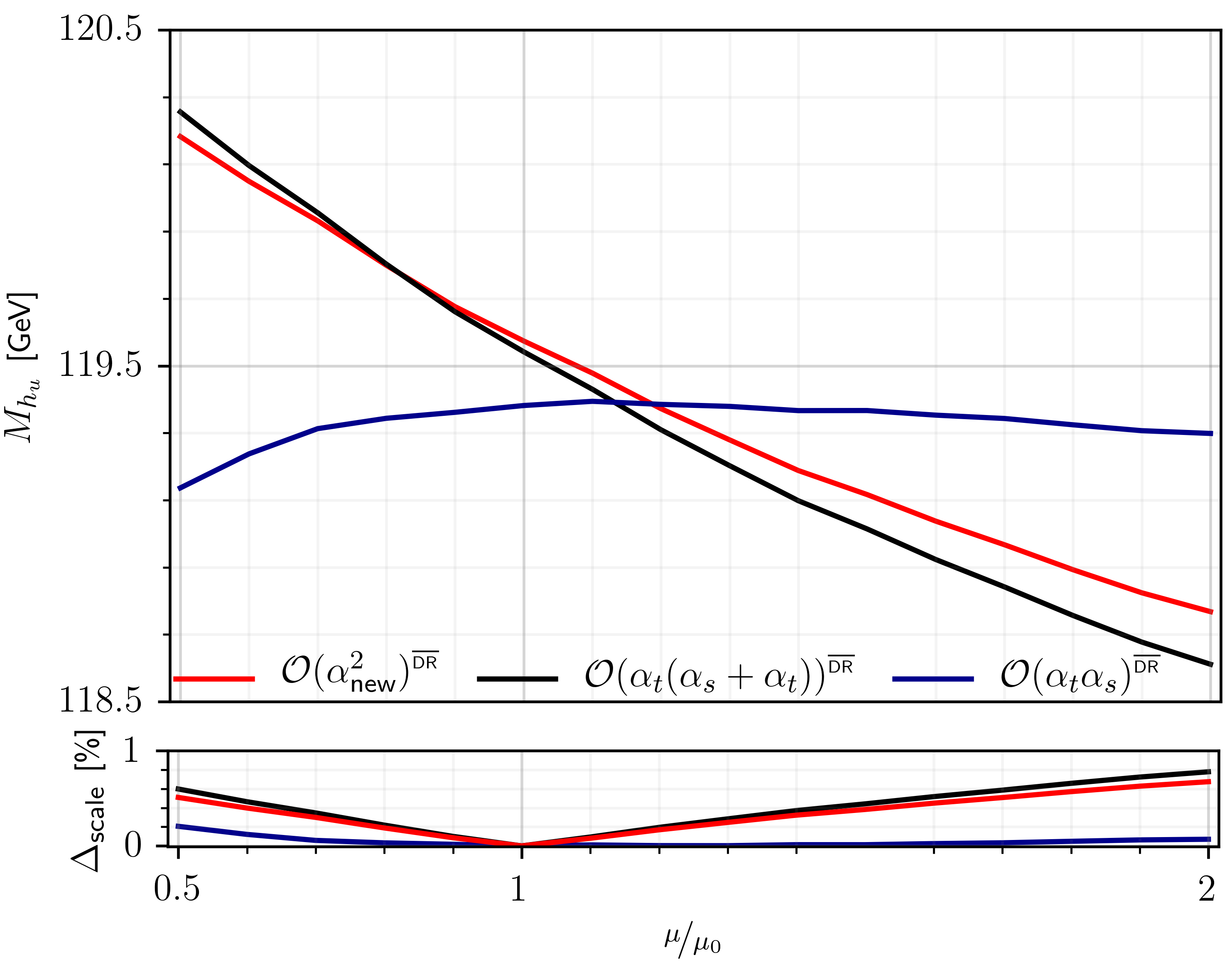}
        \includegraphics[width=0.49\textwidth]{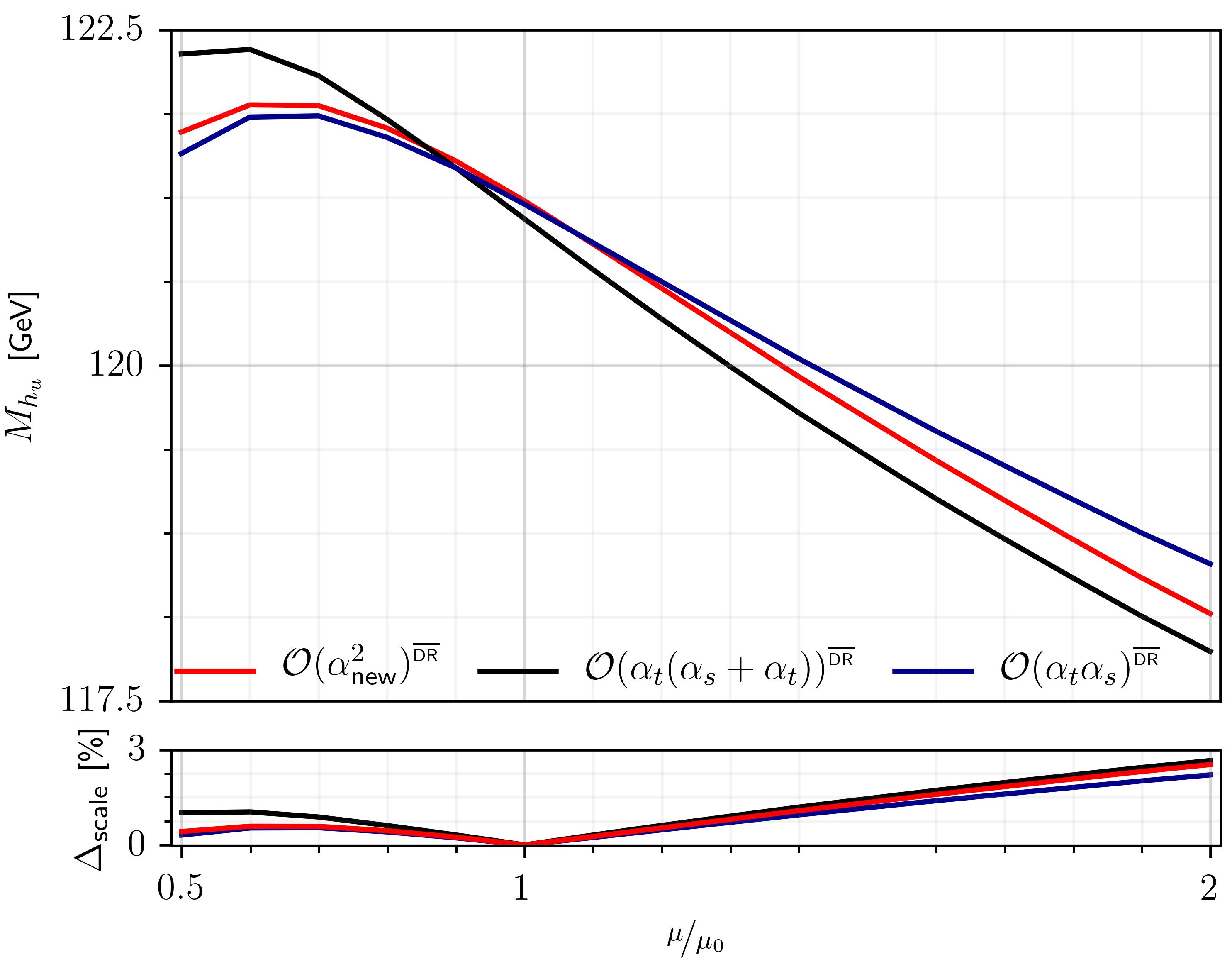}
    \caption{Upper panels: The $h_u$-like Higgs
      mass prediction at ${\cal O}(\alpha_t \alpha_s)$ (blue), ${\cal
        O}(\alpha_t (\alpha_s + \alpha_t)$ (black), and ${\cal
        O}(\alpha_{\text{new}}^2)$ (red) with $\overline{\mbox{DR}}$ renormalisation in
      the top/stop sector as a function of the renormalisation scale
      $\mu$ normalized to the default scale $\mu_0= M_{\text{SUSY}}$,
      {\it cf.}~Eq.~(\ref{eq:renscale}), for {\tt P1OS} (left) and {\tt
        P2OS} (right).  
      Lower panels: The scale dependence $\Delta_{\text{scale}}$,
      Eq.~(\ref{eq:defscaledep}), in percent. 
    }
    \label{fig:scaledep}
\end{figure}
We now turn to the discussion of the renormalisation scale dependence
of the two-loop corrected Higgs boson mass which can be taken as a
rough estimate of the uncertainty due to missing higher-order
corrections. The RGEs implemented in {\tt NMSSMCALC} are used as described in appendix E of \cite{Dao:2019qaz}. 
We define the scale dependence of the loop-corrected
Higgs mass $M_h$ at a given loop order as the relative change of the
mass value at the scale $\mu$ with respect to our default scale $\mu_0=
M_{\text{SUSY}}$, hence
\beq
\Delta_{\text{scale}} = \frac{|M_h (\mu) - M_h (\mu_0)|}{M_h (\mu_0)}
\, . \;
\label{eq:defscaledep}
\eeq
Figure~\ref{fig:scaledep} (upper) shows the scale dependence of our sample
points {\tt P1OS} (left) and {\tt P2OS} (right) at the three two-loop
orders ${\cal O}(\alpha_{\text{new}}^2)$ (red), 
${\cal O}(\alpha_t (\alpha_s + \alpha_t)$ (black), and 
${\cal O}(\alpha_t \alpha_s)$ (blue) with 
$\overline{\mbox{DR}}$ renormalisation in
the top/stop sector as a function of a variation of the
renormalisation scale $\mu$ between one half and twice the
default scale $\mu_0$. The lower plots show the relative scale dependence
$\Delta_{\text{scale}}$. For both points the scale dependence is
rather small and remains below 3\% for all three two-loop
corrections. For {\tt P1OS} we observe the smallest dependence 
for ${\cal O}(\alpha_t \alpha_s)$. It increases after including the 
${\cal O}(\alpha_t (\alpha_s + \alpha_t))$ corrections and is reduced
again with our new corrections ${\cal O}(\alpha_{\text{new}}^2)$. 
For {\tt P2OS} we find a similar relation between the different contributions.
It has to be noted here that
the renormalisation group equations applied in the generation of the
plot include all two-loop contributions while for consistency at the
three different two-loop orders only the respective contribution
corresponding to the given included loop corrections should be taken
into account in the renormalisation group equations so that the
comparison of the three curves should be taken with caution. Still,
overall we see that the inclusion of the two-loop corrections leads to
rather small remaining scale dependences.
However, the inclusion of the new corrections from the Higgs- and
electroweakino sector does only lead to a
minor reduction of the scale dependence. Therefore, we argue that the
largest uncertainty comes from the top/stop sector requiring either
higher orders or the resummation of large logarithms.

\subsection{Numerical Comparison of the Three Regulation Schemes}
\label{sec:gbccomparison}
\begin{figure}[t]
    \centering
    \includegraphics[width=0.49\textwidth]{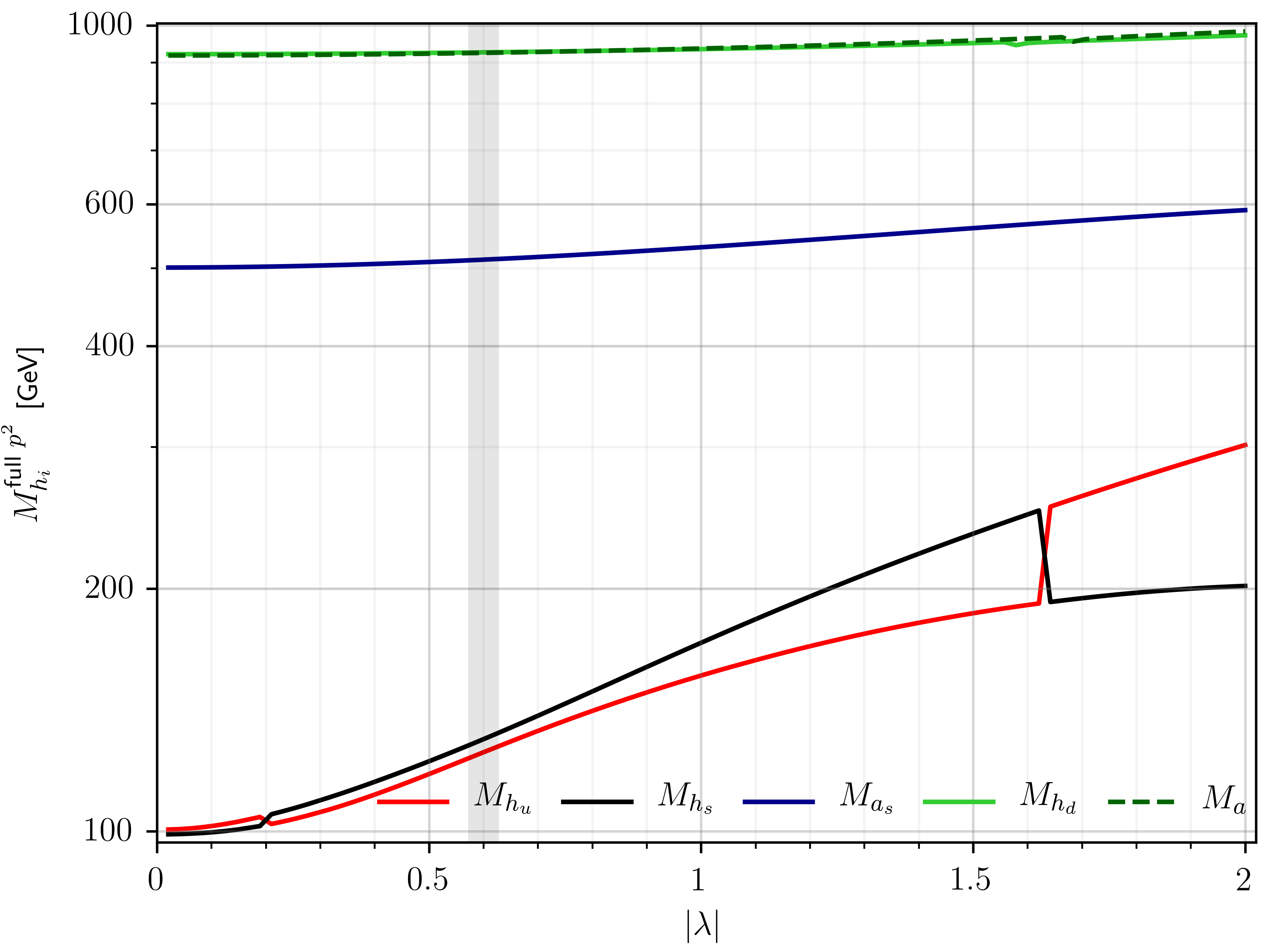}
    \includegraphics[width=0.49\textwidth]{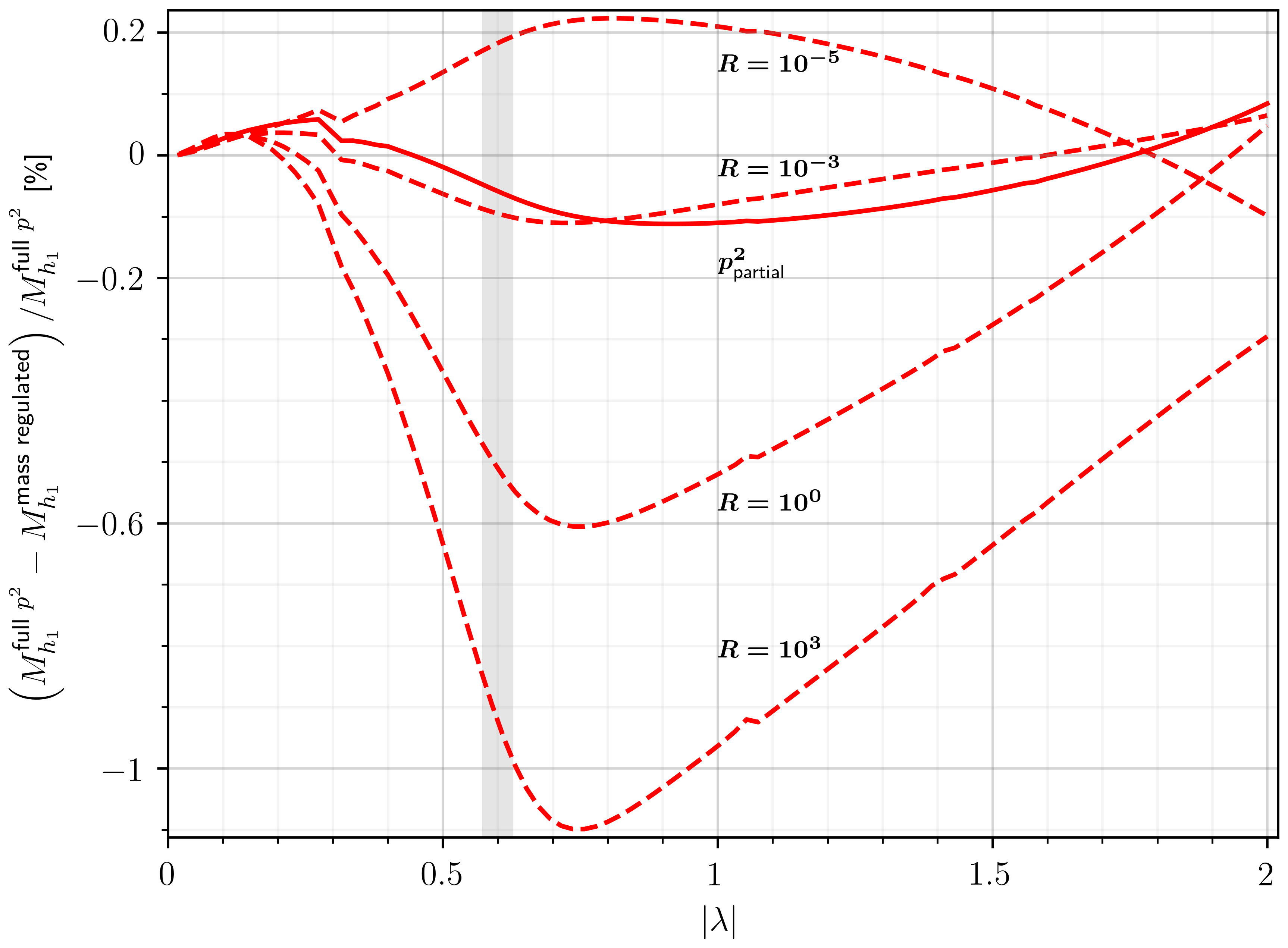}
    \caption{Left: Mass spectrum $M_{h_i}^{\text{full}-p^2}$ for the point {\tt P2OS} at ${\cal
        O}(\alpha_{\text{new}}^2)$ when including the external
        momentum in all diagrams of the order 
        $\order{(\alpha_t+\alpha_\lambda+\alpha_\kappa)^2}$, as a 
        function of $\lambda$.
        Right: Comparison of $M_{h_1}^{full-p^2}$ for the lightest
        state $h_1$ with the mass values obtained in the pure
        mass-regulated method (for $R=10^{-5},10^{-3},1$ and $10^3$, dashed
        lines, see text for definition)
        as well as with the partial momentum expansion (solid line).
        The grey shaded region is compatible with experimental Higgs boson data.
    }
    \label{fig:mhp2compP1OSnew}
\end{figure}

In Sec.~\ref{sec:self-energies} we discussed three regulation schemes
to cure the Goldstone boson catastrophe, which we compare in this
section for the parameter point {\tt P2OS}. In
Fig.~\ref{fig:mhp2compP1OSnew} (left) we show the 
loop-corrected mass values at ${\cal O}(\alpha_{\text{new}}^2)$
including the full external momentum dependence in all diagrams of
order $\order{(\alpha_t+\alpha_\lambda+\alpha_\kappa)^2}$, as
described in Sec.~\ref{sec:full}, as a function of $\lambda$. The momentum
is approximated by the fixed value $p^2=(m_{h_i}^2+m_{h_j}^2)/2$,
where $m_{h_{i,j}}$ are the 
tree-level Higgs boson masses, for the calculation of the new self-energy
corrections $\hat{\Sigma}^{(2)}_{ij}(p^2)$. 
At $\lambda = 0.2$ we observe a cross-over. Here $h_u$ (red line) and $h_s$ (black)
are close in mass and change their roles with respect to the mass
ordering. Below this $\lambda$ value 
$M_{h_u}$ is the second-lightest Higgs boson and above it is the lightest
one. A second cross-over is observed at $\lambda=1.63$ where $h_u$ and
$h_s$ strongly mix and change their roles, so that above this
$\lambda$ value the $h_u$-like Higgs boson is the second-lightest
Higgs state again. We furthermore see that the mass of the singlet-like Higgs
boson $h_s$ (black) shows a stronger
dependence on the singlet-doublet coupling $\lambda$ in contrast to
those of the doublet-like Higgs bosons $h_d$ (green) and $a$ (dark
green, dashed). Due to the strong mixing between $h_u$ and $h_s$ also the
loop-corrected $h_u$-like Higgs mass shows a significant dependence on
$\lambda$.\s

In Fig.~\ref{fig:mhp2compP1OSnew} (right) we compare the loop-corrected mass
$M_{h_1}^{\text{full}-p^2}$ at ${\cal O}(\alpha_{\text{new}}^2)$ including external
momentum in all diagrams of the order
$\order{(\alpha_t+\alpha_\lambda+\alpha_\kappa)^2}$ with the ${\cal
  O}(\alpha_{\text{new}}^2)$ result obtained with partial
momentum dependence as described
Sec.~\ref{sec:partialp}, and denoted by $p_{\text{partial}}^2$ in the
plot. We furthermore compare $M_{h_1}^{\text{full}-p^2}$ with the results of
the mass computation where the IR divergence is 
regulated by a mass regulator, {\it cf.}~Sec.~\ref{sec:gbcmassreg}. Here, we
consider four cases for the regulator mass $M_R^2$, namely 
$R=M_R^2/\mu_0^2 = 10^{-5},$ $10^{-3}$, 1 and $10^3$. Note
that we included the full momentum dependence in all diagrams of the considered order whereas the partial
 momentum approximation and the regulator mass is only applied in the
GBC subset which is always  proportional to $\lambda$ and $\kappa$.
As expected partial  momentum inclusion approximates the full momentum
dependence best, as well as the regulator mass result for $R=10^{-3}$. For regulator
masses departing more and more from $10^{-3} \mu_0^2$ the
difference increases. The difference for all approximations shows a
strong dependence on $\lambda$ but agrees for $\lambda=0$ as expected since the IR effects are 
related to $\lambda$. The largest deviation from
$M_{h_1}^{\text{full}-p^2}$ amounts to $\sim 1.5$~GeV in the mass regulated
scheme with $R=10^3$. In contrast, the maximum deviation remains below
150 MeV for $R=10^{-3}$ for
this parameter point and $\lambda$ varying between 0 and
2. The kinks in the plot are not an artifact of the numerical
integration but are due to the strong dependence of the neutral
Higgs mixing matrix on $\lambda$ and the resulting change of certain sub-dominant
admixtures when departing from the region allowed by the current
collider constraints (grey shaded region).\s

The result that the full- and partial-momentum corrections do not
deviate by more than 1-200 MeV shows that our incomplete calculation of the
full-momentum corrections has to be taken with care as
these corrections are of similar size as the momentum
corrections of $\order{\alpha_s\alpha_t}$ one might expect comparing
to results from the MSSM 
\cite{Degrassi:2014pfa,Borowka:2014wla,Domingo:2020wiy} (which are not calculated in
this work) even for very large values of $\lambda$. Therefore, the
calculation of the momentum-dependent 
corrections at the order $\order{\alpha_s\alpha_t}$ is still an open task to be done. \s

\begin{figure}[t]
    \centering
    \includegraphics[width=0.59\textwidth]{tbeta_cbar.png}\\
    \includegraphics[width=0.49\textwidth]{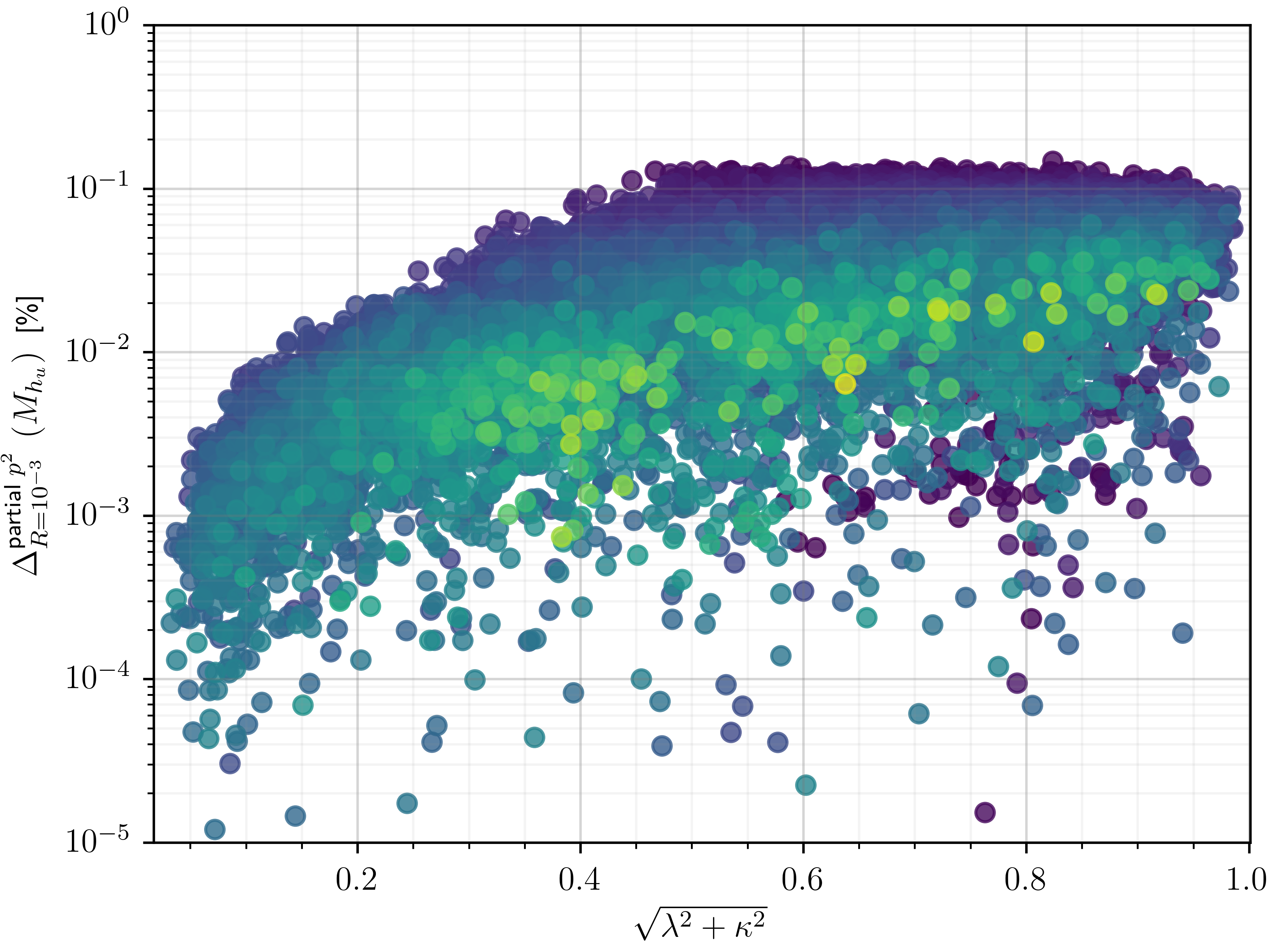}
    \includegraphics[width=0.49\textwidth]{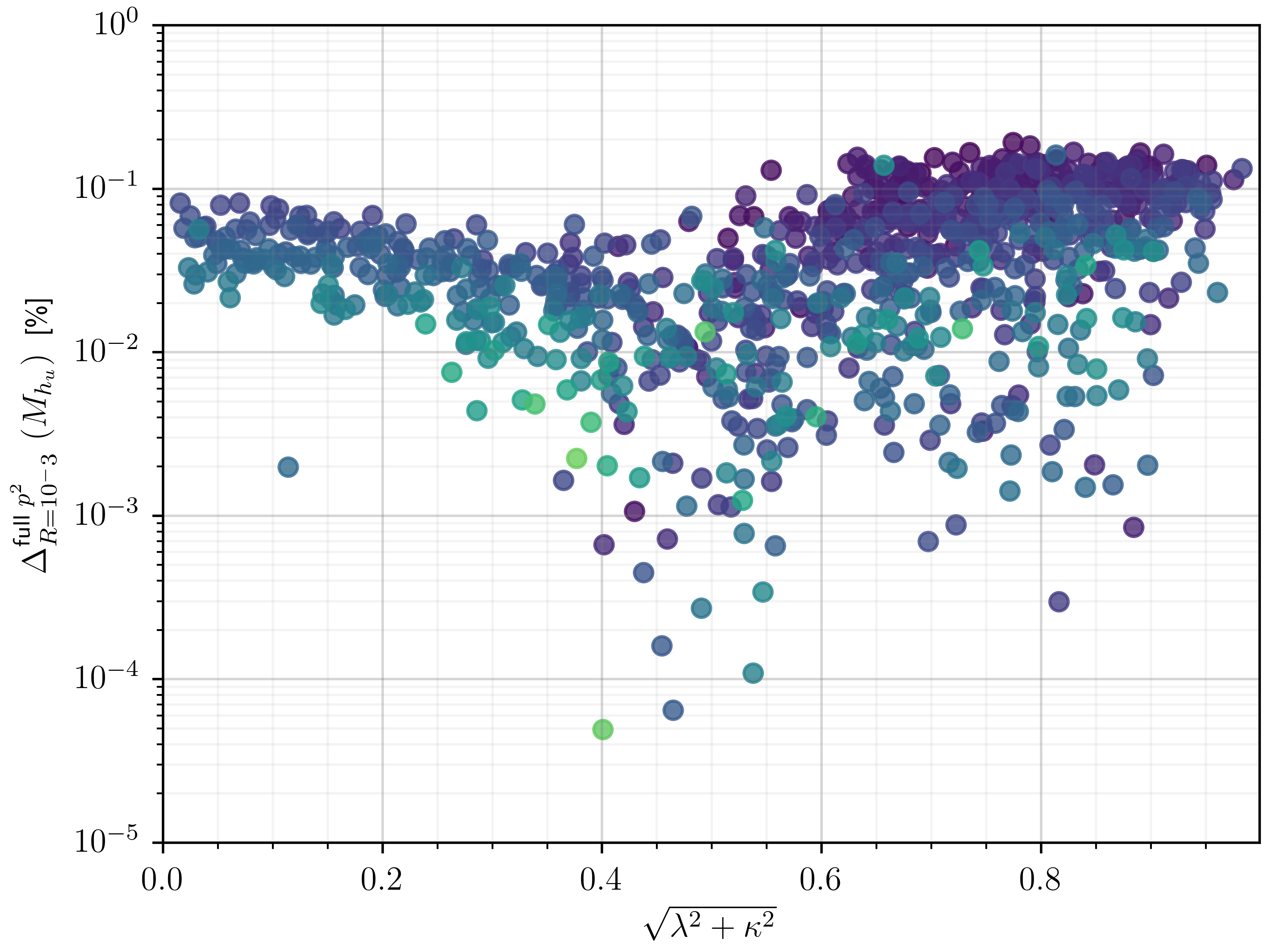}
    \caption{
        Left: Relative difference in the Higgs boson mass prediction $M_{h_u}$ when using 
        partial external momentum or purely mass-regulated
        IR-divergences with $R=10^{-3}$. Right: Same but comparing the
        purely mass regulated result with the full-momentum one for a
        randomly selected subset of 1000 points.
}
    \label{fig:PartialP2Scatter}
\end{figure}

In order to get a more general picture we compare the results in the
various approximations for all of our allowed points.
Figure~\ref{fig:PartialP2Scatter} (left) shows the relative
difference $\Delta^{\text{partial}-p^2}_{R=10^{-3}}(M_{h_u})$ between the
${\cal O}(\alpha_{\text{new}}^2)$-corrected $M_{h_u}$ value obtained
in the partial momentum approximation and in the mass-regulated
computation with a regulator mass squared of $10^{-3}
\mu_0^2$ for all allowed 
parameter points. 
Note that these two different treatments only affect the IR-divergent diagrams.
The results are shown as a function of the NMSSM-specific parameter
$\sqrt{\lambda^2+\kappa^2}$ and the color code indicates
$\tan\beta$. The maximum relative difference increases with
$\sqrt{\lambda^2+\kappa^2}$  and saturates at values close to 1
permille. This behaviour is expected, since the
IR-regulated diagrams - and therefore the dependence on the IR
regulator mass - are always proportional to $\lambda$ and $\kappa$.
The right plot displays the relative difference in $M_{h_u}$
when including full momentum dependence in the ${\cal O}((\alpha_t +
\alpha_\lambda + \alpha_\kappa)^2)$ contributions and the
mass-regulated result for $R=10^{-3}$ when considering a subset of one
thousand random points\footnote{This was done to save computational
resources. We confirmed that this subset still contains all important features
of the original sample.}.
Note that the full momentum dependence is also taken into account in
all IR-finite diagrams of the order
$\order{(\alpha_t+\alpha_\lambda+\alpha_\kappa)^2}$. 
The maximum deviation shows
only a weak dependence on $\sqrt{\lambda^2+\kappa^2}$ and reaches
at most 2 permille for large values of $\sqrt{\lambda^2+\kappa^2}$. 
The behaviour shows that the
full-momentum corrections at $\order{\alpha_t^2}$ ({\it i.e.}~for
$\lambda,\kappa\to0$ in the right plot of \cref{fig:PartialP2Scatter})  
and those at $\order{(\alpha_t+\alpha_\lambda+\alpha_\kappa)^2}$ for finite $\lambda$
are equally important.
For the singlet-like masses of our parameter
sample we found similar results. This allows us to conclude that
the regulation of the GBC with a regulator mass of $R=10^{-3}$ is a good compromise between
accuracy and computational costs\footnote{The computation
  time considerably increases when the full momentum dependence is
  included.} with a difference that remains in the subpercentage
range. Also $R$ is not too small in this case to 
lead to numerically instable results. Comparing
\cref{fig:PartialP2Scatter} (left) with \cref{fig:overall1} (left), we
conclude that the error made for $R=10^{-3}$ compared to the result
obtained with partial momentum dependence is always around one order
of magnitude smaller than the overall size of the new two-loop
corrections. Therefore, all parameter samples that we
present in this paper have been obtained with a regulator mass of
$R=10^{-3}$, unless stated otherwise. The value $R=10^{-3}$
is also the default setting that we have implemented in our new version of
{\tt NMSSMCALC} that has been extended to the here presented new
two-loop corrections.

\subsection{CP-Violating Phases}
\label{sec:CPV}
\begin{figure}[t!]
    \centering
    \includegraphics[width=0.49\textwidth]{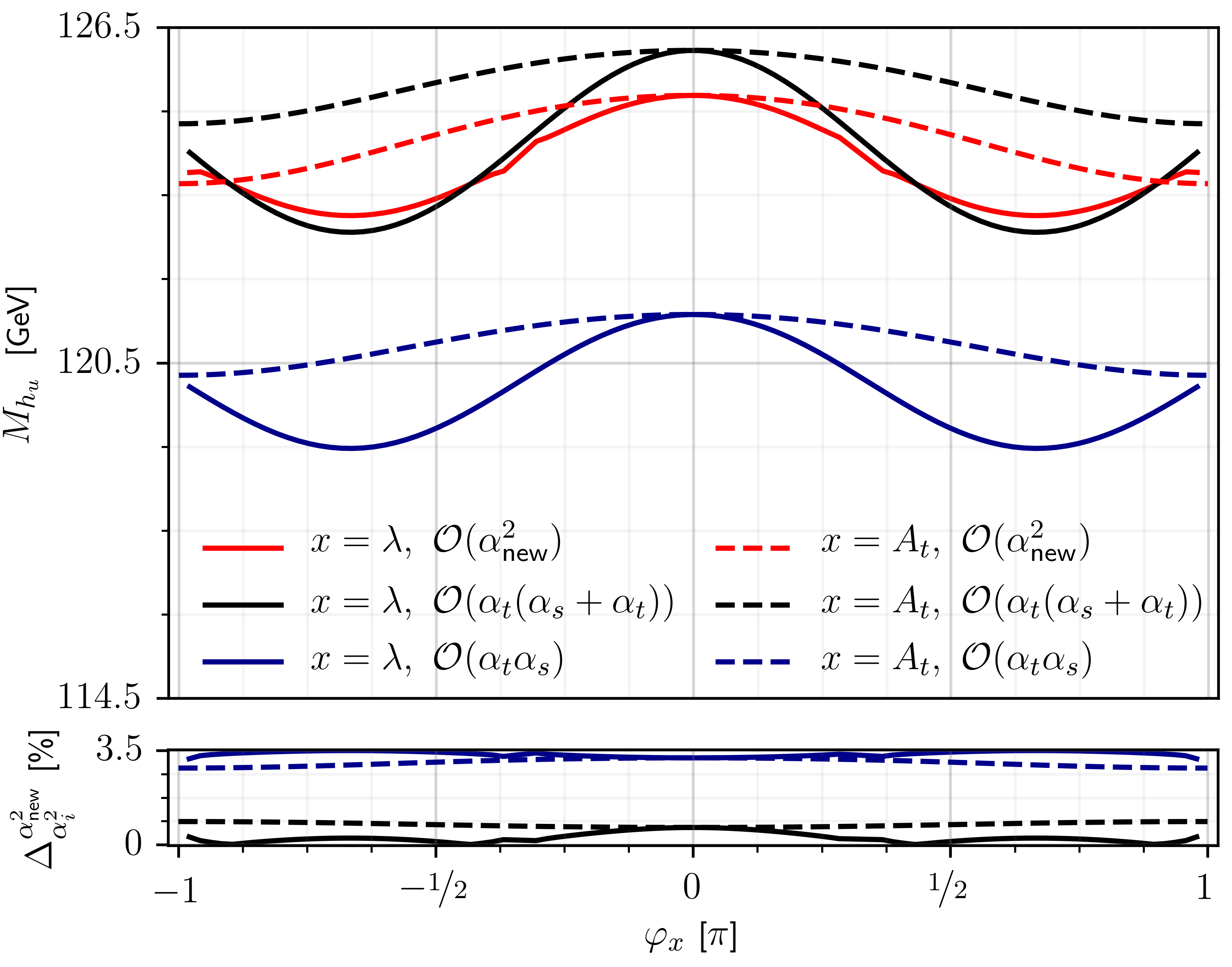}
    \includegraphics[width=0.49\textwidth]{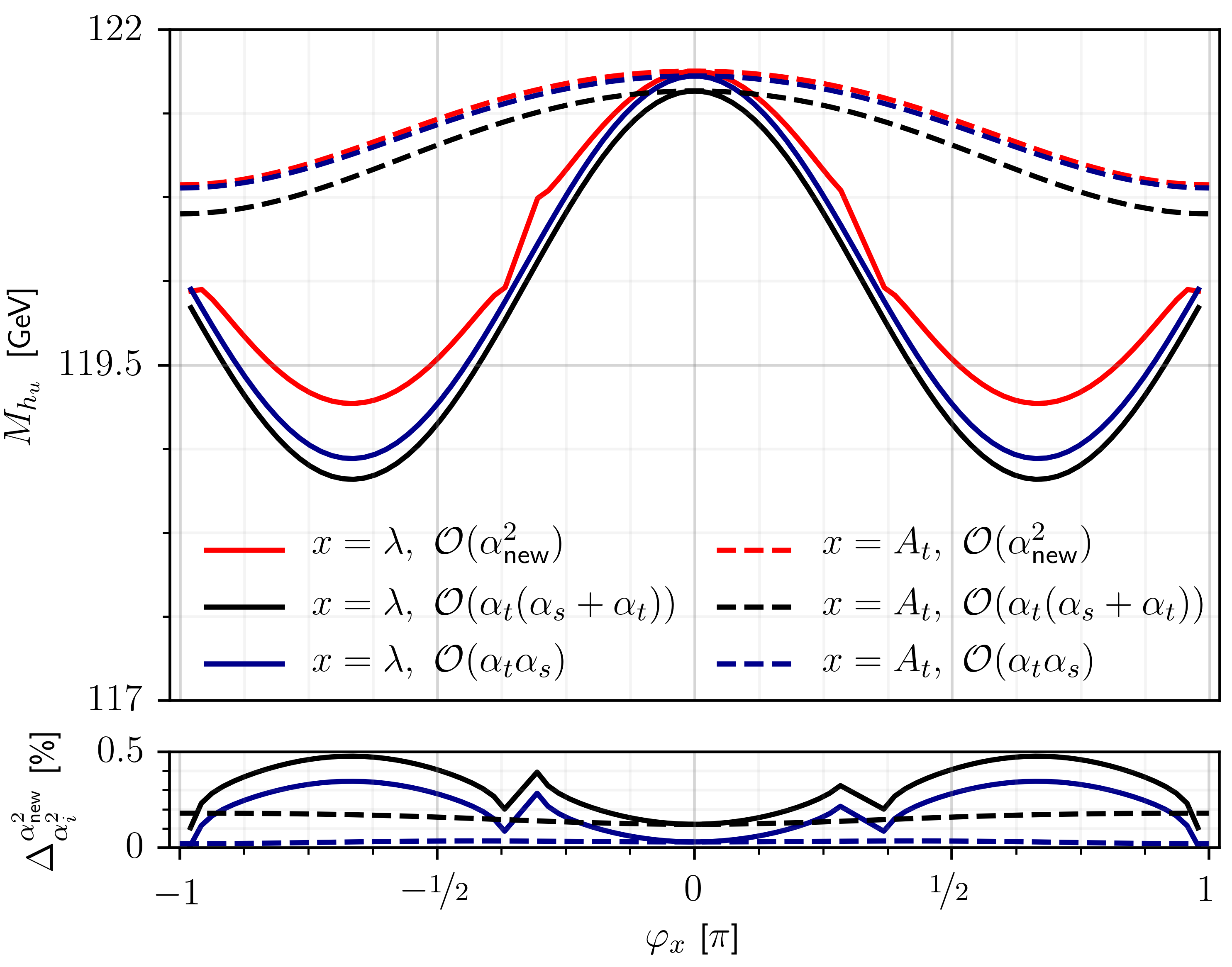}
\caption{
    Upper panels: Loop-corrected $h_u$-like Higgs boson mass $M_{h_u}$
    for {\tt P2OS} as function of $\varphi_\lambda$ (solid) and
    $\varphi_{A_t}$ (dashed) at order $\order{\alpha_\text{new}}$ (red),
     $\order{\alpha_t(\alpha_s+\alpha_t)}$ (black), and
     $\order{\alpha_t\alpha_s}$ (blue) for OS (left) and
     $\overline{\mbox{DR}}$ (right) renormalisation in the top/stop
     sector. Lower panels: Relative
     size of the new two-loop corrections with respect to the
     previously calculated two-loop results
     $\alpha_i=\alpha_t(\alpha_s+\alpha_t)$ (black) and
     $\alpha_t\alpha_s$ (blue). 
     The phases are not varied simultaneously. For details, see text.
    }
    \label{fig:phases}
\end{figure}
In this section, we discuss the influence of the CP-violating phases
on the loop corrections to the Higgs masses. In Fig.~\ref{fig:phases} (upper)
we show by choosing the parameter point {\tt P2OS} as starting point
the loop-corrected mass $M_{h_u}$ of the $h_u$-like Higgs 
boson\footnote{The kinks in the plot appear at parameter
  configurations where the $h_u$-like and $h_s$-like Higgs bosons swap
their mass ordering and are therefore nearly
degenerate.} as a function of the
CP-violating phases $\varphi_\lambda$ (full) and $\varphi_{A_t}$ (dashed) at
${\cal O} (\alpha_{\text{new}}^2)$ (red), ${\cal O}(\alpha_t
(\alpha_s+ \alpha_t)$ (black), and ${\cal O}(\alpha_t \alpha_s)$
(blue) for OS (left) and $\overline{\mbox{DR}}$ (right) renormalisation
  in the top/stop sector. In the phase variation only one phase is
  varied at a time. The phase $\varphi_\lambda$ is varied such that the
  CP-violating phase $\varphi_y$ (Eq.~(\ref{eq:phase1})) appearing already at tree level is
  kept zero, more specifically
  $\varphi_\lambda=2\varphi_s=2/3 \varphi_{\mu_{\text{eff.}}}$ 
and $\varphi_\kappa=\varphi_u=0$. We thereby ensure to study only
radiatively induced CP-violating effects. Otherwise all CP-violating
phases are kept zero. Note also that for illustrative reasons, the
phases are varied in ranges beyond their allowed validity 
by the EDM constraints.\footnote{Actually, non-zero $\varphi_\lambda$ are
  excluded by the EDM constraints for these parameter points while
  $\varphi_{A_t}$ is allowed in the whole
  range.} The lower inserts quantify the 
effect $\Delta^{\alpha_{\text{new}}^2}_{\alpha_i^2}$ of the newly
calculated loop corrections with respect to the ${\cal O}(\alpha_t
(\alpha_s+\alpha_t))$ (black) and the ${\cal O}(\alpha_t \alpha_s)$
(blue) corrections. Both in OS and
$\overline{\mbox{DR}}$ renormalisation all three two-loop corrections
show the same behaviour with respect to a variation of
$\varphi_{A_t}$. This can also be inferred from the lower panels where
the dashed lines are almost flat. The variation of the loop
corrections with a change of the phases is more pronounced in the OS
scheme than in the $\overline{\mbox{DR}}$ scheme, also the behaviour
of the new corrections differs more from the other two two-loop
orders. Still the curves for the relative deviation
$\Delta^{\alpha_{\text{new}}^2}_{\alpha_i^2}$ do not vary much with a
change of the phases. Overall the impact of the CP-violating phase on
the new corrections with respect to the previous ones is small.
\section{Conclusions and Outlook}
\label{sec:conclusions}
We have computed the $\order{(\alpha_t+\alpha_\lambda+\alpha_\kappa)^2}$ two-loop
corrections to the Higgs boson masses of the CP-violating NMSSM in the
Feynman-diagrammatic approach in the gaugeless limit and at vanishing
external momentum. While these limits give a good approximation and
  simplify computations significantly, they may induce infrared divergences 
in two-loop Feynman diagrams with multiple massless Goldstone bosons. 
 We have shown that using OS conditions for tadpoles at one-loop order
makes the two-loop tadpoles, the charged Higgs, the $W$ and the $Z$
boson self-energies at vanishing
external momentum at
$\order{(\alpha_t+\alpha_\lambda+\alpha_\kappa)^2}$ IR finite, however,
IR divergences remain in the neutral Higgs boson self-energies.
For the treatment of the IR divergences we have followed
three different approaches - the introduction of a regulator mass, the
application of a small momentum expansion, and the inclusion of the
full momentum dependence in all Feynman diagrams of
$\order{(\alpha_t+\alpha_\lambda+\alpha_\kappa)^2}$.
By comparing the three methods, we found that 
the  regulator mass approach reproduces the momentum-dependent results well for
squared regulator masses that amount to a permille of the
renormalisation scale squared. Due to the robustness of this approach
we have implemented the 
new corrections using this value for the regulator mass as default 
in the published version of {\tt NMSSMCALC}. In order to
quantify the impact of our newly computed corrections we have
performed a scan in the NMSSM parameter range and kept only those
points that are compatible with experimental constraints. We found that
our corrections increase with $\lambda$ and $\kappa$ as expected.
For $\lambda$ and $\kappa$ values compatible with perturbativity below
the GUT scale, the corrections are less than $3\%$  relative to the already available
$\order{\alpha_t(\alpha_t+\alpha_s)}$ corrections. Our new corrections reduce slightly 
the theoretical uncertainties due to missing higher-order corrections
that we estimated by changing the renormalization scheme in the
top/stop sector and by varying the renormalization scale. We have also 
shown that the impact of the  new corrections on the Higgs mixings, 
which manifest themselves in the couplings between the Higgs bosons
and the SM particles, is significant and strongly affects the
compatibility with the Higgs data. The impact of
the CP-violating phases on the new corrections has been found to be
small.  With our calculation we further improve
the precision on the NMSSM Higgs boson masses and mixings. The next steps to be
taken are the inclusion of the full gauge dependence and of non-zero
momentum in the computation of the two-loop corrections.

\section*{Acknowledgements}
M.M. and M.G. acknowledge support by the Deutsche
Forschungsgemeinschaft (DFG, German Research Foundation) under grant
396021762 - TRR 257. M.G. acknowledges financial support by the
Graduiertenkolleg GRK 1694: ``Elementarteilchenphysik bei h\"ochster
Energie und h\"ochster Pr\"azision''. T.N.D.~is funded by the Vietnam
National Foundation for Science and Technology Development (NAFOSTED)
under grant number 103.01-2020.17. H.R. was partly supported by the
German Federal Ministry for Education 
and Research (BMBF) under contract no.\ 05H18VFCA1 and partly  by  the
Deutsche Forschungsgemeinschaft (DFG, German Research
Foundation)---project no.\ 442089526.

\begin{appendix}
%
\section{Mass Regulated One- and Two-Loop Functions}
\label{app:irsafeloop}
This appendix complements the analytically known one- and two-loop
self-energy integrals with a set of IR-regularised functions. We implicitly assume vanishing external momentum
unless stated otherwise. In this limit, all two-(one-\nolinebreak) loop integrals can be
written in terms of two-(one-)loop tadpole integrals $I(x,y,z)$
($A(x)$). The analytical solution of $I$ has been studied e.g. in Refs.
\cite{Ford:1992pn,Martin:2001vx} in detail. While $I(x,y,z)$ and $A(x)$
themselves are IR-finite, their derivatives w.r.t. squared masses
become divergent in the IR regime. 
In this regime, we replace vanishing scalar masses with a mass regulator
$0\to M_R^2$ in the loop functions and expand
around the small regulator mass while keeping all terms of order
$\order{\overline{\log}^{n\leq 2}M_R^2}$ and $\order{M_R^{-n\leq 0}}$.
\s

The notation and definition of the integrals closely
follow those of Refs. \cite{Martin:2003qz,Martin:2005qm}.
In addition, we introduce the scalar three-point integral {\bf C}
\begin{equation}
    \textbf{C}(x,y,z)  = \frac{\textbf{B}(y,x) - \textbf{B}(z,x)}{y-z} 
\end{equation}
in terms of the two-point integral {\bf B}, allowing us to keep track on the spurious
IR-divergences $\textbf{C}(x,0,0)$ which are not connected to the GBC but
cancelled between counterterm-inserted diagrams and genuine two-loop
diagrams involving the \textbf{V}-integral introduced later.
\s

The required IR-save one-loop functions are
\begin{align}
    \textbf{C}(x,0,0) &= \del \textbf{B}(0,x) \\
    \textbf{C}(0,y,0) &= \frac{\textbf{B}(y,0) - \textbf{B}(0,0)}{x} \\\label{eq:B00}
    \textbf{B}(0,0)   &= -\lnMR + \epsilon \frac{(\zeta_2 + \lnbar^2 M_R^2)}{2} + \frac{1}{\epsilon} \\
    \del \textbf{B}(0,0) &= -\frac{1}{2 M_R^2} + \epsilon \frac{\lnMR}{2 M_R^2} \\
    \del \textbf{B}(0,y) &= \frac{B(0,y)}{y} + \frac{\lnMR}{y} + \frac{\epsilon}{y}\left( 1 -\lnbar y + \lnbar^2 y - \frac{\lnbar^2 M_R^2}{2}\right)
    \, .
\end{align}
Diagrams that factorise into products of one-loop functions can contribute with finite terms like \textit{e.g.}
$\textbf{A}(M_R^2)\textbf{C}(M_R^2,M_R^2,M_R^2)=  \frac{1}{2}+\lnMR +
\frac{1}{2\epsilon}+ \order{M_R^2}$ using a strict expansion in
$M_R^2$, where {\bf A} denotes the one-point
function. However, these diagrams would vanish when including the full momentum dependence.
Therefore, we always set the one-point function $\textbf{A}(0)=0$ before
we start with the expansion in a small regulator mass.\s

We continue with the IR-regulated two-loop functions. For convenience,
we define the following abbreviations for derivatives of the tadpole
integral $I$:
\begin{align}
    \del I(x,y,z) &\equiv \frac{\del}{\del x^\prime} I(x^\prime,y,z)|_{x^\prime=x}\\
    \del^2 I(x,y,z) &\equiv \frac{\del^2}{\del x^\prime \del y^\prime} I(x^\prime,y^\prime,z)|_{x^\prime=x, y^\prime=y}
    \, .
\end{align}
For the NMSSM, we need the following special cases
\begin{align}
    \del I(0,x,y)   &=  \lnMR B(x,y) - \overline{T}(0,x,y) \\
    \del^2 I(0,0,z) & = \frac{z (2 \lnbar z-5)-2 I(0,0,z)}{z^2}+\frac{\lnMR (-2 \lnbar z+\lnMR+2)}{z}\\
    \del^2 I(0,y,0) & = \frac{\lnbar y}{y}-\frac{\lnMR}{y} \\
    \del^2 I(0,y,y) & = \frac{\lnbar y-\lnMR+1}{2 y}\\
    \del^2 I(0,y,z) & = z\frac{2 I(0,y,z)-\lnbar y (4 y+z)+(\lnbar y-2) \lnbar z (y+z)}{(y-z)^3} \\
                    &\,\,\,\,  + \frac{y^2 \lnbar y + 5 z (y+z)}{(y-z)^3}  +\frac{\lnMR (z \lnbar y-z \lnbar z-y+z)}{(y-z)^2}
    \, .
\end{align}
The function $\overline{T}$ is IR-finite, has been introduced in
Ref. \cite{Martin:2003qz} and is identical to the function
$R_{SS}(x,y)$ used in Refs. \cite{Kumar:2016ltb,Braathen:2016cqe}. For
completeness, we recall here only the
expressions needed in the regularisation procedure:
\begin{align}
    \overline{T}(0,x,y) = & \frac{(x+y) I(0,x,y)+2 (A(x)-y) (A(y)-x)+x^2+y^2}{(x-y)^2} \\
    \overline{T}(0,x,x) &= -\frac{1}{2} \lnbar x^2-\lnbar x-\frac{3}{2}
    \, .
\end{align}
With this set of functions we can define all remaining two-loop functions
in an IR-regulated way. We start with the UV-divergent 
$\textbf{U}$-integral at vanishing external momentum, 
\begin{align}\label{eq:Uxyzu}
    \textbf{U}(x,y,z,u) &= U(x,y,z,u) + \frac{\textbf{B}(x,y)}{\epsilon} + \frac{1}{2}\left(\frac{1}{\epsilon} - \frac{1}{\epsilon^2} \right)\\
    U(x,y,z,u)  &\stackrel{p^2=0}{=} \frac{I(z,u,y) - I(z,u,x)}{x-y}\\
    U(x,x,y,z)  &= -\del I(x,y,z) \;.
\end{align}
Therefore we need to regulate 
\begin{equation}
    \textbf{U}(0,0,x,y) = - \del I(0,x,y) + \frac{\zeta_2 + \overline{\log}^2M_R^2 }{2} + \frac{1- 2\lnMR}{2 \epsilon} + \frac{1}{2\epsilon^2}
    \, .
\end{equation}
We have verified that the UV-IR mixing terms
$\order{\lnMR/\epsilon}$ cancel in the sub-loop
renormalisation with counterterm-inserted diagrams that involve a
vertex counterterm and 
an IR-divergent $\textbf{B}(0,0)$ integral.\s

The $\textbf{V}$-integral
\begin{equation}
    \textbf{V}(x,y,z,u) = -\frac{\del}{\del y} \textbf{U}(x,y,z,u)
\end{equation}
can also contribute to the UV-IR mixing
terms because its single pole can be written as
\begin{equation}
    \textbf{V}(x,0,z,u)|_{\epsilon^{-1}} = 
    -\frac{\del}{\del y} \textbf{U}(x,y,z,u)|_{y=0}|_{\epsilon^{-1}}
    = -\frac{\textbf{C}(x,0,0)}{\epsilon} \;.
\end{equation}
It precisely cancels against counterterm-inserted
diagrams involving a Higgs 
boson mass counterterm and a three-point function. Furthermore, we
need the following special cases of vanishing arguments,
\begin{align}
    \textbf{V}(0,0,0,u) &= -\frac{\lnbar u}{2 M_R^2}-\frac{\lnbar u}{2 u}-\frac{\lnMR}{2 M_R^2}+\frac{\lnMR}{2 u}+\frac{1}{2 M_R^2}(1+\frac{1}{\epsilon}) \\ 
    \textbf{V}(0,0,z,z) &= -\frac{\lnbar(z)}{2 M_R^2}-\frac{\lnMR}{2 M_R^2}+\frac{1}{2 M_R^2 \epsilon } \\ 
    \textbf{V}(0,0,z,u) &= \frac{1}{2(z-u)^4} \left\{  -u \lnbar u \left[2 z \lnbar z (u+z)+2 (\lnMR-3) u z-(2 \lnMR+3) z^2+u^2\right] \right. \nonumber \\
                        & \qquad\qquad\qquad\qquad + z \lnbar z \left((2 \lnMR+3) u^2-2 (\lnMR-3) u z-z^2\right) \nonumber \\
                        & \qquad\qquad\qquad\qquad \left. -4 u z I(z,0,u) + (u+z) \left(\lnMR (z-u)^2-10 u z\right)  \right\} \nonumber \\
                        & \quad + \frac{u \lnbar u- z \lnbar z-u+z}{2 M_R^2 (z-u)} -\frac{\lnMR}{2 M_R^2} + \frac{1}{2 M_R^2 \epsilon }\\
    \textbf{V}(x,0,z,u) &= \frac{I(x,z,u)-I(0,z,u)-x \del I(0,z,u)}{x^2} \\
                        & \quad +\frac{2 \lnbar x-\lnbar^2 x  +\lnbar^2 M_R^2 - 2}{2 x}+\frac{\lnbar x-\lnMR-1}{x \epsilon }
                        \, .
\end{align}
The UV-finite two-loop master integral $\textbf{M}$ at vanishing
external momentum requires IR-re\-gu\-la\-ri\-sa\-tion for the following cases,
\begin{align}
    \textbf{M}(x,0,z,0,v) &= \frac{\del I(0,x,v)-\del I(0,z,v)}{x-z} \\
    \textbf{M}(x,0,x,0,v) &= \del^2 I(0,x,v) \,.
\end{align}
Finally, the derivatives w.r.t. external momentum are needed for
the calculation of the wave function renormalisation constants. They are regularised by
\begin{align}
    \del_{p^2} \textbf{U}(0,0,x,y)|_{\text{UV-div}} &= \frac{\del_{p^2} \textbf{B}(0,0)}{\epsilon} \\
    \del_{p^2} \textbf{C}(x,0,0) &= \frac{5+ 2 \lnMR - \lnbar x}{2 x^2} \\
    \del_{p^2} \textbf{C}(0,0,0) &= -\frac{1}{12 M_R^4} \\
    \del_{p^2} \textbf{B}(0,0) &= \frac{1}{6 M_R^2} \,,
\end{align}
while the finite part of the $\textbf{U}$-integral is not needed as we
renormalise all wave functions in the $\DRbar$ scheme.
\s

Keeping the $\textbf{M},\, \textbf{U},\, \textbf{V}$ and $\textbf{S}$
integrals even in the zero-momentum approximation (and expressing them
in terms of the $I$ integrals during the numerical evaluation) makes
the transition to finite/partial external momentum dependence
straightforward as we simply replace them by the functions computed
with \TSIL.
\s

As a closing remark we want to mention that only the $\order{\lnMR}$ and
$\order{\lnbar^2 M_R^2}$ divergences are actually physical as they
correspond to $\order{\epsilon_\text{IR}^{-1}}$ and
$\order{\epsilon_\text{IR}^{-2}}$ terms in dimensional
regularisation. As a further cross-check, we explicitly checked that
indeed all $\order{M_R^{-n\leq-2}}$-terms cancel exactly in the sum of all two-loop diagrams.

\section{IR-Divergent Topologies}
\label{sec:irtopos}
In this appendix we give all IR-divergent cases for the two-loop
tadpole and self-energy topologies ind \cref{tab:irtoposTAD} and
\cref{tab:irtoposSELF}.
The first column shows the topology with labels on each generic
propagator. The second column lists special cases of vanishing masses
in the propagators that lead to IR-divergent loop functions. The third
column collects the various sets discussed in \cref{sec:gbcmassreg}. 
The fourth column lists the IR-divergent loop functions that appear in all
possible field-insertions after applying the {\tt TARCER}'s algorithm.
The last column indicates whether the IR-divergence cancels in the
final result or if it requires the inclusion of external momentum. \s

\begin{table}[ht]
    \centering
    \begin{tabular}{c|ccccc}
        \# & topology & \makecell{conditions for\\ IR-divergence} & set & \makecell{IR-divergent\\ functions} & \makecell{momentum\\ regularisable?} \\ \hline 
        1 & \tablegraph{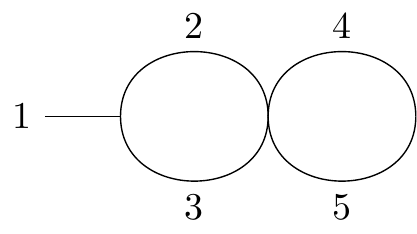} & $m_2=m_3=0$ &  A  & $\textbf{B}(0,0)$  & no \\
        2 & \tablegraph{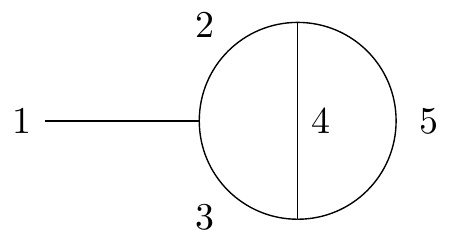} & $m_2=m_3=0$ &  A  & \makecell{$\textbf{U}(0,0,m_4^2,m_5^2), \textbf{B}(0,0)$} & \makecell{no} \\
        3 & \tablegraph{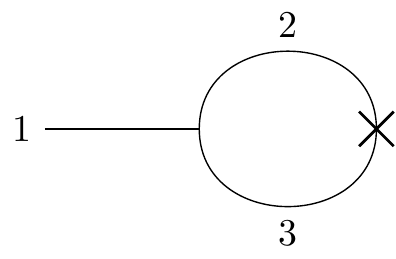} & $m_2=m_3=0$ &  A  & $\textbf{B}(0,0)$  & no
    \end{tabular}
    \caption{All IR-divergent two-loop tadpole topologies generated with \FeynArts.
    Note that IR divergences are only caused by vanishing scalar
masses while all remaining lines can be scalars or fermions if the
couplings allow for it. This is the case for topology 2 at the lines 3
and 4 which lead to the additional occurrence of a $\textbf{B}(0,0)$
after applying {\tt TarcerRecurse}.}
    \label{tab:irtoposTAD}
\end{table}\noindent
While the tadpole diagrams cannot be treated with external
momentum, the self-energy diagrams indeed need external momentum in a
few cases. In addition, there are cases (such as for instance $m_3=m_4=0$
in topology 4), where loop integrals do not require
momentum-regularisation. In these cases, the IR-divergence was
found to cancel against other diagrams that are connected by the BPHZ
theorem \cite{Zimmermann:1969jj} (such as topologies 4, 7 and 11). Therefore, the
subset of topologies 4,7 and 11 with at least one massive Higgs in the
outer loop forms an IR-finite set. Similarly to the tadpole diagrams,
the topologies 8, 10 and 13 form an IR-finite subset as well.
After these considerations, only the diagrams in \cref{fig:selfirfin}
are regularised by momentum. 
\begin{table}[t!]
    \centering
    \begin{tabular}{c|ccccc}
        \# & topology & \makecell{conditions for\\ IR-divergence} & set  & \makecell{IR-divergent\\ functions} & \makecell{momentum\\ regularisable?} \\ \hline
         \vspace{-0.1cm}
        4 & \tablegraph{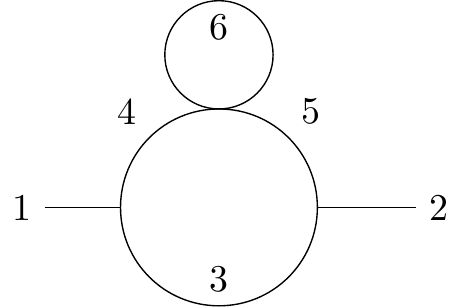} & \makecell{$m_{3,4,5}=0$ \\ $m_{4}=m_{5}=0$ \\ $m_3=m_{4/5}=0$} & \makecell{E\\ B\\ C}  & \makecell{$\textbf{C}(0, 0, 0)$\\ $\textbf{C}(0, m_{4/3}^2, 0)$ \\ $\textbf{C}(m_5^2,0,0)$} & \makecell{yes\\ no\\no} \\
         \vspace{-0.1cm}
        5 & \tablegraph{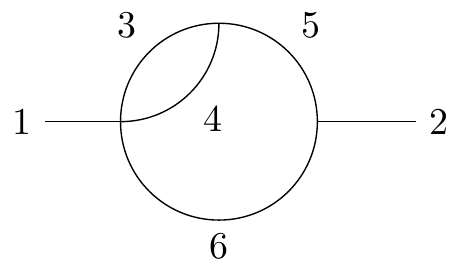} & $m_5=m_6=0$ & \makecell{D} & \makecell{$\textbf{U}(0,0,m_3^2,m_4^2)$} & \makecell{yes} \\
         \vspace{-0.1cm}
        6 & \tablegraph{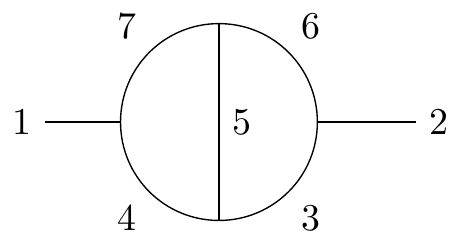} & \makecell{$m_4=m_7=0$ \\ $m_3=m_6=0 \,\quad$} & \makecell{D \\D}  &  \makecell{$\textbf{M}(m_3^2,m_4^2,m_6^2,m_7^2,m_5^2),$\\$\textbf{B}(0,0)$}  & yes \\[5ex]
         \vspace{-0.1cm}
        7 & \tablegraph{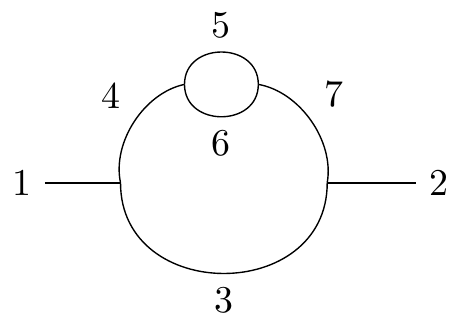} & \makecell{$m_{3,4,7}=0$\\ $m_4=m_7=0$ \\  $m_3=m_{4/7}=0$ } & \makecell{E\\ B\\ C} & \makecell{$\textbf{V}(0,0,m_5^2,m_6^2), \textbf{C}(0,0,0)$ \\ $\textbf{V}(m_3^2,0,m_5^2,m_6^2),\textbf{C}(m_3^2,0,0)$\\ $\textbf{U}(0,0,m_5^2,m_6^2),\textbf{C}(0,m_{7/4}^2,0)$ }  & \makecell{yes\\ no\\ no}\\
         \vspace{-0.1cm}
        8 & \tablegraph{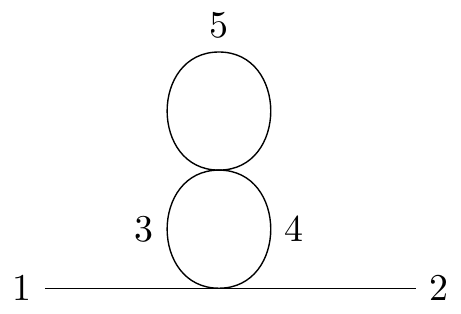} & $m_3=m_4=0$  & A & \makecell{$\textbf{B}(0,0)$} & no \\
         \vspace{-0.1cm}
        9 & \tablegraph{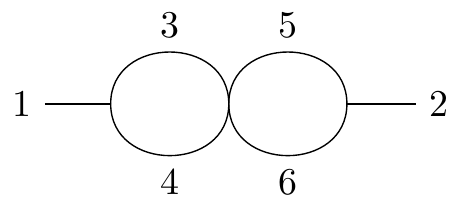} & \makecell{$m_3=m_4=0$ \\ $m_5=m_6=0 \,\quad$} & \makecell{D \\ D} & \makecell{$\textbf{B}(0,0)$} & yes \\
         \vspace{-0.1cm}
        10 & \tablegraph{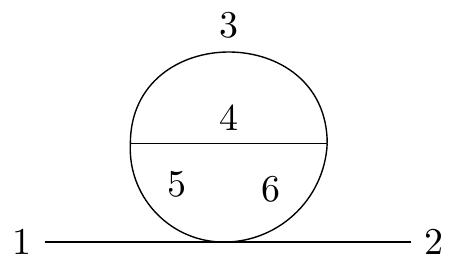} & $m_5=m_6=0$  & A & $\makecell{\textbf{U}(0,0, m_4^2, m_3^2), \textbf{B}(0,0)}$ & no \\
         \vspace{-0.1cm}
        11 & \tablegraph{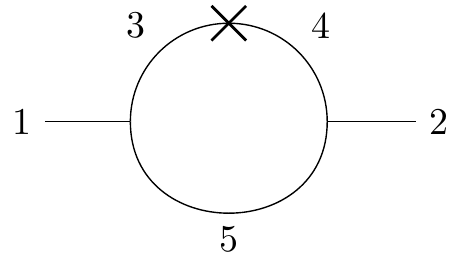} & \makecell{$\,\,\quad m_{3,4,5}=0$\\ $m_{3}=m_{4}=0$ \\ $m_{3/4}=m_5=0 \,\quad$} & \makecell{E\\ B\\ C} & \makecell{$\textbf{C}(0, 0, 0)$\\ $\textbf{C}(0, m_{4/3}^2, 0)$ \\ $\textbf{C}(m_5^2,0,0)$} & \makecell{yes\\ no\\no} \\
         \vspace{-0.1cm}
        12 & \tablegraph{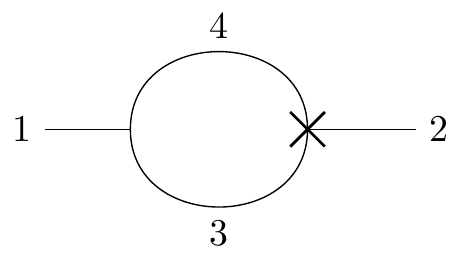} & $m_3=m_4=0$ & D & $\textbf{B}(0,0)$ & yes\\
         \vspace{-0.1cm}
        13 & \tablegraph{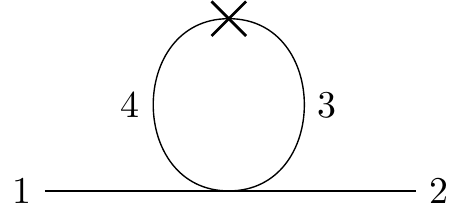} & $m_3=m_4=0$ & A & $\textbf{B}(0,0)$ & no
    \end{tabular}
    \caption{All two-loop self-energy topologies generated with
      \FeynArts. Those IR divergences which are not regularisable by momentum should cancel in the total sum. For topologies 5 and 12 mirror diagrams exist which can be derived by renaming the indices accordingly.}
    \label{tab:irtoposSELF}
\end{table}

\section{Two-Loop Diagrams}
\subsection{Tadpoles}
\label{app:tads}
\begin{figure}[t!]
    \centering
    \subfloat{\includegraphics[scale=0.6]{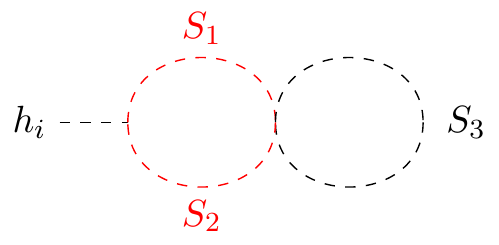}}\hfill
    \subfloat{\includegraphics[scale=0.6]{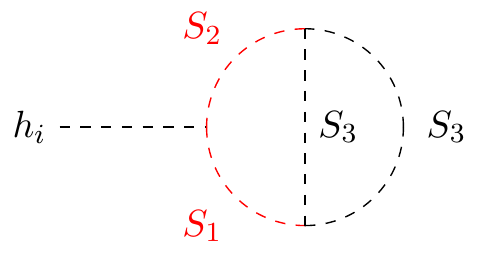}}\hfill
    \subfloat{\includegraphics[scale=0.6]{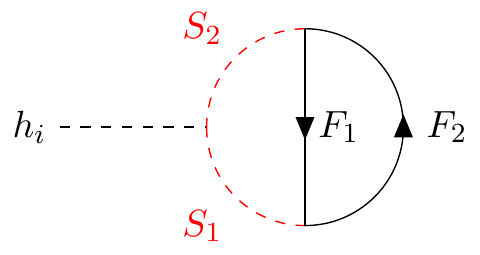}}\hfill
    \subfloat{\includegraphics[scale=0.6]{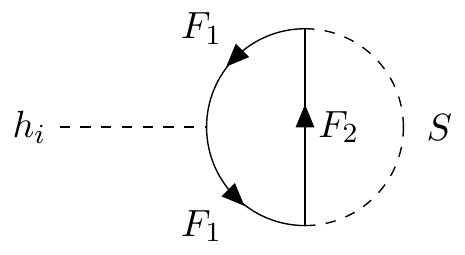}}\\
    \subfloat{\includegraphics[scale=0.6]{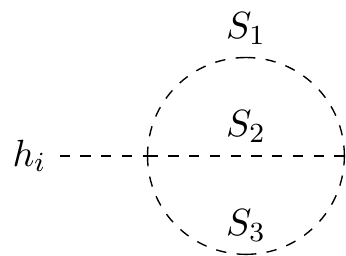}}\hfill
    \subfloat{\includegraphics[scale=0.6]{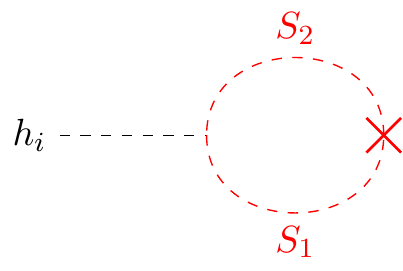}}\hfill
    \subfloat{\includegraphics[scale=0.6]{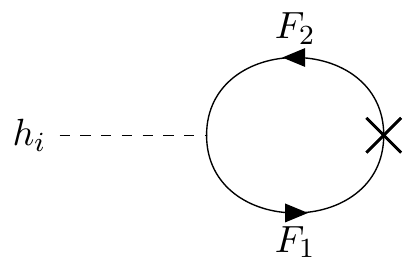}}\hfill
    \subfloat{\includegraphics[scale=0.6]{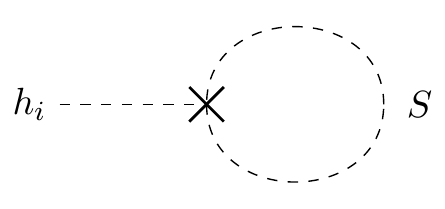}}\hfill
    \subfloat{\includegraphics[scale=0.6]{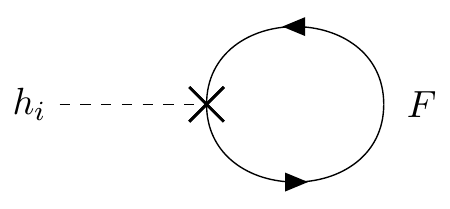}}
    \caption{
        Generic two-loop tadpole diagrams considered in this work
        where $S$ denotes scalars and $F$ denotes fermions. 
        Diagrams with red propagators become IR divergent for massless
        Goldstone bosons.
    }
    \label{fig:twolooptads}
\end{figure}

In \cref{fig:twolooptads} we show all types of tadpole diagrams that
enter at our considered loop order. 
In the summation over internal fields we consider only those two-loop diagrams
that obey the constraint 
\begin{equation}
    n_{h_{i}}+n_{h^{\pm}_j}+n_{\chi_k}+n_{\chi^\pm_l}\ge 2 \, ,
    \label{eq:taddiagselect}
\end{equation}
where $n_\Phi$ is the number of internal propagators of the field type $\Phi$.
Diagrams with one loop are taken into account if they contain at
  least one Higgs boson, electroweakino, stop or top field. 
In addition, we also consider the diagrams given at
$\order{\alpha_t^2}$, \textit{cf.} Ref. \cite{Dao:2019qaz} Figs. 14
and 15, but also taking into account terms of the order
$\order{\alpha_t\alpha_\lambda}$.\s

The two-loop tadpole diagrams with red colored propagators shown in
\cref{fig:twolooptads} suffer from IR divergences for
$\Phi=G^0,G^{\pm}$. Using the mass regularised loop functions from \cref{app:irsafeloop} we find that 
the sum of all contributions is indeed IR-finite and does not depend on the regulator mass. 

\subsection{Self-Energies}
\label{app:selfdiags}
\begin{figure}[t!]
    \subfloat[]{\includegraphics[scale=0.6]{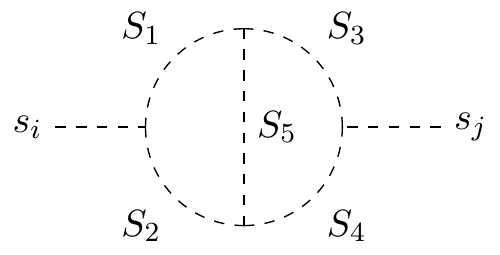}}
    \subfloat[]{\includegraphics[scale=0.6]{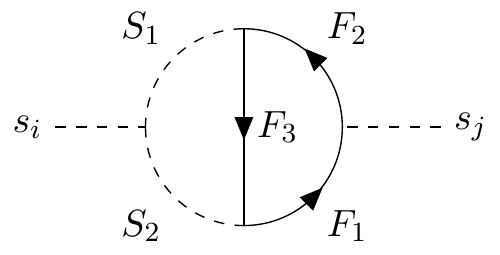}}
    \subfloat[]{\includegraphics[scale=0.6]{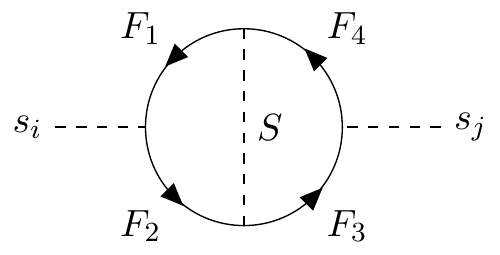}}
    \subfloat[]{\includegraphics[scale=0.6]{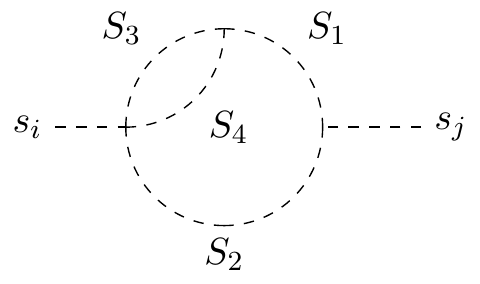}}
    \subfloat[]{\includegraphics[scale=0.6]{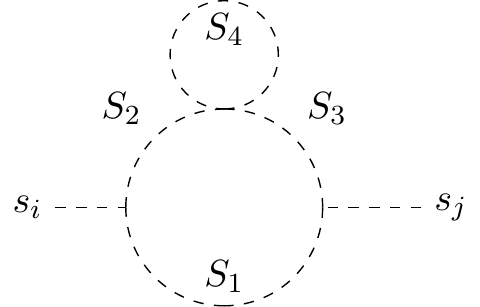}}\\
    \subfloat[]{\includegraphics[scale=0.6]{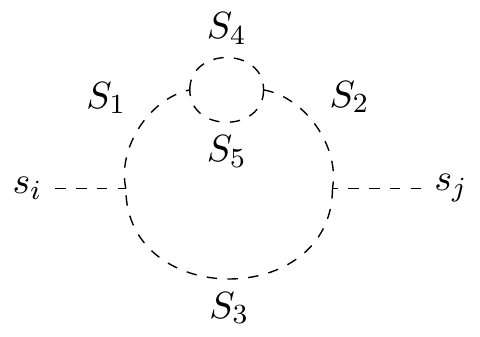}}
    \subfloat[]{\includegraphics[scale=0.6]{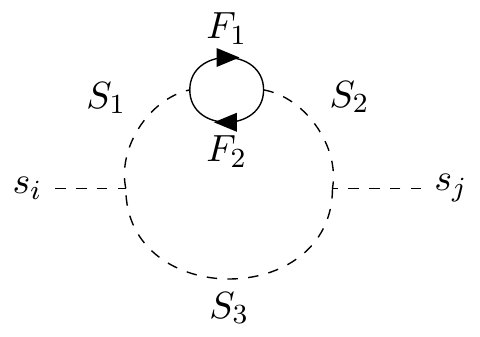}}
    \subfloat[]{\includegraphics[scale=0.6]{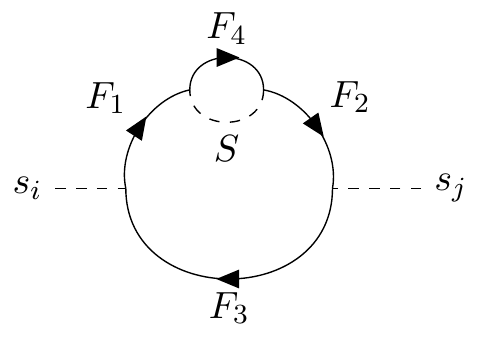}}
    \subfloat[]{\includegraphics[scale=0.6]{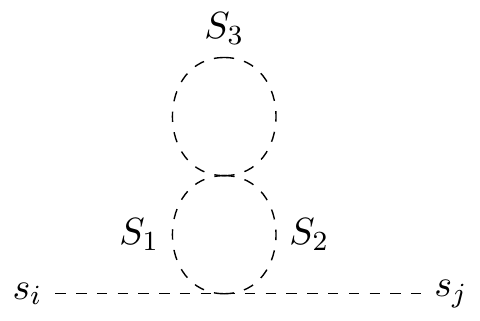}}
    \subfloat[]{\includegraphics[scale=0.6]{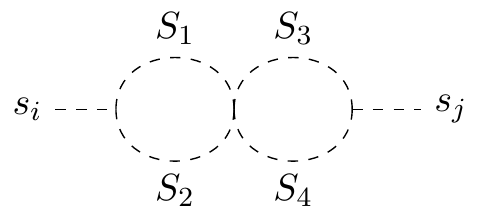}}\\
    \subfloat[]{\includegraphics[scale=0.6]{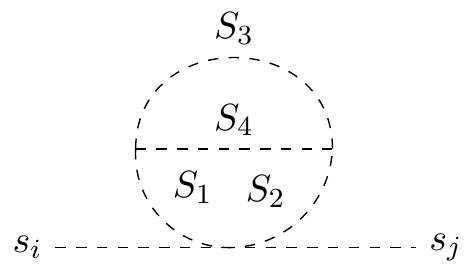}}
    \subfloat[]{\includegraphics[scale=0.6]{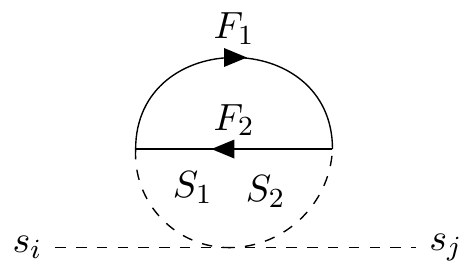}}
    \subfloat[]{\includegraphics[scale=0.6]{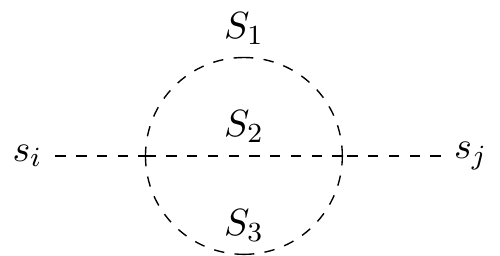}}
    \subfloat[]{\includegraphics[scale=0.6]{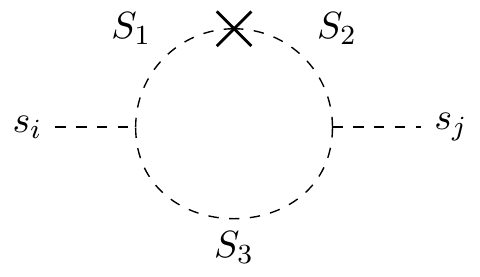}}
    \subfloat[]{\includegraphics[scale=0.6]{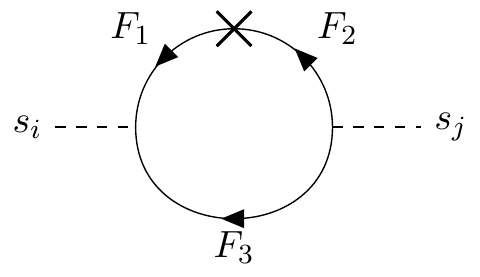}}\\
    \subfloat[]{\includegraphics[scale=0.6]{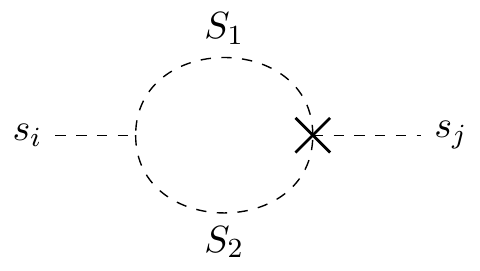}}
    \subfloat[]{\includegraphics[scale=0.6]{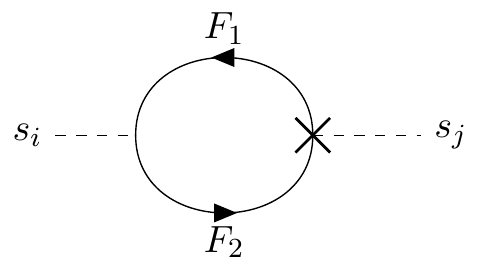}}
    \subfloat[]{\includegraphics[scale=0.6]{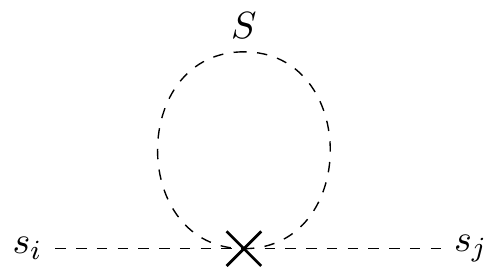}}
    \subfloat[]{\includegraphics[scale=0.6]{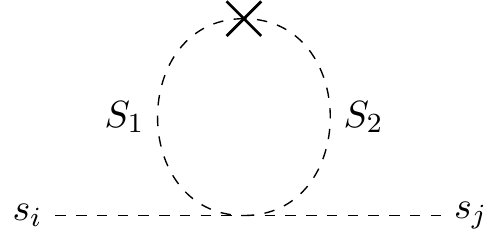}}
    \caption{All generic two-loop self-energy diagrams with internal
        scalars $S$ and fermions $F$. The external fields can be
        neutral/charged scalars as well as vector bosons (for the
        latter some diagrams such as (j) and (m) are not present due
        to gauge invariance and the $p^2=0$ approximation). The
        internal fields are selected according to
        \cref{eq:taddiagselect}.
    }
    \label{fig:allselfgeneric}
\end{figure}

In \cref{fig:allselfgeneric} we show all generic self-energy diagrams
appearing in the calculation of the
$\order{(\alpha_t+\alpha_\lambda+\alpha_\kappa)^2}$-corrected Higgs
boson masses. The external fields ($s_{i}$) can be the neutral
($h_{i})$ or charged ($h^-_{2}$) Higgs 
bosons as well as massive vector bosons ($Z$ or $W^-$). \s

The summation over the internal scalars and fermions running in the
loops is performed with the same selection principle as for the
tadpoles, \cref{eq:taddiagselect}.
The diagrams (b), (d), (p) and (q) pick up an additional factor
of two to account for their mirrored versions.

\subsection{Momentum Regulated Diagrams}
\begin{figure}[h]
    \centering
    \includegraphics[width=0.235\textwidth]{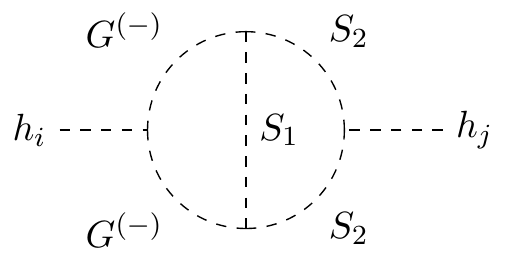}
    \includegraphics[width=0.235\textwidth]{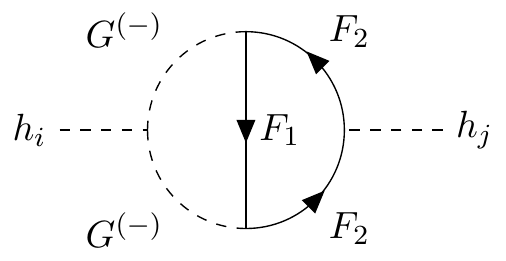}
    \includegraphics[width=0.235\textwidth]{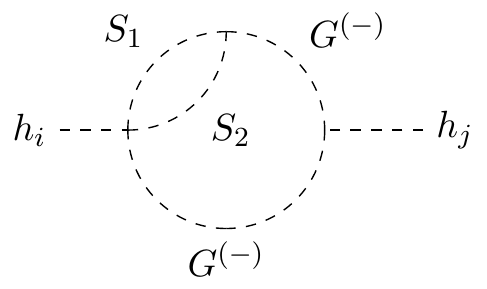}
    \includegraphics[width=0.235\textwidth]{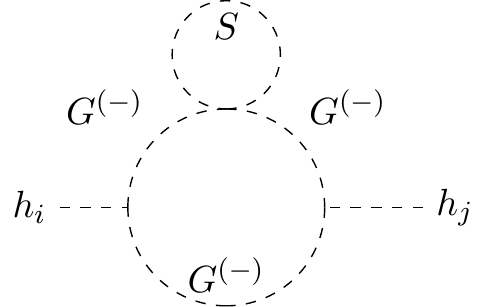}
    \includegraphics[width=0.235\textwidth]{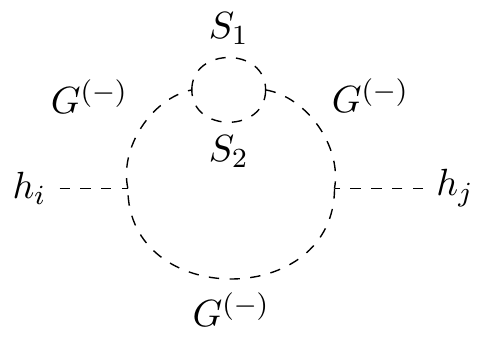}
    \includegraphics[width=0.235\textwidth]{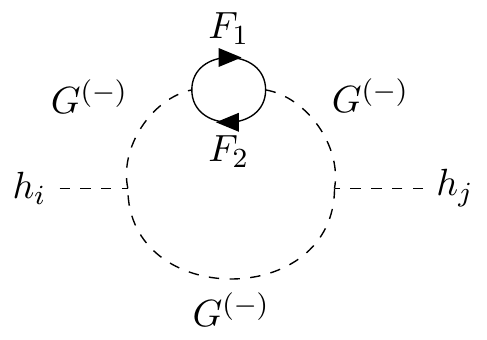}
    \includegraphics[width=0.235\textwidth]{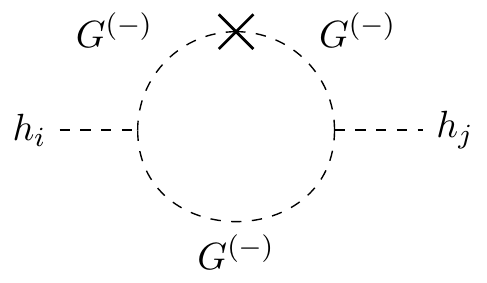}
    \includegraphics[width=0.235\textwidth]{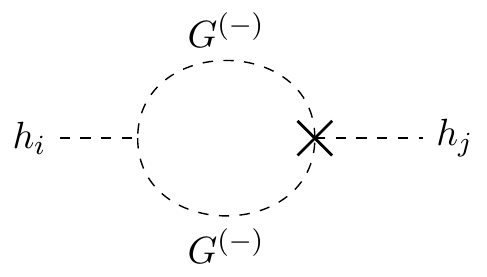}
    \includegraphics[width=0.71\textwidth]{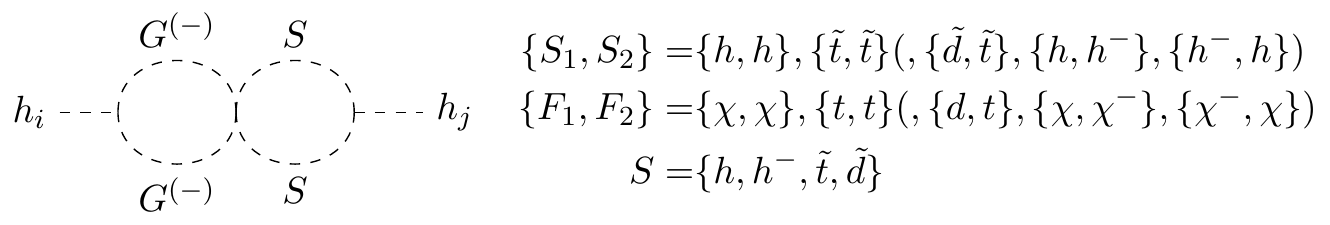}
    \caption{
        All two-loop self-energy diagrams with a residual
        dependence on the IR mass regulator, \textit{cf.}
        App.~\ref{app:irsafeloop}, \cref{tab:irtoposSELF}. Note that
        the first and the last diagrams 
        feature $\overline{\log}^2 M_R^2$ divergences if $S_2=G^{(-)}$,
        respectively $S=G^{(-)}$.
    }
    \label{fig:selfirfin}
\end{figure}

For the sake of simplicity, we also list the full set of Feynman
diagrams in \cref{fig:selfirfin} which feature a residual dependence
on the IR mass regulator ({\it i.e.}~the IR divergences in all other
diagrams do cancel). \cref{fig:selfirfin} can also be inferred from
the information given in \cref{tab:irtoposSELF}.

\end{appendix}


\bibliographystyle{ieeetr}

\end{document}